\documentclass[11pt,a4paper]{article}
\usepackage{jheppub}
\usepackage[T1]{fontenc}
\usepackage{amsmath,amssymb}
\usepackage{tikz}
\usepackage{tikz-cd}
\usepackage{ctable}

\newcommand{\tr}{{\operatorname{tr}}}

\title{\boldmath{Perturbations of General Relativity to All Orders and the General $n^{\rm th}$ Order Terms}}

\author[a]{Kyoungho Cho}
\author[b]{Kwangeon Kim}
\author[a,c]{Kanghoon Lee}

\affiliation[a]{Asia Pacific Center for Theoretical Physics, Postech, Pohang 37673, Korea}
\affiliation[b]{Department of Physics, Yonsei University, Seoul 03722, Korea}
\affiliation[c]{Department of Physics, Postech, Pohang 37673, Korea}

\emailAdd{kyoungho.cho@apctp.org}
\emailAdd{kim64656@yonsei.ac.kr}
\emailAdd{kanghoon.lee1@gmail.com}

\abstract{
We derive all-order expressions for perturbations of the Einstein-Hilbert action and the Einstein equation with the general $n$-th order terms. To this end, we employ Cheung and Remmen's perturbation conventions both in tensor density and the usual metric tensor formalisms, including the Einstein-dilaton theory. Remarkably, we find minimal building blocks that generate the entire perturbations for each of our formulations. We show that the number of terms of perturbations grows linearly as the order of perturbations increases. We regard our results as the reference and discuss how to derive perturbations in other conventions from the reference. As a consistency check, we compute graviton scattering amplitudes using the perturbiner method based on the perturbative Einstein equation. Finally we discuss how to generalise the results to curved backgrounds and incorporate additional matter.
}

\begin{document}
\preprint{APCTP Pre2022 - 021}
\maketitle

\section{Introduction}\label{Sec:1}

Perturbative general relativity (GR) is an indispensable tool for describing many gravitational phenomena. Various observations, such as gravitational waves \cite{LIGOScientific:2016aoc,LIGOScientific:2017bnn,LIGOScientific:2017vwq} and cosmology \cite{Planck:2018vyg,Planck:2018jri}, have confirmed that perturbative GR accurately describes gravity in a weak field regime. However, perturbative GR is highly complicated; perturbative expansions of the Einstein-Hilbert action and the Einstein equations generate too many terms to handle. On the other hand, despite the complexity of their perturbations, graviton scattering amplitudes have remarkably simple expressions (See \cite{Elvang:2013cua} and references therein). Thus one might deduce an alternative formulation that makes the perturbations simple.

Recently, Cheung and Remmen (CR) showed that perturbations of the Einstein-Hilbert (EH) action are greatly simplified using an alternative convention of metric perturbations with a specific gauge choice \cite{Cheung:2017kzx}. They introduced a tensor density as $\sigma^{\mu\nu} = \sqrt{-g} g^{\mu\nu}$ and recast the EH action in terms of this density. A crucial ingredient of the simplification is their convention of metric perturbations, which were defined from the inverse tensor density $\sigma^{\mu\nu}$ instead of $\sigma_{\mu\nu}$. In fact, this convention has a long history. It has been used in several contexts \cite{Deser:1969wk,Capper:1973pv,Deser:1987uk,Tomboulis:2017fim}, and recently it has been applied to calculations of gravitational waves and quantum gravity \cite{Cheung:2020sdj,Cheung:2020gyp,Kissler:2020rpf,Britto:2021pud}. However, despite the efficiency and utility of this convention, it has been applied to limited cases only, such as the perturbative expansion of EH action in the tensor density $\sigma$.

In this article we construct a general framework for studying the perturbative GR based on the CR convention and derive new all-order results, especially focusing on the equations of motion (EoM), the Einstein equation. Although the CR convention is defined for tensor density, we refer to perturbations based on the ``inverse quantity'' in general. We first derive all order expressions for perturbations of the EH action in terms of the tensor density $\sigma$ regardless of gauge choice and its associated EoM. We yield the general $n$-th order terms of the expansion, which one can hardly expect such exact all-order expressions due to the complicated structure of perturbative GR. Remarkably, we show that only three monomials generate the entire perturbations. Thus we treat them as building blocks of perturbative GR. We further consider perturbations of the Riemannian geometry using a metric $g$ rather than tensor density $\sigma$. We define linearised perturbations based on the inverse metric $g^{-1}$, similar to the tensor density case, and derive perturbations of the Ricci scalar and tensor to all orders. We then discuss how to generalise to the other perturbation conventions from our results that we regard as the \emph{reference convention}. 

Next, we generalise to the Einstein-dilaton theory by introducing a new tensor density that consists of the dilaton field. We also define a dilaton field inspired by the double field theory \cite{Siegel:1993xq,Hohm:2010jy,Hohm:2010pp} incorporating the measure $\sqrt{-g}$. Again, we employ the CR convention for the new tensor density and define perturbations of the dilaton. It is straightforward to derive perturbations of the ED theory to all orders in perturbations since the structure is almost identical to the pure GR case. We construct a new first-order formalism which is equivalent to the usual ED theory.

As a consistency check, we calculate graviton scattering amplitudes using the off-shell recursion relation \cite{Cheung:2016say, Cheung:2017kzx,Gomez:2021shh,Cheung:2021zvb,Cho:2021nim,Armstrong:2022jsa}. Here we employ the perturbiner method \cite{Rosly:1996vr,Rosly:1997ap,Selivanov:1997aq,Selivanov:1997an,Selivanov:1997ts}, which produces the off-shell recursions from EoM instead of the structure of Feynman vertices. The off-shell recursion relation determines the graviton off-shell currents directly related to the graviton scattering amplitudes. We solve the recursions explicitly and reproduce the known results up to five-point graviton scattering amplitudes. Since the perturbiner method relies on the form of the EoM, this gives a nontrivial check that our perturbations of the Einstein equations are correct.

Finally, we extend the results for which we assumed a flat background to arbitrarily curved backgrounds. This generalisation is achieved straightforwardly by replacing the ordinary derivative with covariant derivatives for the background metric and inserting the Ricci tensor appropriately. We also investigate additional matter fields, their energy-momentum tensors, and their equations of motion. We study three examples: a cosmological constant, a scalar field with potential and Maxwell theory. We also compare perturbations between the $\sigma$ and metric-based formulations.

\paragraph{Notation}~\\~
We use matrix notation to hide cumbersome dummy indices by embracing matrix multiplications in square brackets. Wherever necessary, we will restore the dummy indices. For arbitrary matrices $A^{\mu}{}_{\nu}$, $B^{\mu}{}_{\nu}$ and $C^{\mu}{}_{\nu}$, we denote their matrix product as
\begin{equation}
  A^{\mu}{}_{\rho} B^{\rho}{}_{\nu} = \big[A B\big]^{\mu}{}_{\nu}
  \qquad
  A^{\mu}{}_{\rho} B^{\rho}{}_{\sigma} C^{\sigma}{}_{\nu}= \big[A B C\big]^{\mu}{}_{\nu}\,,
\label{}\end{equation}
and the trace of matrices as
\begin{equation}
  A^{\mu}{}_{\rho} B^{\rho}{}_{\mu} = \tr\big[A B\big]\,,
  \qquad
  A^{\mu}{}_{\rho} B^{\rho}{}_{\sigma} C^{\sigma}{}_{\mu} = \tr\big[A B C\big]\,.
\label{}\end{equation}
For traces of a single element $A$, we denote 
\begin{equation}
  \tr\big[A\big] = A\,,
  \qquad
  \tr\big[\partial_{\mu} A\big] = \partial_{\mu}A\,,
  \qquad
  \tr\big[\partial_{\mu}\partial_{\nu} A\big] = \partial_{\mu}\partial_{\nu}A\,.
\label{}\end{equation}
We may insert partial derivatives into the square bracket; for instance,
\begin{equation}
  A^{\mu}{}_{\rho} \partial_{\sigma} B^{\rho}{}_{\nu} = \big[A \partial_{\sigma} B\big]^{\mu}{}_{\nu}\,,
  \qquad
  A^{\mu}{}_{\kappa} \partial_{\rho} B^{\kappa}{}_{\lambda} C^{\lambda}{}_{\nu} = \big[A \partial_{\rho} B C\big]^{\mu}{}_{\nu}\,.
\label{building_blocks}\end{equation}


\section{ Perturbations of Pure Gravity}\label{Sec:2}

In this section we will consider perturbations of pure gravity without matters in terms of two approaches: tensor density $\sigma$ and metric tensor $g$ formalisms of general relativity (GR). We define metric perturbations with respect to the inverse tensor density $\sigma^{-1}$ and the inverse metric $g^{-1}$, which is the opposite of the usual convention \cite{Cheung:2017kzx,Deser:1969wk,Capper:1973pv,Deser:1987uk,Tomboulis:2017fim}. We present perturbations of the Einstein-Hilbert action and the Einstein equation to all orders in fluctuations and the general $n$-th order terms. We show that three building blocks generate the entire perturbations. We further discuss how to derive other perturbation conventions from our results.

\subsection{Tensor density formalism}\label{Sec:2.1}

Let us start by recasting the Einstein-Hilbert (EH) action in $D$-dimensional spacetime to a useful form after integration by parts
\begin{equation}
\begin{aligned}
  S_{\rm EH}
  =
  \frac{1}{2\kappa^{2}} \int\mathrm{d}^{D} x e^{-2\hat{d}} \bigg[&\
    \frac{1}{4} g^{\mu\nu} \partial_{\mu} g^{\rho\sigma} \partial_{\nu} g_{\rho\sigma}
  - \frac{1}{2} g^{\mu\nu} \partial_{\mu} g^{\rho\sigma} \partial_{\rho} g_{\nu\sigma}
  + 2 g^{\mu\nu} \partial_{\mu}\partial_{\nu} \hat{d}
  \\&
  - e^{2\hat{d}} \partial_{\mu}\partial_{\nu}\big(e^{-2\hat{d}}g^{\mu\nu}\big)\
  \bigg]\,,
\end{aligned}\label{EH_action_metric}
\end{equation}
where $\kappa^{2} = 8\pi G$, $G$ is the Newton constant. Here we denote the integration measure $\sqrt{-g}$ to $e^{-2\hat{d}}$, which is more convenient to handle than the square root of the determinant,
\begin{equation}
  e^{-2\hat{d}} = \sqrt{-g} \,, \qquad \text{or} \qquad  \hat{d}= - \frac{1}{4} \ln |\det g| \,.
\label{DFT_dilaton}\end{equation}
The partial derivative of $\hat{d}$ is given by
\begin{equation}
  \partial_{\mu}\hat{d} = - \frac{1}{4} g^{\rho\sigma} \partial_{\mu} g_{\rho\sigma}\,.
\label{DFT_dilaton2}\end{equation}

We now introduce a tensor density $\sigma$ by absorbing the overall factor $\sqrt{-g}$ into the metric
\begin{equation}
  \sigma^{\mu\nu} = \sqrt{-g} g^{\mu\nu }= e^{-2\hat{d}} g^{\mu\nu}\,,
  \qquad
  \sigma_{\mu\nu} = \frac{1}{\sqrt{-g}} g_{\mu\nu}= e^{2\hat{d}} g_{\mu\nu}\,,
\label{def_sigma1}\end{equation}
and $\sigma^{\mu\nu}$ is the inverse of $\sigma_{\mu\nu}$. Substituting the field redefinition into the action \eqref{EH_action_metric}, we have
\begin{equation}
\begin{aligned}
  S_{\rm EH} = \int \mathrm{d}^{D} x \bigg[&\
  	  \frac{1}{4} \sigma^{\mu\nu} \partial_{\mu} \sigma^{\rho\sigma} \partial_{\nu} \sigma_{\rho\sigma}
  	- \frac{1}{2} \sigma^{\mu\nu} \partial_{\mu} \sigma^{\rho\sigma} \partial_{\rho} \sigma_{\nu\sigma}
   	+ (D-2) \sigma^{\mu\nu}\partial_{\mu}\hat{d} \partial_{\nu}\hat{d}
 	\\
  	&
  	+ \partial_{\mu} \Big( 2\sigma^{\mu\nu} \partial_{\nu} \hat{d} - \partial_{\nu}\sigma^{\mu\nu}\Big)
  	\bigg]\,.
\end{aligned}\label{EH_action_sigma}
\end{equation}
where the last term is a total derivative that we ignore. We may rewrite $\partial_{\mu}\hat{d}$ in \eqref{DFT_dilaton2} using the tensor densities
\begin{equation}
  \partial_{\mu} \hat{d} = \frac{1}{2(D-2)} \sigma^{\rho\sigma} \partial_{\mu} \sigma_{\rho\sigma}\,.
\label{def_d}\end{equation}

Note that we may have an even simpler action by replacing the $\partial_{\mu}\hat{d}$ to an auxiliary field $d_{\mu}$ 
\begin{equation}
\begin{aligned}
  \tilde{S}_{\rm EH} = \int \mathrm{d}^{D} x \bigg[&\
  	  \frac{1}{4} \sigma^{\mu\nu} \partial_{\mu} \sigma^{\rho\sigma} \partial_{\nu} \sigma_{\rho\sigma}
  	- \frac{1}{2} \sigma^{\mu\nu} \partial_{\mu} \sigma^{\rho\sigma} \partial_{\rho} \sigma_{\nu\sigma}
   	- (D-2) \sigma^{\mu\nu} d_{\mu} d_{\nu} 
   	\\&
   	+\sigma^{\mu\nu} \big(\sigma^{\rho\sigma}\partial_{\mu}\sigma_{\rho\sigma}\big) d_{\nu}
   	\bigg]\,.
\end{aligned}\label{sigma_action_dmu}
\end{equation}
It is straightforward to show that the on-shell value of $d_{\mu}$ is $\partial_{\mu}\hat{d}$. If we substitute the on-shell value into the above action $\tilde{S}_{\rm EH}$, the action reduces to $S_{\rm EH}$ \eqref{EH_action_sigma}. As we will see in section \ref{Sec:3}, this action is closely related to the Einstein-dilaton theory.

We now derive the EoM by varying the action \eqref{EH_action_sigma} with respect to $\sigma$. There are two options in choosing the variables that we are varying: $\sigma_{\mu\nu}$ and $\sigma^{\mu\nu}$. As we will see in \eqref{perturb_sigma}, we define graviton perturbations using $\sigma^{\mu\nu}$ rather than $\sigma_{\mu\nu}$. In this convention, perturbations of $\sigma^{\mu\nu}$ are immensely simpler than perturbations of $\sigma_{\mu\nu}$, since $\sigma_{\mu\nu}$ carries an infinite expansion. Thus we have to reduce the numbers of $\sigma_{\mu\nu}$ as much as possible to obtain simpler results. To this end, we variate the action with respect to $\sigma_{\mu\nu}$, which is opposite to the usual convention\footnote{The minus signature in \eqref{def_Gmunu} arise from the relation $\delta \sigma^{\mu\nu} \sigma_{\mu\nu} = - \sigma^{\mu\nu} \delta \sigma_{\mu\nu}$. Then $\mathcal{G}^{\mu\nu}$ is related to $\mathcal{G}_{\mu\nu}$ by $\mathcal{G}^{\mu\nu} = \sigma^{\mu\rho} \sigma^{\nu\sigma} \mathcal{G}_{\rho\sigma}$, where
\begin{equation*}
\begin{aligned}
  \mathcal{G}_{\mu\nu}=&\ \frac{1}{4} \partial_{\mu} \sigma^{\rho\sigma} \partial_{\nu} \sigma_{\rho\sigma} 
  - \frac{1}{2} \sigma_{\mu\kappa} \sigma_{\nu\lambda}\partial_{\sigma}\sigma^{\rho\kappa} \partial_{\rho} \sigma^{\sigma\lambda} 
  - \frac{1}{2} \sigma^{\rho\sigma}\sigma^{\kappa\lambda} \partial_{\rho} \sigma_{\kappa(\mu} \partial_{|\sigma|} \sigma_{\nu)\lambda}
  - \frac{1}{2} \partial_{\rho}\sigma^{\rho\sigma} \big( \partial_{\sigma} \sigma_{\mu\nu} -2 \partial_{(\mu}\sigma_{\nu)\sigma}\big)
  \\&
  - \frac{1}{2} \sigma^{\rho\sigma} \big(\partial_{\rho}\partial_{\sigma} \sigma_{\mu\nu} -2\partial_{\rho}\partial_{(\mu}\sigma_{\nu)\sigma}\big)
  + \sigma_{\mu\nu}  \partial_{\rho}\big(
  	 \sigma^{\rho\sigma}\partial_{\sigma}\hat{d} \big)
  + (D-2) \partial_{\mu}\hat{d} \partial_{\nu}\hat{d}= 0 \,.
\end{aligned}\label{}
\end{equation*}
}
\begin{equation}
  \delta S_{\rm EH} = \int \mathrm{d}^{D}x\ \Big[ - \delta \sigma_{\mu\nu} \mathcal{G}^{\mu\nu} \Big]\,,
\label{def_Gmunu}\end{equation}
where 
\begin{equation}
\begin{aligned}
  \mathcal{G}^{\mu\nu}
  =&\
    \frac{1}{2} \sigma^{\rho\sigma} \Big[\partial_{\rho}\partial_{\sigma}\sigma^{\mu\nu}
  + \partial_{\rho}\sigma^{\kappa\mu} \partial_{\sigma}\sigma_{\kappa\lambda} \sigma^{\nu\lambda} \Big]
  - \sigma^{\rho(\mu} \Big[
    \partial_{\rho}\partial_{\sigma}\sigma^{\nu)\sigma}
  + \partial_{\rho} \sigma^{|\kappa\lambda} \partial_{\kappa}\sigma_{\lambda\sigma} \sigma^{\sigma|\nu)}
  \Big]
  \\&
  + \sigma^{\mu\kappa}\sigma^{\nu\lambda} \bigg[\
    \frac{1}{4} \partial_{\kappa}\sigma^{\rho\sigma}\partial_{\lambda}\sigma_{\rho\sigma}
  + (D-2) \partial_{\kappa} \hat{d} \partial_{\lambda} \hat{d}\
  \bigg]
  \\&
  + \frac{1}{2} \Big[
  	  \partial_{\rho}\sigma^{\rho\sigma} \partial_{\sigma}\sigma^{\mu\nu}
  	- \partial_{\sigma}\sigma^{\rho\mu} \partial_{\rho}\sigma^{\sigma\nu}
  \Big]
  + \sigma^{\mu\nu} \Big[\partial_{\kappa} \big(\sigma^{\kappa\lambda} \partial_{\lambda}\hat{d} \big) \Big]
  \,.
\end{aligned}\label{Gmunu}
\end{equation}
Thus $\mathcal{G}^{\mu\nu}= 0$ is the EoM for $\sigma$.

\subsection{Perturbations of EH action and Einstein equation}

There are many ways to represent metric perturbations depending on physical motivations or computational advantages. The conventional choice of linear fluctuations around a trivial background $\eta_{\mu\nu}$ is given by $g_{\mu\nu} = \eta_{\mu\nu} + h_{\mu\nu}$, where $h_{\mu\nu}$ are small fluctuations, $|h_{\mu\nu}| \ll 1$. Here we adopt an opposite convention defined based on $\sigma^{\mu\nu}$ rather than $\sigma_{\mu\nu}$, which is recently studied by Cheung and Remmen \cite{Cheung:2017kzx}
\begin{equation}
  \sigma^{\mu\nu} = \eta^{\mu\nu} -h^{\mu\nu}\,,
  \qquad
  \sigma_{\mu\nu} = \eta_{\mu\nu} + \sum_{n=1}^{\infty} (h^{n})_{\mu\nu} \,.
\label{perturb_sigma}\end{equation}
 As usual, the indices are raised and lowered by the flat background metric $\eta^{\mu\nu}$ and $\eta_{\mu\nu}$, respectively. However, the indices of $\sigma^{\mu\nu}$ are not raised by $\eta^{\mu\nu}$. We set this expansion as our reference convention. We will discuss in section \ref{Sec:2.3} how other conventions are generated by a mapping from the reference.

One crucial observation from the structure of \eqref{EH_action_sigma} and \eqref{Gmunu} is that the entire set of perturbations is generated by the following three terms only:
\begin{equation}
\begin{aligned}
  \partial_{\mu} \sigma^{\kappa \rho} \partial_{\nu} \sigma_{\rho\sigma} \sigma^{\sigma\lambda}\,,
  \qquad
  \partial_{\mu} \sigma^{\rho\sigma} \partial_{\nu} \sigma_{\rho\sigma}\,,
  \qquad
  \partial_{\mu}\hat{d} = \frac{1}{2(D-2)}\sigma^{\rho\sigma} \partial_{\mu} \sigma_{\rho\sigma} \,.
\end{aligned}\label{building_blocks0}
\end{equation}
Furthermore, their perturbations are simple enough to present all-order expressions, which are our fundamental relations,
\begin{equation}
\begin{aligned}
  \partial_{\mu} \sigma^{\kappa \rho} \partial_{\nu} \sigma_{\rho\sigma} \sigma^{\sigma\lambda}
  &= - \sum_{n=0}^{\infty} \Big[(\partial_{\mu}h) h^{n} (\partial_{\nu} h)\Big]^{\kappa\lambda} \,,
  \\
  \partial_{\mu} \sigma^{\rho\sigma} \partial_{\nu}\sigma_{\rho\sigma} &= - \sum_{n=0}^{\infty}\sum_{m=0}^{n} \tr\big[ h^{m} (\partial_{\mu}h) h^{n-m} (\partial_{\nu} h)\big] \,,
  \\
  2 (D-2) \partial_{\mu}\hat{d}&=\sigma^{\rho\sigma}\partial_{\mu}\sigma_{\rho\sigma} = \sum_{n=0}^{\infty} \tr\big[ h^{n} \partial_{\mu} h\big] \,.
\end{aligned}\label{building_blocks}
\end{equation}

Based on the results in the previous subsection, we now construct perturbations of the EH action and EoM to all orders in $h$ around a trivial background $\eta_{\mu\nu}$. For simplifying the calculation, we introduce a set of abbreviations $X^{\mu\nu}_{\rho\sigma}$, $Y_{\mu\nu}$, $Z^{\mu\nu}$ and $W$ in \eqref{EH_action_aux} according to the structure of indices. See appendix \ref{App:A.1} for the detailed derivation. Here we denote the results only.

First, we consider perturbations of the EH action \eqref{EH_action_sigma}. Substituting the perturbations of  the abbreviations in \eqref{pert_auxiliary1} into the Lagrangian \eqref{Lag_aux}, we obtain
\begin{equation}
\begin{aligned}
  L_{\rm EH} =&
  \sum_{n=0}^{\infty} \bigg[
    - \frac{1}{4} (\eta-h)^{\mu\nu} \sum_{m=0}^{n} \tr\big[ h^{m} \partial_{\mu}h h^{n-m} \partial_{\nu} h\big]
  	+ \frac{1}{2} \big[\partial_{\nu}h h^{n} \partial_{\mu}h\big]^{\mu\nu}
  \\&\qquad\
  	+ \frac{1}{4(D-2)}(\eta-h)^{\mu\nu} \sum_{m=0}^{n} \tr\big[h^{m}\partial_{\mu}h\big] \tr\big[h^{n-m}\partial_{\nu}h\big]
  \bigg] \,.
\end{aligned}\label{pert_action_1}
\end{equation}
One can find their explicit general $n$-th order terms in \eqref{n_th_order_Lag_sigma}. We also present perturbations of $\mathcal{G}^{\mu\nu}$ to all orders in $h$
\begin{equation}
\begin{aligned}
  \mathcal{G}^{\mu\nu} &=
    (\eta-h)^{\rho(\mu} \bigg[
  	\partial_{\rho}\partial_{\sigma}h^{\nu)\sigma}
  	+ \sum_{n=0}^{\infty} \big[\partial_{\rho} h h^{n} \partial_{\sigma}h\big]^{\sigma|\nu)}
  \bigg]
  \\&\quad
  - \frac{1}{2} (\eta-h)^{\rho\sigma} \bigg[\partial_{\rho}\partial_{\sigma}h^{\mu\nu} +\sum_{n=0}^{\infty} \big[\partial_{\rho}h h^{n} \partial_{\sigma} h\big]^{(\mu\nu)}\bigg]
  + \frac{1}{2} \partial_{\rho}h^{\rho\sigma} \partial_{\sigma}h^{\mu\nu}
  - \frac{1}{2} \partial_{\sigma}h^{\rho\mu} \partial_{\rho}h^{\sigma\nu}
  \\&\quad
  + \frac{1}{4}(\eta-h)^{\mu\rho} (\eta-h)^{\nu\sigma}
  \sum_{n=0}^{\infty} \sum_{m=0}^{n}\Bigg[
  	  \frac{\tr\big[h^{m} \partial_{\rho} h\big] \tr\big[h^{n-m} \partial_{\sigma} h\big]}{D-2}
  	{-} \tr\big[h^{m}\partial_{\rho}h h^{n-m} \partial_{\sigma} h\big]
  \Bigg]
  \\&\quad
  + \frac{1}{2(D-2)}(\eta-h)^{\mu\nu} \sum_{n=0}^{\infty} \bigg[
  	  (\eta-h)^{\rho\sigma} \Big(\tr\big[h^{n}\partial_{\rho}\partial_{\sigma}h\big]
  	  + \sum_{m=0}^{n}\tr\big[h^{m}\partial_{\rho}h h^{n-m} \partial_{\sigma}h\big]\Big)
  \\&\qquad\qquad\qquad\qquad\qquad\qquad\quad
  	- \partial_{\rho} h^{\rho\sigma} \tr\big[h^{n}\partial_{\sigma}h\big]\
  \bigg]\,.
\end{aligned}\label{perturb_G}
\end{equation}
The general $n$-th order terms of $\mathcal{G}^{\mu\nu}$ in \eqref{n_th_order_Gmunu1} and \eqref{n_th_order_Gmunu2}. We remark that the number of terms of the perturbed action and $\mathcal{G}^{\mu\nu}$ \eqref{perturb_G} grow linearly as the order of perturbations increases. 

We may also represent the above expansion in terms of a new abbreviation $A_{\mu\nu}{}^{\rho}$ \eqref{Q_field} and its trace $A_{\mu}$ up to a total derivative
\begin{equation}
  L_{\rm EH} =
 (\eta-h)^{\mu\nu} \bigg(\frac{1}{4(D-2)} A_{\mu} A_{\nu}
  - \frac{1}{4} A_{\mu\rho}{}^{\sigma} A_{\nu\sigma}{}^{\rho}\bigg)
  + \frac{1}{2} A_{\mu\rho}{}^{\nu} \partial_{\nu}h^{\mu\rho} \,.
\label{}\end{equation}
This Lagrangian is reminiscent of the first-order formalism introduced by Cheung and Remmen \cite{Cheung:2017kzx}. If we replace the factor $\frac{1}{D-2}$ with $\frac{1}{D-1}$ and treat $A_{\mu\nu}{}^{\rho}$ as an auxiliary field, the Lagrangian reduces to the first order action eq. (18) in \cite{Cheung:2017kzx}. However, this Lagrangian cannot be a first-order formalism. It is straightforward to rewrite $\mathcal{G}^{\mu\nu}$ using $A_{\mu\nu}{}^{\rho}$
\begin{equation}
\begin{aligned}
  \mathcal{G}^{\mu\nu} &= 
  (\eta-h)^{\rho(\mu} \Big(\partial_{\rho}\partial_{\sigma}h^{\nu)\sigma} + A_{\sigma\kappa}{}^{\nu)} \partial_{\rho}h^{\sigma\kappa} \Big) 
  -\frac{1}{2} (\eta-h)^{\rho\sigma} \Big(\partial_{\rho}\partial_{\sigma}h^{\mu\nu} + \partial_{\rho}h^{\mu\kappa}A_{\sigma\kappa}{}^{\nu}\Big) 
  \\&\quad
  + \frac{1}{4} (\eta-h)^{\mu\rho} (\eta-h)^{\mu\sigma} \bigg( \frac{1}{D-2} A_{\rho} A_{\sigma} - A_{\rho\kappa}{}^{\lambda} A_{\sigma\lambda}{}^{\kappa}\bigg)
  \\&\quad
  + \frac{1}{2} \partial_{\rho}h^{\rho\sigma} \partial_{\sigma}h^{\mu\nu}
  - \frac{1}{2} \partial_{\sigma}h^{\rho\mu} \partial_{\rho}h^{\sigma\nu}
  +\frac{1}{2(D-2)} (\eta-h)^{\mu\nu} \partial_{\rho} \Big((\eta-h)^{\rho\sigma} A_{\sigma}\Big)\,.
\end{aligned}\label{}
\end{equation}

As a final remark, we check that the perturbation of $\mathcal{G}^{\mu\nu}$ is consistent with the nonlinear self-interacting energy-momentum tensor in \cite{Tomboulis:2017fim} for the case of $D=4$ case. Their EoM, which is obtained by Belinfante's symmetrisation method, is related to ours by $G^{\mu\nu} - \frac{1}{2} \sigma^{\mu\nu} \sigma_{\rho\sigma} G^{\rho\sigma}=0$. Thus if we substitute the perturbations of the building blocks \eqref{building_blocks} into their EoM and subtract the trace part, one can reproduce \eqref{perturb_G}.

\subsection{Perturbations of Ricci curvatures}

So far, we have considered perturbative GR in terms of the tensor density $\sigma^{\mu\nu}$. However, many GR problems suggest their preferred metric perturbations depending on physical motivations. Thus it would be helpful to derive perturbations of Ricci tensor and scalar using perturbations of metric $g$ directly rather than $\sigma$, which is ambiguous physically.

We now construct perturbations of pure gravity in terms of metric tensor within the Riemannian geometry. Ricci curvatures govern the dynamics of GR in Riemannian geometry. As we discussed in the previous section, we consider the Ricci tensor with upper indices $R^{\mu\nu} = g^{\mu\rho}g^{\nu\sigma} R_{\rho\sigma}$ because our convention of metric perturbations based on the inverse metric as in the tensor density case. The explicit form of $R^{\mu\nu}$ is given by
\begin{equation}
\begin{aligned}
  R^{\mu\nu} =&
    \frac{1}{2} g^{\rho\sigma} \bigg[\partial_{\rho}\partial_{\sigma} g^{\mu\nu}
    - 2 \partial_{\rho} g^{\mu\nu} \partial_{\sigma}\hat{d}
    + g^{\lambda(\mu} \partial_{\rho}g^{\nu)\kappa} \partial_{\sigma}g_{\kappa\lambda}
    	\bigg]
  \\&
  - g^{\rho(\mu} \bigg[
  	  \partial_{\rho} \partial_{\sigma}g^{\nu)\sigma}
  	- 2\partial_{\rho} g^{\nu)\sigma} \partial_{\sigma}\hat{d}
  	+ g^{\nu)\lambda}\partial_{\rho} g^{\sigma\kappa} \partial_{\sigma} g_{\kappa\lambda}
  \bigg]
  \\&
  + g^{\mu\kappa} g^{\nu\lambda} \bigg[
  	  \frac{1}{4} \partial_{\kappa}g^{\rho\sigma} \partial_{\lambda} g_{\rho\sigma}
  	+ 2 \partial_{\kappa}\partial_{\lambda} \hat{d}
  \bigg]
  + \frac{1}{2} \Big[\partial_{\rho}g^{\rho\sigma} \partial_{\sigma} g^{\mu\nu}
  -  \partial_{\rho} g^{\sigma\mu} \partial_{\sigma} g^{\rho\nu}\Big]
  \,,
\end{aligned}\label{Ricci_tensor_g}
\end{equation}
where $\hat{d}$ is defined in \eqref{DFT_dilaton} and \eqref{DFT_dilaton2}. Similarly, the Ricci scalar $R$ is the trace of the Ricci tensor $R = g^{\mu\nu} R_{\mu\nu}$
\begin{equation}
\begin{aligned}
  R = \frac{1}{4} g^{\mu\nu} \partial_{\mu} g^{\rho\sigma} \partial_{\nu} g_{\rho\sigma}
  - \frac{1}{2} g^{\mu\nu} \partial_\mu g^{\rho\sigma} \partial_\rho g_{\nu\sigma}
  + 2g^{\mu\nu} \partial_{\mu} \partial_{\nu} \hat{d}
  - e^{2\hat{d}}\partial_\mu \partial_{\nu} \Big(e^{-2\hat{d}}g^{\mu\nu}\Big)\,.
\end{aligned}\label{Ricci_scalar_g}
\end{equation}

We now derive perturbations of the Ricci curvatures around a flat metric $\eta_{\mu\nu}$. As for the tensor density case, we define metric perturbations based on the inverse metric
\begin{equation}
  g^{\mu\nu} = \eta^{\mu\nu} - h^{\mu\nu}\,,
  \qquad
  g_{\mu\nu} = \eta_{\mu\nu} + \sum_{n=1}^{\infty} (h^{n})_{\mu\nu}\,.
\label{metric_perturb}\end{equation}
Note that although we use the same notation ``$h$'' as the fluctuations of $\sigma^{\mu\nu}$, these are different quantities. Since the structures of perturbations of $\sigma$ and $g$ are the same, the following three terms are still building blocks of perturbations
\begin{equation}
  \partial_{\mu}g^{\kappa \rho} \partial_{\nu} g_{\rho\sigma} g^{\sigma\lambda}\,,
  \qquad
  \partial_{\mu} g^{\rho\sigma} \partial_{\nu}g_{\rho\sigma}\,,
  \qquad
  \partial_{\mu}\hat{d} \,.
\label{building_blocks_g}\end{equation}
and their expansions are the same as for the tensor density formalism \eqref{building_blocks}
\begin{equation}
\begin{aligned}
  \partial_{\mu}g^{\kappa \rho} \partial_{\nu} g_{\rho\sigma} g^{\sigma\lambda}
  &= - \sum_{n=0}^{\infty} \Big[(\partial_{\mu}h) h^{n} (\partial_{\nu} h)\Big]^{\kappa\lambda}\,,
  \\
  \partial_{\mu} g^{\rho\sigma} \partial_{\nu}g_{\rho\sigma} &= - \sum_{n=0}^{\infty}\sum_{m=0}^{n} \tr\big[ h^{m} (\partial_{\mu}h) h^{n-m} (\partial_{\nu} h)\big]\,,
  \\
  \partial_{\mu}\hat{d} &= -\frac{1}{4} \sum_{n=0}^{\infty} \tr\big[ h^{n} \partial_{\mu} h\big] \,.
\end{aligned}\label{building_blocks_g_pertub}
\end{equation}

Similar to the tensor density formalism, we define a set of abbreviations $\check{X}^{\mu\nu}_{\rho\sigma}$, $\check{Y}_{\mu\nu}$, and $\check{Z}^{\mu\nu}$ denoted in \eqref{abbreviation_g}. Note that the three fields are enough for the metric formulation. Substituting their perturbative expansions in \eqref{perturb_auxiliary_g} into \eqref{Ricci_tensor_aux} and \eqref{Ricci_scalar_aux}, we may derive perturbations of the Ricci scalar $R$ to all orders in $h$
\begin{equation}
\begin{aligned}
  R =&\
    \partial_{\mu} \partial_{\nu} h^{\mu \nu}{}_{}
  + \sum_{n=0}^{\infty}\bigg[\ \frac{1}{2} \big[\partial_{\nu} h h^{n} \partial_{\mu}h\big]^{\mu\nu}
  + \partial_\mu h^{\mu\nu} \operatorname{tr}[h^{n} \partial_\nu h]
  - (\eta -h)^{\mu\nu} \tr\big[h^{n} \partial_{\mu}\partial_{\nu} h \big]
  \bigg]
  \\&
  - \frac{1}{4} \sum_{n=0}^{\infty}\sum_{m=0}^{n} \bigg[\ (\eta - h)^{\mu\nu} \Big(
  	  5 \tr\big[h^m \partial_{\mu} h h^{n-m} \partial_{\nu} h \big]
  	+ \tr[h^{m} \partial_\mu h] \tr[h^{n-m} \partial_{\nu}h]
  \Big)
  \bigg]\,.
\end{aligned}\label{perturb_Ricci_Scalar}
\end{equation}
We may rearrange the expansion of $R$ order by order in $h$ and derive the general term of the perturbative expansion
\begin{equation}
  R = \sum_{n=1}^{\infty} R_{n}\,,
\label{expansion_R}\end{equation}
where
\begin{equation}
\begin{aligned}
  R_{1} =&\
  \partial_{\mu} \partial_{\nu} h^{\mu \nu} - \square h \,,
\end{aligned}\label{}
\end{equation}
and we have the general form for $n\geq 2$
\begin{equation}
\begin{aligned}
  R_{n} =&\
    \frac{1}{2} \big[\partial_{\nu} h h^{n-2} \partial_{\mu}h\big]^{\mu\nu}
  + \partial_\mu h^{\mu\nu} \operatorname{tr}[h^{n-2} \partial_\nu h]
  + \tr\big[h^{n-1} \Box h \big]
  + h^{\mu\nu} \tr\big[h^{n-2} \partial_\mu \partial_\nu h \big]
  \\&
  - \sum_{p=0}^{n-2} \bigg(
  	  \frac{5}{4} \tr\big[h^p \partial^{\mu} h h^{n-p-2} \partial_{\mu} h \big]
  	+ \frac{1}{4} \tr[h^{p} \partial^\mu h] \tr[h^{n-p-2} \partial_\mu h]
  \bigg)
  \\&
  + \sum_{p=0}^{n-3} h^{\mu \nu} \bigg(
    \frac{5}{4}\tr \big[h^p \partial_{\mu} h h^{n-p-3} \partial_\nu h \big]
  + \frac{1}{4} \tr[h^{p} \partial_\mu h] \tr[h^{n-p-3} \partial_{\nu}h]
  \bigg) \,.
\end{aligned}\label{}
\end{equation}

We next turn to perturbations of the Ricci tensor $R^{\mu\nu}$ defined in \eqref{Ricci_tensor_aux}. Using the relations in \eqref{perturb_auxiliary_g}, we have all order expressions for $R^{\mu\nu}$  
\begin{equation}
\begin{aligned}
  R^{\mu\nu} =&
  - \frac{1}{2} (\eta-h)^{\rho\sigma} \bigg[ \partial_{\rho}\partial_{\sigma} h^{\mu\nu}
  +\sum_{n=0}^{\infty}\Big(\big[\partial_{\rho}h h^{n} \partial_{\sigma}h\big]^{(\mu\nu)}
  	 + \frac{1}{2} \partial_{\sigma}h^{\mu\nu} \tr\big[h^{n}\partial_{\rho}h\big]\Big)\bigg]
  \\&
  + (\eta-h)^{\rho(\mu} \bigg[ \partial_{\rho}\partial_{\sigma}h^{\nu)\sigma}+\sum_{n=0}^{\infty} \Big(\big[\partial_{\rho}h h^{n} \partial_{\sigma}h\big]^{|\sigma|\nu)}
  	 + \frac{1}{2} \partial_{\sigma}h^{|\sigma|\nu)} \tr\big[h^{n}\partial_{\rho}h\big]\Big) \bigg]
  \\&
  - (\eta-h)^{\mu\rho}(\eta-h)^{\nu\sigma} \sum_{n=0}^{\infty} \bigg[
  	 \frac{1}{2} \tr\big[h^{n}\partial_{\rho}\partial_{\sigma}h\big] + \frac{3}{4} \sum_{m=0}^{n} \tr\big[h^{m}\partial_{\rho} h h^{n-m} \partial_{\sigma}h\big]
  \bigg]
  \\&
    + \frac{1}{2} \partial_{\rho} h^{\rho\sigma} \partial_{\sigma}h^{\mu\nu}
  - \frac{1}{2} \partial_{\rho} h^{\sigma\mu} \partial_{\sigma}h^{\rho\nu}
\,.
\end{aligned}\label{perturb_Ricci_Tensor}
\end{equation}
One can find the general $n$-th order terms of $R^{\mu\nu}$ in \eqref{general_n_Rmunu1} and \eqref{general_n_Rmunu2}. 

Since the Einstein equation, $G^{\mu\nu} = R^{\mu\nu} - \frac{1}{2} g^{\mu\nu} R = 0$, which is given by the Ricci scalar and tensor, we can easily represent the perturbations of Einstein equation by combining the above results \eqref{perturb_Ricci_Scalar} and \eqref{perturb_Ricci_Tensor}.

\subsection{Other conventions of metric perturbations}\label{Sec:2.3}

So far we have constructed perturbative GR using the reference convention \eqref{perturb_sigma}. However, some physical problems require specific forms of metric perturbations. Here we discuss how to derive perturbations in a different convention from the results in the reference convention. We investigate the construction from two examples: ordinary perturbations and exponential perturbations.

First we remark that the fluctuations of $\sigma^{\mu\nu}$ are given by a single term $h^{\mu\nu}$ in the reference convention. Thus, if we denote the graviton fluctuations in the other conventions as $h'^{\mu\nu}$, it is straightforward to find a map between $h$ and $h'$ by comparing $\sigma^{\mu\nu}$ for each case, 
\begin{equation}
  h^{\mu\nu} = h^{\mu\nu}[h']\,.
\label{}\end{equation}
One can generate perturbations in other conventions by substituting into this map the previous results in the reference convention. 

\paragraph{Usual metric perturbations}~\\
The usual metric perturbations are given by
\begin{equation}
  \sigma^{\mu\nu} = \eta^{\mu\nu} - \sum_{n=1}^{\infty} (-1)^{n+1} (h'^{n})^{\mu\nu}\,,
\label{}\end{equation}
and we have the mapping
\begin{equation}
  h^{\mu\nu} = h'^{\mu\nu} - (h'^{2})^{\mu\nu} + (h'^{3})^{\mu\nu} + \cdots = \sum_{n=1}^{\infty} (-1)^{n+1} (h'^{n})^{\mu\nu} \,.
\label{mapping_usual}\end{equation}

\paragraph{Exponential metric perturbations}
~\\
Another popular expansion is the exponential form
\begin{equation}
  \sigma^{\mu\nu} = \big[e^{-h'}\big]^{\mu\nu}  = \eta^{\mu\nu} - \sum_{n=1}^{\infty} \frac{(-1)^{n+1}}{n!} (h'^{n})^{\mu\nu} \,.
\label{exponential_perturbations}\end{equation}
Then the mapping between the fluctuations $h$ and $h'$ is given by
\begin{equation}
  h^{\mu\nu} =  h'^{\mu\nu} - \frac{1}{2} (h'^{2})^{\mu\nu} + \frac{1}{3!} (h'^{3})^{\mu\nu} + \cdots = \sum_{n=1}^{\infty} \frac{(-1)^{n+1}}{n!} (h'^{n})^{\mu\nu} \,.
\label{mapping_exponential}\end{equation}

Employing the mappings from the reference convention, one may construct perturbations of the action and EoM in other metric perturbation conventions. We consider the same examples as before, the usual metric perturbations
\begin{equation}
  g^{\mu\nu} = \eta^{\mu\nu} - \sum_{n=1}^{\infty} (-1)^{n+1} (h'^{n})^{\mu\nu} \,,
\label{usual_metric_pert}\end{equation}
and the exponential expansion
\begin{equation}
  g^{\mu\nu} = \big[e^{-h'}\big]^{\mu\nu} = \eta^{\mu\nu} - \sum_{n=1}^{\infty} \frac{(-1)^{n+1}}{n!} (h'^{n})^{\mu\nu} \,.
\label{exp_metric_pert}\end{equation}
Recall that we use the same notation for the fluctuations of $\sigma^{-1}$ and $g^{-1}$. Thus the mapping between $h'$ and $h$ for each convention is the same as before, \eqref{mapping_usual} and \eqref{mapping_exponential}. Since the definitions of $Z$ and $\check{Z}$ are identical, their perturbations are also the same. We present the explicit perturbations of $\check{X}^{\mu\nu}_{\rho\sigma}$ and $\check{Y}_{\mu\nu}$ up to fourth order in appendix \ref{App:A.2}.

\section{Perturbations of Einstein-Dilaton Theory}\label{Sec:3}

In this section we consider perturbations of the Einstein-dilaton theory by applying the same strategy as we have developed in the previous section. We define a new tensor density, which is related by a Weyl rescaling to the previous tensor density, and a modified dilaton field $d$, which incorporates $\sqrt{-g}$ with the usual dilaton field $\phi$. Again, we also adopt CR perturbation convention with respect to $\sigma$. In this setup, Einstein-dilaton theory offers a more compact expansion than pure GR. We further introduce a first-order formalism for ED theory.
.
\subsection{Action and EoM}
The action for Einstein-dilaton theory in the string frame is given by
\begin{equation}
  S_{\rm ED} = \int \mathrm{d}^{D} x \sqrt{-g} e^{-2\phi} \Big[R + 4 g^{\mu\nu}\partial_{\mu}\phi \partial_{\nu}\phi\Big]\,,
\label{ED_action_g}\end{equation}
where $\phi$ is the dilaton field. Recall that through the field redefinition of metric \eqref{def_sigma1} to the tensor density $\sigma$, the integration measure $\sqrt{-g}$ is disappeared in the EH action. Similarly we define a field redefinition of metric to a new tensor density $\sigma$ (with a slight abuse of notation) to absorb the $e^{-2\phi}\sqrt{-g}$ overall factor in $S_{\rm ED}$ 
\begin{equation}
  \sigma^{\mu\nu} = e^{-2\phi}\sqrt{-g} g^{\mu\nu}\,,
  \qquad
  \sigma_{\mu\nu} = \frac{1}{\sqrt{-g}} e^{2\phi}g_{\mu\nu}\,,
\label{}\end{equation}
where $\sigma^{\mu\nu}$ is the inverse of $\sigma_{\mu\nu}$. Thus the new tensor density is related to the previous tensor density by the Weyl transformation $e^{-2d}$. We also make a field redefinition for the usual dilaton $\phi$ by incorporating $\sqrt{-g}$
\begin{equation}
  e^{-2d} = e^{-2\phi}\sqrt{-g}\,, \qquad \text{or} \qquad d = \phi -\frac{1}{2} \ln \sqrt{-g}\,.
\label{}\end{equation}
This definition is not new; it is a field content in the double field theory \cite{Siegel:1993xq,Hohm:2010jy,Hohm:2010pp}, which is a reformulation of the half-maximal supergravities with manifest T-duality. $d$ is a scalar field under the $\mathit{O}(D,D)$ T-duality. Note that here $d$ is not the same as $\hat{d}$ \eqref{DFT_dilaton} but a field redefinition of the usual dilaton $\phi$. On the other hand, $\hat{d}$ is just an abbreviation for $\sqrt{-g}$.

We may recast the ED theory action $S_{\rm ED}$ \eqref{ED_action_g} in terms of $\sigma$ and $d$
\begin{equation}
\begin{aligned}
  \tilde{S}_{\rm ED} = \int \mathrm{d}^{D} x \bigg[&\
  \frac{1}{4} \sigma^{\mu\nu} \partial_{\mu} \sigma^{\rho\sigma} \partial_{\nu} \sigma_{\rho\sigma}
  	- \frac{1}{2} \sigma^{\mu\nu} \partial_{\mu} \sigma^{\rho\sigma} \partial_{\rho} \sigma_{\nu\sigma}
  	+ \partial_{\mu} \Big( 2\sigma^{\mu\nu} \partial_{\nu} d - \partial_{\nu}\sigma^{\mu\nu}\Big)
  	\\
  	&
  	- (D-2) \sigma^{\mu\nu} \partial_{\mu}d \partial_{\nu}d
  	+ \sigma^{\mu\nu}\big( \sigma^{\rho\sigma} \partial_{\mu} \sigma_{\rho\sigma} \big) \partial_{\nu}d\,
  \bigg]\,.
\end{aligned}\label{ED_action_sigma_d}
\end{equation}
One can show the equivalence with the $S_{\rm ED}$ \eqref{ED_action_g} by substituting the definitions of $\sigma$ and $d$. The EoM of $d$ is an inhomogeneous partial differential equation as
\begin{equation}
  \partial_{\mu} \big(\sigma^{\mu\nu}\partial_{\nu}d\big) = \frac{1}{2(D-2)} \partial_{\mu} \big(\sigma^{\mu\nu} \sigma^{\rho\sigma} \partial_{\nu} \sigma_{\rho\sigma}\big)\,,
\label{}\end{equation}
and the solutions are divided into a homogeneous solution and a particular solution such that
\begin{equation}
  \partial_{\mu} d = \partial_{\mu} \phi + \frac{1}{2(D-2)} \sigma^{\rho\sigma} \partial_{\mu} \sigma_{\rho\sigma} = \partial_{\mu} \phi +\partial_{\mu}\hat{d}\,,
\label{EoM_d}\end{equation}
where $\phi$ denotes the homogeneous solution satisfying
\begin{equation}
   \partial_{\mu} \big(\sigma^{\mu\nu}\partial_{\nu} \phi\big) = 0\,.
\label{}\end{equation}
Note that this equation is the EoM of the dilaton in the Einstein frame. Furthermore, if we substitute \eqref{EoM_d} to the $\tilde{S}_{\rm ED}$ \eqref{ED_action_sigma_d}, it reduces to the usual ED theory action in the Einstein frame. Suppose we ignore the homogeneous part $\phi$ by hand. In that case, \eqref{EoM_d} reduces to $\partial_{\mu} d = \partial_{\mu}\hat{d}$, we obtain the previous EH action \eqref{EH_action_sigma} after substituting the on-shell value of $d$ into the action. 
 Therefore, the distinction between Einstein-dilaton theory and pure GR arises after imposing a boundary condition that decides whether we keep $\phi$ or not.

We now consider the EoM of $\sigma_{\mu\nu}$ and $d$ by varying the action
\begin{equation}
  \delta \tilde{S}_{\rm ED} = \int \mathrm{d}^{D}x\ \Big( -\delta \sigma_{\mu\nu} \tilde{G}^{\mu\nu} + \delta d \,F\Big) \,,
\label{}\end{equation}
where%
\begin{equation}
\begin{aligned}
  \tilde{G}^{\mu\nu}
  =&\
    \frac{1}{2} \sigma^{\rho\sigma} \Big[
      \partial_{\rho}\partial_{\sigma}\sigma^{\mu\nu}
    + \sigma^{\lambda(\mu} \partial_{\rho}\sigma^{\nu)\kappa} \partial_{\sigma}\sigma_{\kappa\lambda}
    \Big]
  - \sigma^{\rho(\mu} \Big[
  	  \partial_{\rho}\partial_{\sigma}\sigma^{\nu)\sigma}
  	+ \sigma^{\nu)\sigma} \partial_{\rho} \sigma^{\kappa\lambda} \partial_{\kappa}\sigma_{\lambda\sigma}
  	\Big]
  \\&
  +\sigma^{\kappa (\mu}\sigma^{\nu)\lambda} \bigg[
    \frac{1}{4}\partial_{\kappa}\sigma^{\rho\sigma}\partial_{\lambda}\sigma_{\rho\sigma}
  + \partial_{\kappa} d\ \sigma^{\rho\sigma} \partial_{\lambda}\sigma_{\rho\sigma}
  - (D-2) \partial_{\kappa}d \partial_{\lambda}d
  \bigg]
  \\&
  + \frac{1}{2} \Big[
    \partial_{\rho}\sigma^{\rho\sigma} \partial_{\sigma}\sigma^{\mu\nu}
  - \partial_{\sigma}\sigma^{\rho\mu} \partial_{\rho}\sigma^{\sigma\nu}
  \Big]
  + \sigma^{\mu\nu} \partial_{\rho} \big( \sigma^{\rho\sigma}\partial_{\sigma}d\big)
  \,,
\end{aligned}\label{tilde_Gmunu}
\end{equation}
and
\begin{equation}
  F = 2(D-2) \partial_{\mu} \big(\sigma^{\mu\nu}\partial_{\nu}d\big)
  - \partial_{\mu} \big(\sigma^{\mu\nu} \sigma^{\rho\sigma} \partial_{\nu} \sigma_{\rho\sigma}\big)\,.
\label{EoM_F}\end{equation}
%

\subsection{Perturbations of the action and EoM}

Let us consider perturbations of the ED theory. As before we employ the CR convention for $\sigma$ using the same convention with the pure GR case $\sigma^{\mu\nu} = \eta^{\mu\nu} - h^{\mu\nu}$, even though these are different to each other. Perturbations of $d$ are given by a simple linear fluctuation $f$ because $d$ is an independent field
\begin{equation}
  d = d_{0} + f \,,
\label{}\end{equation}
where $d_{0}$ is a trivial backgound field. Here we choose $d_{0}=0$. Note that perturbations of $\hat{d}$, which is related to the on-shell value of $d$ \eqref{EoM_d}, are much more complicated than $d$. Thus perturbations of $\tilde{S}_{\rm ED}$ and its EoM are simpler than perturbations of pure GR, although ED theory looks more complicated. 

Again, we introduce a set of abbreviations $X^{\mu\nu}_{\rho\sigma}$, $\tilde{Y}_{\mu\nu}$, $Z^{\mu\nu}$ and $\tilde{W}$ for simplifying calculations. Here $X^{\mu\nu}_{\rho\sigma}$ and $Z^{\mu\nu}$ are the same as the pure GR case, and $\tilde{Y}_{\mu\nu}$ and $\tilde{W}$ are replaced as in \eqref{aux_tilde}. Substituting the perturbations of the abbreviations \eqref{pert_auxiliary_2} into \eqref{ED_action_XY}, we can derive perturbations of the Lagrangian $\tilde{L}_{\rm ED}$ to all orders in fluctuations
\begin{equation}
\begin{aligned}
  \tilde{L}_{\rm ED} =&
  \sum_{n=0}^{\infty} \bigg[
    - \frac{1}{4} (\eta-h)^{\mu\nu} \sum_{m=0}^{n} \tr\big[ h^{m} \partial_{\mu}h h^{n-m} \partial_{\nu} h\big]
  	+ \frac{1}{2} \big[\partial_{\nu}h h^{n} \partial_{\mu}h\big]^{\mu\nu}
  \\&\qquad\
  	+ (\eta-h)^{\mu\nu} \partial_{\mu} f\ \tr\big[h^{n}\partial_{\nu}h\big]
  \bigg]
  - (D-2) (\eta-h)^{\mu\nu} \partial_{\mu}f \partial_{\nu}f\,.
\end{aligned}\label{pert_action_1}
\end{equation}

Similarly we may derive perturbations of $\tilde{G}^{\mu\nu}$ and $F$ using \eqref{tildeGmunu}
\begin{equation}
\begin{aligned}
  \tilde{G}^{\mu\nu} &=
    (\eta-h)^{\rho(\mu} \Bigg[
  	\partial_{\rho}\partial_{\sigma}h^{\nu)\sigma}
  	+ \sum_{n=0}^{\infty} \big[\partial_{\rho} h h^{n} \partial_{\sigma}h\big]^{\sigma|\nu)}
  \Bigg]
  \\&\quad
  - \frac{1}{2} (\eta-h)^{\rho\sigma} \Bigg[\partial_{\rho}\partial_{\sigma}h^{\mu\nu} +\sum_{n=0}^{\infty} \big[\partial_{\rho}h h^{n} \partial_{\sigma} h\big]^{\mu\nu}\Bigg]
  \\&\quad
  + (\eta-h)^{\mu\rho} (\eta-h)^{\nu\sigma} \Bigg[
  \sum_{n=0}^{\infty} \bigg( \tr\big[h^{n} \partial_{(\rho} h\big]  \partial_{\sigma)}f - \frac{1}{4}\sum_{m=0}^{n} \tr\big[h^{m}\partial_{\rho}h h^{n-m} \partial_{\sigma} h\big]\bigg)
  \\&\qquad\qquad\qquad\qquad\qquad\quad
  - (D-2)\partial_{\rho}f \partial_{\sigma} f
  \Bigg]
  \\&\quad
  + (\eta-h)^{\mu\nu} \Big[ (\eta-h)^{\rho\sigma} \partial_{\rho} \partial_{\sigma} f- \partial_{\rho} h^{\rho\sigma} \partial_{\sigma} f  \Big]
  - \frac{1}{2} \partial_{\sigma}h^{\rho\mu} \partial_{\rho}h^{\sigma\nu}
  + \frac{1}{2} \partial_{\rho}h^{\rho\sigma} \partial_{\sigma}h^{\mu\nu}\,,
\end{aligned}\label{}
\end{equation}
and
\begin{equation}
  F = 2(D-2) \partial_{\mu}\Big( (\eta-h)^{\mu\nu} \partial_{\nu} f \Big)
  - \partial_{\mu} \bigg( (\eta-h)^{\mu\nu} \sum_{n=0}^{\infty} \tr \big[h^{n} \partial_{\nu}h\big] \bigg) \,.
\label{pert_EoM_d}\end{equation}
Again, one can find their general $n$-th order terms in \eqref{pert_auxiliary_2}.

\subsection{First-order formalism for Einstein-dilaton theory} \label{Sec:B}

We now consider the first-order formalism for the ED action \eqref{ED_action_sigma_d} following the idea of \cite{Cheung:2017kzx}. The basic idea is to rewrite the action in a complete square form and find an equivalent form introducing an auxiliary field from the Gaussian integration formula. One can recast \eqref{ED_action_sigma_d} as a complete square form as
\begin{equation}
  S_{\rm ED} = \int \mathrm{d}^{D}x \bigg[\frac{1}{2}
  \Big(\partial_{\mu}\sigma_{\kappa\lambda} - 2 \sigma_{\mu(\kappa}\partial_{\lambda)} d\Big) T^{\mu\kappa\lambda,\nu\rho\sigma} \big(\partial_{\nu}\sigma_{\rho\sigma} - 2 \sigma_{\nu(\rho}\partial_{\sigma)} d\big)
  - (D-1) \sigma^{\mu\nu} \partial_{\mu}d  \partial_{\nu}d
  \bigg]\,,
\label{}\end{equation}
where
\begin{equation}
  T^{\mu\kappa\lambda,\nu\rho\sigma} =
  - \frac{1}{2} \sigma^{\mu\nu} \sigma^{\rho(\kappa} \sigma^{\lambda)\sigma}
  + \frac{1}{2} \sigma^{\mu\rho} \sigma^{\nu(\kappa} \sigma^{\lambda)\sigma}
  + \frac{1}{2} \sigma^{\mu\sigma} \sigma^{\nu(\kappa} \sigma^{\lambda)\rho}\,.
\label{}\end{equation}
Using the Gaussian integral formula, we may rewrite the action in terms of an auxiliary field $Q^{\mu}{}_{\kappa\lambda}$
\begin{equation}
\begin{aligned}
  \tilde{S} = \int \mathrm{d}^{D}x \bigg[&- Q^{\mu}{}_{\kappa\lambda}
    (T^{-1})_{\mu}{}^{\kappa\lambda}{}_{\nu}{}^{\rho\sigma} Q^{\nu}{}_{\rho\sigma}
  + Q^{\mu}{}_{\kappa\lambda} \Big(
  \partial_{\mu} \sigma^{\kappa\lambda}
  - 2 \delta_{\mu}{}^{(\kappa}\sigma^{\lambda)\rho} \partial_{\rho}d
  \Big)
  \\&
  - (D-1) \sigma^{\mu\nu} \partial_{\mu}d  \partial_{\nu}d \
  \bigg]\,.
\end{aligned}\label{}
\end{equation}

Here $(T^{-1})_{\mu}{}^{\kappa\lambda}{}_{,\nu}{}^{\rho\sigma} = \sigma^{\kappa\alpha} \sigma^{\lambda\beta} \sigma^{\rho\gamma} \sigma^{\sigma\delta}(T^{-1})_{\mu\alpha\beta,\nu\gamma\delta}$, and $(T^{-1})_{\mu\kappa\lambda,\nu\rho\sigma}$ is the inverse of $T^{\mu\kappa\lambda,\nu\rho\sigma}$, which satisfies
\begin{equation}
  T^{\mu\kappa\lambda,\tau\alpha\beta} (T^{-1})_{\tau\alpha\beta,\nu\rho\sigma} = \delta_{\nu}{}^{\mu} \delta_{\rho}{}^{(\kappa}\delta_{\sigma	}{}^{\lambda)}\,.
\label{}\end{equation}
One can find the explicit form of the $T^{-1}$ as
\begin{equation}
  (T^{-1})_{\mu}{}^{\kappa\lambda}{}_{,\nu}{}^{\rho\sigma} =
    \delta_{\mu}{}^{\rho} \delta_{\nu}{}^{(\kappa} \sigma^{\lambda)\sigma}
  + \delta_{\mu}{}^{\sigma} \delta_{\nu}{}^{(\kappa} \sigma^{\lambda)\rho}
  \,.
\label{}\end{equation}
Then the action reduces to
\begin{equation}
\begin{aligned}
  \tilde{S} = \int \mathrm{d}^{D}x \bigg[&
  - 2 \sigma^{\rho\sigma} Q^{\mu}{}_{\nu\rho} Q^{\nu}{}_{\mu\sigma}
  + Q^{\mu}{}_{\kappa\lambda} \Big(
  \partial_{\mu} \sigma^{\kappa\lambda}
  - 2 \delta_{\mu}{}^{(\kappa}\sigma^{\lambda)\rho} \partial_{\rho}d
  \Big)
  \\&
  - (D-1) \sigma^{\mu\nu} \partial_{\mu}d  \partial_{\nu}d \
  \bigg]\,.
\end{aligned}\label{First_order_ED_action}
\end{equation}
The EoM of each field are as follows:
\begin{equation}
\begin{aligned}
  2 Q^{\rho}{}_{\sigma\mu} Q^{\sigma}{}_{\rho\nu} + \partial_{\rho} Q^{\rho}{}_{\mu\nu} + 2 Q^{\rho}{}_{\rho(\mu} \partial_{\nu)}d + (D-1) \partial_{\mu}d \partial_{\nu}d =0\,,
  \\
  -4 Q^{(\nu}{}_{\mu\sigma} \sigma^{\rho)\sigma}
  + \partial_{\mu} \sigma^{\nu\rho}
  - 2 \delta_{\mu}{}^{(\nu}\sigma^{\rho)\sigma} \partial_{\sigma}d =0\,,
  \\
  (D-1) \partial_{\mu}\big(\sigma^{\mu\nu}\partial_{\nu}d\big) + \partial_{\mu} \big(\sigma^{\mu\nu} Q^{\rho}{}_{\rho\nu}\big) =0\,.
\end{aligned}\label{}
\end{equation}

The on-shell value of $Q_{\mu}{}^{\kappa\lambda}$ is simply given by
\begin{equation}
\begin{aligned}
  Q^{\mu}{}_{\kappa\lambda} &= \frac{1}{2} T^{\mu}{}_{\kappa\lambda,}{}^{\nu}{}_{\rho\sigma} \Big(
  \partial_{\nu} \sigma^{\rho\sigma}
  - 2 \delta_{\nu}{}^{(\rho}\sigma^{\sigma)\tau} \partial_{\tau}d
  \Big)
  \\
  &=
    \frac{1}{2} \Gamma^{\mu}_{\kappa\lambda} [\sigma]
  - \frac{1}{2} \sigma_{\kappa\lambda} \sigma^{\mu\rho} \partial_{\rho}d\,.
\end{aligned}\label{}
\end{equation}
where $\Gamma^{\mu}_{\kappa\lambda} [\sigma]$ is the Christoffel connection with metrics replaced by $\sigma$. It is straightforward to derive the first-order formalism for \eqref{sigma_action_dmu} by replacing $\partial_{\mu}d \to d_{\mu}$ from \eqref{First_order_ED_action} since they have the same structure.


\section{Graviton Scattering Amplitudes via the Off-shell Recursion Relation} \label{Sec:4}

In this section, we will calculate graviton scattering amplitudes as a consistency check of our results. There are several methods to compute scattering amplitudes, but here we employ the off-shell recursion relation by directly defining graviton off-shell currents associated with graviton scattering amplitudes. We employ the so-called perturbiner method that generates the off-shell recursion relation from classical EoM \cite{Rosly:1996vr,Rosly:1997ap,Selivanov:1997aq,Selivanov:1997an,Selivanov:1997ts} instead of the recursive structure of Feynman diagrams. Recently, this method was derived from the quantum effective action formalism \cite{Lee:2022aiu} and extended to loop level \cite{Lee:2022aiu,Gomez:2022dzk}. Since this method relies on the detailed form of the perturbed Einstein equation, it can be used for checking the results in the previous sections.

The fundamental object of this method is the perturbiner expansion, which is a generating function of off-shell currents. It is defined by a multi-particle expansion of the graviton field $h^{\mu\nu}$ in a plane-wave basis. Recently, the perturbiner expansion is derived from the quantum effective action formalism \cite{Lee:2022aiu}. The graviton off-shell recursion relation can be obtained by substituting the perturbiner expansion into the Einstein equation. We can solve the recursion relation order by order and determine graviton scattering amplitudes from the off-shell currents. Thus it directly connects tree-level scattering amplitudes to the solution of the equations of motion.

The perturbiner expansion for the graviton is defined by
\begin{equation}
  h^{\mu \nu} = \sum_{i} J_{i}^{\mu \nu} e^{ik_i \cdot x}
  + \sum_{i<j} J_{ij}^{\mu \nu} e^{i k_{ij} \cdot x}
  + \sum_{i<j<k} J_{ijk}^{\mu \nu} e^{i k_{ijk} \cdot x} + \cdots,
\label{}\end{equation}
where $J^{\mu \nu}_{i}, J_{ij}^{\mu \nu}, \cdots$ are the off-shell graviton currents and $i,j,k, \cdots$ are letters which label each single particle. Here $k^{ijk \cdots}_{\mu} = k^{i}_{\mu} + k^{j}_{\mu} +k^{k}_{\mu} + \cdots$. We may simplify the expansion using the multiparticle indices $\mathcal{P}, \mathcal{Q}, \mathcal{R}\cdots$ that are the so-called ``ordered words'' \cite{Mafra:2016ltu,Mizera:2018jbh}
 \begin{align}
   & h^{\mu \nu}= \sum_{\mathcal{P}} J_{\mathcal{P}}^{\mu \nu} e^{ik_\mathcal{P} \cdot x}.
 \end{align}
The ordered words consist of the letters, such as $\mathcal{P} = i, j, k, \cdots$ with a definite ordering, $i< j<k<\cdots$. We also call the length of the words their ``rank', denoted as $|\mathcal{P}|$. Here
\begin{align}
  k^{\mathcal{P}}_{\mu} = k_{\mu}^i + k^{j}_{\mu} + k^{k}_{\mu} + \cdots.
\end{align}
The $(n+1)$-point graviton scattering amplitudes $M(1,2, \cdots, n+1)$ are related to the rank-$n$ current $J^{12\cdots n}_{\mu \nu}$
\begin{align}
  M(1,2, \cdots, n+1) = \lim_{s_{1\cdots n} \rightarrow 0} s_{12 \cdots n} \epsilon^{\mu \nu}_{n+1} J_{12\cdots n}^{\mu \nu}\,,
\label{scttering_amp_current}\end{align}
where $\epsilon^{\mu \nu}_{n+1}$ is a polarization tensor for the $(n + 1)$-th external graviton and $s_{1\cdots n}$ is the Mandelstam variable defined by
\begin{equation}
  s_{12\cdots n} = - k^{2}_{12\cdots n}\,.
\label{}\end{equation}
%

\subsection{Off-Shell recursions}
We now derive the off-shell recursions from the perturbiner expansion. Since the diffeomorphism in GR is a gauge symmetry, we first need a gauge fixing to define the propagator. Here we employ the de Donder gauge condition $\partial_{\mu} h^{\mu\nu} =0$, and the corresponding gauge fixing term is given by
\begin{equation}
  L_{\rm gf} = -\frac{1}{2} \partial_{\mu} h^{\mu\rho} \partial^{\nu} h_{\nu\rho}\,.
\label{gauge_fixing_term}\end{equation}
We obtain a gauge fixed Lagrangian by adding $L_{\rm gf}$ into the quadratic order Lagrangian \eqref{2nd_L}
\begin{equation}
  L'_{2} = L_{2} + L_{\rm gf} = -\frac{1}{4} \partial^{\rho} h^{\mu\nu} \partial_{\rho} h_{\mu\nu}
  + \frac{1}{4(D-2)} \partial^{\mu}h \partial_{\mu} h\,.
\label{}\end{equation}
We may rewrite $L'_{2}$ using the kinetic operator $K^{\mu\nu,\rho\sigma}[x,y]$
\begin{equation}
  S'_{2} = \int \mathrm{d}^{D}x\mathrm{d}^{D}y \bigg[ - \frac{1}{2} h_{\mu\nu}(x) K^{\mu\nu,\rho\sigma}[x,y] h_{\rho\sigma}(y) \bigg]\,,
\label{}\end{equation}
where
\begin{equation}
  K^{\mu\nu,\rho\sigma}[x,y] = - \frac{1}{2} \bigg( \eta^{(\mu|\rho|} \eta^{\nu)\sigma} - \frac{1}{D-2}\eta^{\mu\nu} \eta^{\rho\sigma}\bigg) \Box_{y}\delta^{D} (x-y) \,.
\label{}\end{equation}
Then we can read off the graviton propagator from the inverse of $K^{\mu\nu,\rho\sigma}[x,y]$, giving
\begin{equation}
  D_{\mu\nu,\rho\sigma}[x,y] = \int \frac{\mathrm{d}^{D} p}{(2\pi)^{D}} \frac{2 \eta_{(\mu|\rho|} \eta_{\nu)\sigma} -\eta_{\mu\nu} \eta_{\rho\sigma}}{p^{2}} e^{ip\cdot (x-y)} \,.
\label{}\end{equation}

In the perturbiner method, the recursion relation for the off-shell currents arises from the EoM. Since the EoM $\mathcal{G}^{\mu\nu}=0$ is represented by $X^{\mu\nu}_{\rho\sigma}$, $Y_{\mu\nu}$, $Z^{\mu\nu}$ and $W$, it is convenient to define their off-shell currents as
\begin{equation}
\begin{aligned}
  X^{\mu \nu}_{\rho \sigma} &= \sum_{\mathcal{P}} \big[\mathcal{X}_{\mathcal{P}} \big]^{\mu\nu}_{\rho\sigma} e^{ik_{\mathcal{P}} \cdot x}\,,
  &\quad
  Y_{\mu\nu} &= \sum_{\mathcal{P}} \mathcal{Y}_{\mathcal{P}}^{\mu\nu} e^{ik_{\mathcal{P}} \cdot x}\,,
  \\
  Z^{\mu \nu}&= \sum_{\mathcal{P}} \mathcal{Z}_{\mathcal{P}}^{\mu \nu} e^{ik_{\mathcal{P}}\cdot x}\,,
  &\quad
  W &= \sum_{\mathcal{P}} \mathcal{W}_{\mathcal{P}} e^{ik_{\mathcal{P}} \cdot x}\,,
\end{aligned}\label{}
\end{equation}
where $[\mathcal{X}_{\mathcal{P}}]^{\mu\nu}_{\rho\sigma}$, $\mathcal{Y}_{\mathcal{P}}^{\rho \sigma}$, $\mathcal{Z}_{\mathcal{P}}^{\mu \nu}$ and $\mathcal{W}_{\mathcal{P}}$ are rank-$|\mathcal{P}|$ currents. Note that we do not distinguish the position of the spacetime indices in the off-shell currents. 

Substituting these perturbiner expansions above into \eqref{pert_auxiliary2}, we represent the currents $[\mathcal{X}_{\mathcal{P}}]^{\mu\nu}_{\rho\sigma}$, $\mathcal{Y}_{\mathcal{P}}^{\rho \sigma}$, $\mathcal{Z}_{\mathcal{P}}^{\mu \nu}$ and $\mathcal{W}_{\mathcal{P}}$ as
\begin{equation}
\begin{aligned}
  \big[\mathcal{X}_{\mathcal{P}}\big]^{\mu \nu}_{\rho \sigma} &=
  - k^{\rho}_{\mathcal{P}} k^{\sigma}_{\mathcal{P}} J^{\mu \nu}_{\mathcal{P}}
  + \big[\hat{\mathcal{X}}_{\mathcal{P}}\big]^{\mu \nu}_{\rho \sigma} \,,
  \qquad
  \\
  \mathcal{Y}_{\mathcal{P}}^{\mu\nu} &=
   \frac{1}{4} \sum_{\mathcal{P}= \mathcal{Q} \cup \mathcal{R}} \left[
   	  \frac{1}{(D-2)} \mathcal{A}_{\mathcal{Q}}^{\mu} \mathcal{A}_{\mathcal{R}}^{\nu}
   	- \mathcal{A}_{\mathcal{Q}}^{\mu \kappa \lambda} \mathcal{A}^{\nu \lambda \kappa}_{\mathcal{R}}
   	\right]\,,
  \\
  \mathcal{Z}_{\mathcal{P}}^{\mu \nu} &= -\frac{1}{2} \sum_{\mathcal{P}= \mathcal{Q} \cup \mathcal{R}} k^{\rho}_{\mathcal{Q}} k^{\sigma}_{\mathcal{R}} \left[ J^{\rho \sigma}_{\mathcal{Q}} J^{\mu \nu}_{\mathcal{R}} - J^{\sigma \mu}_{\mathcal{Q}} J^{\rho \nu}_{\mathcal{R}} \right]\,,
  \\
  \mathcal{W}_{\mathcal{P}} &= - \frac{1}{2(D-2)} k^{\rho}_{\mathcal{P}}  k^{\rho}_{\mathcal{P}} J^{\sigma \sigma}_{\mathcal{P}} + \hat{\mathcal{W}}_{\mathcal{P}}
   \,, \\
\end{aligned}\label{def_subcurrents}
\end{equation}
where we have singled out the linear-order perturbations in $\big[\mathcal{X}_{\mathcal{P}}\big]^{\mu \nu}_{\rho \sigma}$ and $\mathcal{W}_{\mathcal{P}}$ for later convenience, and their higher order perturbations, which are denoted by $\big[\hat{\mathcal{X}}_{\mathcal{P}}\big]^{\mu \nu}_{\rho \sigma}$ and $\hat{\mathcal{W}}_{\mathcal{P}}$, are as follows:
\begin{equation}
\begin{aligned}
  \big[\hat{\mathcal{X}}_{\mathcal{P}}\big]^{\mu \nu}_{\rho \sigma} &= \sum_{\mathcal{P}=\mathcal{Q} \cup \mathcal{R}} i k^{\rho}_{Q} J^{\mu \kappa}_{\mathcal{Q}} \mathcal{A}_{\mathcal{R}}^{\sigma \kappa \nu}\,,
  \\
  \hat{\mathcal{W}}_{\mathcal{P}} &=
   \frac{i}{2(D-2)} \left(-
   i  k^{\rho}_{\mathcal{P}}  k^{\rho}_{\mathcal{P}} J^{\sigma \sigma}_{\mathcal{P}}
   +k^{\mu}_{\mathcal{P}} \mathcal{A}_{\mathcal{P}}^{\mu}
   -   \sum_{\mathcal{P}=\mathcal{Q} \cup \mathcal{R}} k^{\mu}_{\mathcal{P}} J^{\mu \nu}_\mathcal{Q} \mathcal{A}_{\mathcal{R}}^{\nu} \right)\,.
\end{aligned}\label{}
\end{equation}
%

We now derive the off-shell recursion relation from the Einstein equation $\mathcal{G}^{\mu\nu}=0$. If we denote the rank-$|\mathcal{P}|$ currents for $\mathcal{G}^{\mu\nu}$ as $\mathcal{G}^{\mu \nu}_{\mathcal P}$, its perturbiner expansion is given by $\mathcal{G}^{\mu \nu} = \sum_{\mathcal P} \mathcal{G}^{\mu \nu}_{\mathcal P} e^{ik_{\mathcal P} \cdot x}$, where
\begin{equation}
\begin{aligned}
 \mathcal{G}^{\mu \nu}_{\mathcal P} &= - k^{\rho}_{\mathcal P} k^{\sigma}_{\mathcal P} \left( \eta^{\rho(\mu} J^{|\sigma| \nu)}_{\mathcal P} - \frac{1}{2}  \eta^{\rho \sigma} J^{(\mu \nu)}_{\mathcal P} \right)
  - \frac{1}{2(D-2)} \eta^{\mu\nu} k^{\rho}_{\mathcal{P}}  k^{\rho}_{\mathcal{P}} J^{\sigma \sigma}_{\mathcal{P}}  
  \\&
  + \eta^{\rho(\mu} [\hat{\mathcal{X}}_{P}]^{|\sigma| \nu)}_{\rho \sigma} - \frac{1}{2} \eta^{\rho \sigma} [\hat{\mathcal{X}}_{P}]^{(\mu \nu)}_{\rho \sigma}
  + \mathcal{Y}^{\mu \nu}_{\mathcal P} + \mathcal{Z}^{\mu \nu}_{\mathcal P} + \eta^{\mu \nu} \hat{\mathcal{W}}_{\mathcal P}
  \\&\quad
  - \sum_{\mathcal P=\mathcal Q \cup \mathcal R} \left[
  	  J^{\rho(\mu}_{\mathcal Q} [\mathcal{X}_{\mathcal R}]^{|\sigma| \nu)}_{\rho \sigma}
  	- \frac{1}{2} J^{\rho \sigma}_{\mathcal Q} [\mathcal{X}_{\mathcal R}]^{(\mu \nu)}_{\rho \sigma}
 	+ J^{\mu \rho}_{\mathcal Q} \mathcal{Y}^{\rho \nu}_{\mathcal R}
 	+ J^{\nu \sigma}_{\mathcal Q} \mathcal{Y}^{\mu \sigma}_{\mathcal R}
 	+ J^{\mu \nu}_{\mathcal Q} \mathcal{W}_{\mathcal R}
  \right]
   \\&\quad
   + \sum_{\mathcal P=\mathcal Q \cup \mathcal R \cup \mathcal S} J^{\mu \rho}_{\mathcal Q} J^{\nu \sigma}_{\mathcal R} \mathcal{Y}^{\rho \sigma}_{\mathcal S} = 0\,.
\end{aligned}\label{}
\end{equation}
Adding a gauge fixing term associated with de Donder gauge \eqref{gauge_fixing_term}, we have
\begin{equation}
\begin{aligned}
  \!
  J_{\mathcal P}^{\mu\nu} = \frac{2}{s_{\mathcal P}} P^{\mu \nu, \kappa \lambda} \Bigg(&
  	  \eta^{\rho(\kappa} \big[\hat{\mathcal{X}}_{\mathcal P}\big]^{|\sigma| \lambda)}_{\rho \sigma}
  	- \frac{1}{2} \eta^{\rho \sigma} [\hat{\mathcal{X}}_{\mathcal P}]^{(\kappa\lambda)}_{\rho \sigma}
  	+ \mathcal{Y}^{\kappa\lambda}_{\mathcal P}
  	+ \mathcal{Z}^{\kappa\lambda}_{\mathcal P}
  	+ \eta^{\kappa\lambda} \hat{\mathcal{W}}_{\mathcal P}
  \\&
  - \sum_{\mathcal P=\mathcal Q \cup \mathcal R} \left[
  	  J^{\rho(\kappa}_{\mathcal Q} [\mathcal{X}_{\mathcal R}]^{|\sigma| \lambda)}_{\rho \sigma}
  	- \frac{1}{2} J^{\rho \sigma}_{\mathcal Q} [\mathcal{X}_{\mathcal R}]^{(\kappa\lambda)}_{\rho \sigma}
  	+ 2J^{\rho(\kappa}_{\mathcal Q} \mathcal{Y}^{\lambda)\rho}_{\mathcal R}
  	+ J^{\kappa\lambda}_{\mathcal Q} \mathcal{W}_{\mathcal R}
  \right]
  \\&
  + \sum_{\mathcal P=\mathcal Q \cup \mathcal R \cup \mathcal S} J^{\kappa \rho}_{\mathcal Q} J^{\lambda \sigma}_{\mathcal R} \mathcal{Y}^{\rho \sigma}_{\mathcal S} \Bigg)\,,
\end{aligned}\label{recursions}
\end{equation}
where $P^{\mu\nu, \kappa \lambda} = \eta^{(\mu |\kappa|} \eta^{\nu) \lambda} - \frac{1}{2} \eta^{\mu \nu} \eta^{\kappa \lambda}$. The initial condition of the recursion relation is given by graviton polarisation tensors
\begin{equation}
  J^{\mu\nu}_{i} = \epsilon^{\mu}_{i} \epsilon^{\nu}_{i}\,.
\label{initial_condition}\end{equation}

It is straightforward to solve the recursion relation \eqref{recursions} order by order in the rank of the words. Substituting the expressions of the subcurrents \eqref{def_subcurrents} into the recursion relation \eqref{recursions}, we can derive a rank-$|\mathcal{P}|$ current $J^{\mu\nu}_{\mathcal{P}}$ using the lower rank currents. We present the explicit form of the off-shell currents up to rank 4. See appendix \ref{App:C} for the explicit form of the graviton currents. The result indicates our perturbations of Einstein equation are correct.


\section{Generalisations to Curved Backgrounds with Matters} \label{Sec:5}

So far we have considered perturbations of pure GR around a flat metric. However, many interesting GR problems are defined in curved spacetime with additional matter, such as physics in cosmology and black hole backgrounds. This section will extend the previous analysis in a flat background to arbitrarily curved backgrounds. We will also investigate perturbations of energy-momentum tensor with three examples.

\subsection{Curved spacetime generalisation}

We now construct perturbations of the EH action and the Einstein equation around an arbitrary background. We start by extending the definition of the perturbations of $\sigma_{\mu\nu}$ and $\sigma^{\mu\nu}$ \eqref{perturb_sigma} to a curved background $\bar{g}_{\mu\nu}$
\begin{equation}
  \sigma^{\mu\nu} = \frac{\sqrt{-g}}{\sqrt{-\bar{g}}} g^{\mu\nu} = e^{-2\hat{f}} g^{\mu\nu} \,,
  \qquad
  \sigma_{\mu\nu} = \frac{\sqrt{-\bar{g}}}{\sqrt{-g}} g_{\mu\nu} = e^{2\hat{f}} g^{\mu\nu}\,,
\label{}\end{equation}
where $\hat{f}$ is fluctuations of $\hat{d}$
\begin{equation}
  e^{-2\hat{d}} = \sqrt{-\bar{g}} e^{-2\hat{f}} \,.
\label{perturbed_d}\end{equation}
We denote perturbations of $\sigma$ as
\begin{equation}
  \sigma^{\mu\nu} = \bar{g}^{\mu\nu} - h^{\mu\nu} \,,
  \qquad
  \sigma_{\mu\nu} = \bar{g}_{\mu\nu} + \sum_{n=1}^{\infty} (h^{n})_{\mu\nu}\,.
\label{}\end{equation}
We introduce a background covariant derivative in terms of the connection $\bar{\Gamma}_{\mu\nu}^{\rho}$ associated with the backgound metric $\bar{g}_{\mu\nu}$
\begin{equation}
  \bar{\nabla}_{\mu} = \partial_{\mu} + \bar{\Gamma}_{\mu}\,.
\label{}\end{equation}
Then $\bar{\nabla}_{\mu}\hat{d}$ is written as
\begin{equation}
  \bar{\nabla}_{\mu} \hat{d} = \frac{1}{2(D-2)} \sigma^{\rho\sigma} \bar{\nabla}_{\mu} \sigma_{\rho\sigma}\,.
\label{def_d_curved}\end{equation}

The EH action in terms of $\sigma$ \eqref{EH_action_sigma} is now written as
\begin{equation}
\begin{aligned}
  S_{\rm EH} = \int \mathrm{d}^{D} x \sqrt{-\bar{g}} \bigg[&\
  	  \frac{1}{4} \sigma^{\mu\nu} \bar{\nabla}_{\mu} \sigma^{\rho\sigma} \bar{\nabla}_{\nu} \sigma_{\rho\sigma}
  	- \frac{1}{2} \sigma^{\mu\nu} \bar{\nabla}_{\mu} \sigma^{\rho\sigma} \bar{\nabla}_{\rho} \sigma_{\nu\sigma}
   	- (D-2) \sigma^{\mu\nu}\bar{\nabla}_{\mu}\hat{d} \bar{\nabla}_{\nu}\hat{d}
   	\\&
   	+ \sigma^{\mu\nu} \bar{R}_{\mu\nu}
  	\bigg]\,,
\end{aligned}\label{ED_action_sigma_curved}
\end{equation}
where $\bar{R}_{\mu\nu}$ is the Ricci tensor for the background metric $\bar{g}_{\mu\nu}$. The corresponding Einstein tensor $G^{\mu\nu}$ is obtained by mapping $\partial_{\mu} \to \bar{\nabla}_{\mu}$ and adding $\bar{R}_{\mu\nu}$ appropriately
\begin{equation}
\begin{aligned}
  G^{\mu\nu}
  =&\
    \frac{1}{2} \sigma^{\rho\sigma} \Big[\bar{\nabla}_{\rho}\bar{\nabla}_{\sigma}\sigma^{\mu\nu}
  + \bar{\nabla}_{\rho}\sigma^{\kappa(\mu} \bar{\nabla}_{\sigma}\sigma_{\kappa\lambda} \sigma^{\nu)\lambda} \Big]
  - \sigma^{\rho(\mu} \Big[
  	  \bar{\nabla}_{\rho}\bar{\nabla}_{\sigma}\sigma^{\nu)\sigma}
  	+ \bar{\nabla}_{\rho} \sigma^{|\kappa\lambda} \bar{\nabla}_{\kappa}\sigma_{\lambda\sigma} \sigma^{\sigma|\nu)}
  \Big]
  \\&
  + \frac{1}{4} \sigma^{\kappa (\mu}\sigma^{\nu)\lambda} \Big[
  	  \bar{\nabla}_{\kappa}\sigma^{\rho\sigma} \bar{\nabla}_{\lambda}\sigma_{\rho\sigma}
  	+ (D-2) \bar{\nabla}_{\kappa} \hat{d} \bar{\nabla}_{\lambda} \hat{d}
  	+ \bar{R}_{\kappa\lambda}
  \Big]
  \\&
  + \frac{1}{2} \Big[
  	  \bar{\nabla}_{\rho}\sigma^{\rho\sigma} \bar{\nabla}_{\sigma}\sigma^{\mu\nu}
  	- \bar{\nabla}_{\sigma}\sigma^{\rho\mu} \bar{\nabla}_{\rho}\sigma^{\sigma\nu}
  \Big]
  + \sigma^{\mu\nu} \bar{\nabla}_{\kappa} \big( \sigma^{\kappa\lambda} \bar{\nabla}_{\lambda}\hat{d} \big)
  \,.
\end{aligned}\label{Gmunu_curved}
\end{equation}
Note that we need to be careful with the ordering of the covariant derivatives. Since the structure of $G^{\mu\nu}$ is the same as the flat background case, it is also straightforward to extend the previous $X^{\mu\nu}_{\rho\sigma}$, $Y_{\mu\nu}$, $Z^{\mu\nu}$ and $W$ (see appendix A) to curved backgrounds by replacing $\partial_{\mu} \to \bar{\nabla}_{\mu}$ such as
\begin{equation}
\begin{aligned}
  X^{\mu\nu}_{\rho\sigma} &= \bar{\nabla}_{\rho}\bar{\nabla}_{\sigma}h^{\mu\nu} +\sum_{n=0}^{\infty} \big[\bar{\nabla}_{\rho}h h^{n} \bar{\nabla}_{\sigma} h\big]^{\mu\nu}\,,
  \\
  Y_{\mu\nu} & =
  \sum_{n=0}^{\infty} \sum_{m=0}^{n} \bigg[ \frac{1}{4(D-2)} \tr\big[h^{m} \bar{\nabla}_{\mu} h\big]  \tr\big[h^{n-m} \bar{\nabla}_{\nu} h\big]
  - \frac{1}{4}\tr\big[h^{m}\bar{\nabla}_{\mu}h h^{n-m} \bar{\nabla}_{\nu} h\big]\bigg]\,,
  \\
  Z^{\mu\nu} &= \frac{1}{2} \bar{\nabla}_{\rho} h^{\rho\sigma} \bar{\nabla}_{\sigma} h^{\mu\nu}
  - \frac{1}{2} \bar{\nabla}_{\rho} h^{\sigma\mu} \bar{\nabla}_{\sigma} h^{\rho\nu}\,,
  \\
  W &= \frac{1}{2(D-2)} \bigg(
  	  (\eta-h)^{\mu\nu}\sum_{n=0}^{\infty} \bigg[ \tr\big[h^{n} \bar{\nabla}_{\mu}\bar{\nabla}_{\nu} h\big]
  	+ \sum_{m=0}^{n} \tr\big[h^{m}\bar{\nabla}_{\mu}h h^{n-m} \bar{\nabla}_{\nu} h\big]\bigg]
  	\\&\qquad\qquad\qquad
  	- \bar{\nabla}_{\mu}h^{\mu\nu}  \sum_{n=0}^{\infty} \tr\big[ h^{n} \bar{\nabla}_{\nu} h\big]
  \bigg)
  \,.
\end{aligned}\label{pert_auxiliary1_curved}
\end{equation}

Using these fields, we may recast $L_{\rm EH}$ \eqref{ED_action_sigma_curved} and $G^{\mu\nu}$ \eqref{Gmunu_curved} as
\begin{equation}
\begin{aligned}
  L_{\rm EH} &= X^{\mu\nu}_{\nu\mu}
  - \frac{1}{2} \sigma_{\mu\nu} \sigma^{\rho\sigma} X^{\mu\nu}_{\rho\sigma}
  + \sigma^{\mu\nu} Y_{\mu\nu}
  + \sigma_{\mu\nu} Z^{\mu\nu}
  + D W
  \\
  &=
  \sigma^{\mu\nu} \bar{R}_{\mu\nu}
  + \frac{1}{2} X^{\mu\nu}_{\nu\mu} + \sigma^{\mu\nu} Y_{\mu\nu}
  + \bar{\nabla}_{\mu}\Big( 2\sigma^{\mu\nu} \bar{\nabla}_{\nu} \hat{d} - \frac{1}{2} \bar{\nabla}_{\nu} \sigma^{\mu\nu}\Big)
  \,,
\end{aligned}\label{}
\end{equation}
and
\begin{equation}
\begin{aligned}
  G^{\mu\nu}=&
  \sigma^{\mu\rho} \sigma^{\nu\sigma} \bar{R}_{\rho\sigma}
  - \frac{1}{2} \sigma^{\rho\sigma} X^{(\mu\nu)}_{\rho\sigma}
  + \sigma^{\rho(\mu} X^{|\sigma|\nu)}_{\rho\sigma}
  + \sigma^{\mu\rho}\sigma^{\nu\sigma} Y_{(\rho\sigma)}
  + Z^{\mu\nu}
  + \sigma^{\mu\nu} W
  \,.
\end{aligned}\label{}
\end{equation}
Their perturbations can be obtained using \eqref{pert_auxiliary1_curved}. This feature is also true for other formulations, $\tilde{S}_{\rm EH}$, $\tilde{G}^{\mu\nu}$, $R$ and $R^{\mu\nu}$, and we will leave out their explicit construction.

\subsection{Perturbations of the energy-momentum tensor}
We now consider perturbations of the matter action $S_{\rm matter}$ and its energy-momentum (EM) tensor both in the $\sigma$-based and the metric-based formulations. The EM tensor $T_{\mu\nu}$ is defined as the variation of a matter action with respect to the metric
\begin{equation}
  T^{\mu\nu} = \frac{2}{\sqrt{-g}} \frac{\delta S_{\rm matter}}{\delta g_{\mu\nu}}\,.
\label{}\end{equation}
To consider the density frame case, we modify the definition to be the variation with respect to $\sigma_{\mu\nu}$. Though it is not equivalent to $T^{\mu\nu}$, we still call it the EM tensor and denote it $\mathcal{T}^{\mu\nu}$
\begin{equation}
  \mathcal{T}^{\mu\nu} = 2 \frac{\delta S_{\rm matter}}{\delta \sigma_{\mu\nu}}\,.
\label{}\end{equation}

Together with the EH action, we introduce the total action
\begin{equation}
  S_{\rm total} = S_{\rm EH} + S_{\rm matter}\,.
\label{}\end{equation}
Varying the total action, we obtain the EoM
\begin{equation}
  G^{\mu\nu} = \kappa^{2} \mathcal{T}^{\mu\nu}\,,
\label{}\end{equation}
which is equivalent to the standard form of the Einstein equation $R^{\mu\nu} - \frac{1}{2} g^{\mu\nu} R = \kappa^{2} T^{\mu\nu}$. We can construct perturbations of $\mathcal{T}^{\mu\nu}$ and $T^{\mu\nu}$ by substituting the perturbations of $\sigma^{\mu\nu}$ and $g^{\mu\nu}$. We will consider three examples: a cosmological constant, a scalar field theory with a potential and Maxwell theory.

The action for the cosmological constant $\Lambda$ is given by
\begin{equation}
\begin{aligned}
  S_{\Lambda} &= \frac{1}{2\kappa^{2}} \int \mathrm{d}^{D}x  |\det \sigma|^{-\frac{1}{D-2}} \Big(-2\Lambda\Big)
  \\
  &= \frac{1}{2\kappa^{2}} \int \mathrm{d}^{D}x \sqrt{-g} \Big(-2\Lambda\Big) \,.
\end{aligned}\label{}
\end{equation}
Here we used the relation, $|\det \sigma| = |\det g|^{-\frac{(D-2)}{2}}$\,. The corresponding EM tensors are
\begin{equation}
  \mathcal{T}^{\mu\nu}_{\Lambda} = \frac{4}{D-2} (-\det \sigma)^{-\frac{1}{D-2}} \sigma^{\mu\nu} \Lambda \,,
  \qquad
  T_{\Lambda}^{\mu\nu} = - \frac{1}{\kappa^{2}} g^{\mu\nu} \Lambda\,.
\label{}\end{equation}
Perturbations of $T_{\Lambda}^{\mu\nu}$ are trivial: $T_{\Lambda}^{\mu\nu} = - \frac{1}{\kappa^{2}} (\eta-h)^{\mu\nu} \Lambda$. On the other hand, we need to determine the perturbations of $|\det \sigma|^{-\frac{1}{D-2}}$ to derive perturbations of the $S_{\Lambda}$ and $\mathcal{T}^{\mu\nu}_{\Lambda}$. These are obtained from
\begin{equation}
\begin{aligned}
    |\det \sigma|^{-\frac{1}{D-2}} &= e^{\frac{1}{D-2} \tr \ln (1-h)}\,.
\end{aligned}\label{pert_det_sigma}
\end{equation}
We may expand this up to fourth order in $h$
\begin{equation}
\begin{aligned}
    |\det \sigma|^{-\frac{1}{D-2}} &=
     1 - \frac{1}{D-2} h + \frac{1}{2(D-2)} \bigg( \frac{h^{2}}{D-2} - \tr\big[h^{2}\big] \bigg)
    \\&\quad
    - \frac{1}{D-2}\bigg( \frac{1}{6(D-2)^{2}}h^{3} - \frac{1}{2(D-2)} h\ \tr\big[h^{2}\big] + \frac{1}{3}\tr\big[h^{3}\big]\bigg)
    \\&\quad
    + \frac{1}{4(D-2)} \bigg(
      \frac{1}{3!(D-2)^{3}} h^{4}
    - \frac{1}{(D-2)^{2}} h^{2}\tr\big[h^{2}\big]
    - \tr\big[h^{4}\big]
    \\&\qquad\qquad\qquad\quad
    + \frac{1}{6(D-2)}\Big(8h \tr\big[h^{3}\big] +3\tr\big[h^{2}\big] \tr\big[h^{2}\big]\Big)
    \bigg)
    +\cdots\,.
\end{aligned}\label{}
\end{equation}
Interestingly, the only nontrivial contributions to the perturbations of EM tensors in our examples arise from powers of $|\det \sigma|$.

Next, we consider a scalar field $\varphi$ with a potential $V[\varphi]$. We present the action in terms of both $\sigma$ and the metric
\begin{equation}
\begin{aligned}
  S_{\rm scalar}
  &= \int \mathrm{d}^{D}x\, \bigg[- \frac{1}{2} \sigma^{\mu\nu} \partial_{\mu} \varphi \partial_{\nu} \varphi
    -|\det \sigma|^{-\frac{1}{D-2}} V[\varphi] \bigg]
  \\
  &= \int \mathrm{d}^{D}x\, \sqrt{-g} \bigg[- \frac{1}{2} g^{\mu\nu} \partial_{\mu} \varphi \partial_{\nu} \varphi - V[\varphi] \bigg]\,.
\end{aligned}\label{}
\end{equation}
The corresponding EM tensors are
\begin{equation}
\begin{aligned}
  \mathcal{T}_{\rm scalar}^{\mu\nu} &=
    \sigma^{\mu\rho}\sigma^{\nu\sigma} \partial_{\rho} \varphi \partial_{\sigma} \varphi
  + \frac{2}{D-2} \sigma^{\mu\nu} |\det \sigma|^{-\frac{1}{D-2}} V[\varphi]\,,
  \\
  T_{\rm scalar}^{\mu\nu} &=
  g^{\mu\rho} g^{\nu\sigma} \partial_{\rho} \varphi \partial_{\sigma} \varphi
  - \frac{1}{2} g^{\mu\nu} g^{\rho\sigma} \partial_{\rho}\varphi \partial_{\sigma}\varphi - g^{\mu\nu} V[\varphi]\,,
\end{aligned}\label{}
\end{equation}
and EoM of $\varphi$ are
\begin{equation}
\begin{aligned}
  &\partial_{\mu} \big(\sigma^{\mu\nu} \partial_{\nu}\varphi \big) - |\det \sigma|^{-\frac{1}{D-2}} \frac{\delta V[\varphi]}{\delta\varphi}=0\,,
  \\
  & \partial_{\mu}\big(\sqrt{-g} g^{\mu\nu} \partial_{\nu}\varphi \big) - \frac{\delta V[\varphi]}{\delta\varphi}=0 \,.
\end{aligned}\label{EoM_salar}
\end{equation}
Again, the perturbations of $T^{\mu\nu}_{\rm scalar}$ and the EoM of $\varphi$ in $g$ are trivial; we can obtain them by substituting the perturbations of $g^{\mu\nu}$. For the case of $\mathcal{T}_{\rm scalar}^{\mu\nu}$, we need to use the perturbations of $|\det \sigma|^{-\frac{1}{D-2}}$ in \eqref{pert_det_sigma}.

Finally, let us consider Maxwell theory with the action
\begin{equation}
\begin{aligned}
  S_{\rm Maxwell} &= \int \mathrm{d}^{D}x \sqrt{-g} \Big[ -\frac{1}{4} g^{\mu\rho} g^{\nu\sigma} F_{\mu\nu} F_{\rho\sigma}\Big]\,,
  \\
  &= \int \mathrm{d}^{D}x \bigg[ -\frac{1}{4} |\det \sigma|^{\frac{1}{D-2}}\sigma^{\mu\rho} \sigma^{\nu\sigma} F_{\mu\nu} F_{\rho\sigma}\bigg]\,.
\end{aligned}\label{}
\end{equation}
The corresponding EM tensors are
\begin{equation}
\begin{aligned}
  \mathcal{T}^{\mu\nu}_{\rm Maxwell} &= |\det \sigma|^{\frac{1}{D-2}} \bigg[ \sigma^{\mu\rho}\sigma^{\nu\sigma} \sigma^{\kappa\lambda} F_{\rho\kappa} F_{\sigma\lambda}
  + \frac{1}{2(D-2)} \sigma^{\mu\nu}\sigma^{\kappa\rho} \sigma^{\lambda\sigma} F_{\kappa\lambda} F_{\rho\sigma}
  \bigg]\,,
  \\
  T^{\mu\nu}_{\rm Maxwell} &=
    g^{\mu\rho}g^{\nu\sigma} g^{\kappa\lambda} F_{\rho\kappa} F_{\sigma\lambda}
  - \frac{1}{4} g^{\mu\nu} g^{\kappa\rho} g^{\lambda\sigma} F_{\kappa\lambda} F_{\rho\sigma} \,.
\end{aligned}\label{}
\end{equation}
and the EoM of the Maxwell field are
\begin{equation}
\begin{aligned}
  &\partial_{\mu} \Big(|\det \sigma|^{\frac{1}{D-2}} \sigma^{\mu\rho} \sigma^{\nu\sigma} F_{\rho\sigma} \Big) = 0\,,
  \\
  &\partial_{\mu} \Big(\sqrt{-g} g^{\mu\rho} g^{\nu\sigma} F_{\rho\sigma} \Big) = 0\,.
\end{aligned}\label{}
\end{equation}
Their perturbations are obtained straightforwardly as in the previous examples.


\section{Conclusion}

This article constructs the perturbative expansion of EH action and the Einstein equations with or without matters to all orders in perturbations and the general $n$-th order terms by exploiting Cheung and Remmen's convention. We considered the perturbative GR based on the tensor density and the metric formalism. In the metric formulation, we constructed perturbations of the Ricci scalar and the Ricci tensor and presented their all-order expressions. Remarkably, we showed that there exist three basic building blocks that generate the entire perturbations both in the tensor density and metric formalisms. We also generalised the pure GR results to the Einstein-dilaton theory and derived all-order expressions of the perturbations of the action and EoM. We further derived a new first-order formalism for ED theory.

We regarded our metric perturbations as the reference scheme. We discussed how to generate other perturbation schemes from the reference scheme. We studied two examples, usual perturbations and exponential perturbations, and presented their explicit expressions up to quartic order in fluctuations in the appendix. Therefore, it is unnecessary to follow other laborious methods of obtaining perturbations of the action and EoM; one may derive the perturbative GR in the other convention by simply applying the desired mapping presented here. Finally, we generalised our perturbations to arbitrarily curved backgrounds with matter fields. 

For a consistency check, we calculated graviton scattering amplitudes using the perturbiner method that uses EoM directly. We defined the perturbiner expansion for the graviton off-shell currents, which are directly related to the scattering amplitudes, and generated the off-shell recursion relations by substituting the perturbiner expansion into our perturbed Einstein equation. Solving the recursions explicitly, we obtained the off-shell currents up to rank 4, and it shows that our results are correct. This result ensures that the perturbations of the EoM are correct. 

We hope our approach is helpful for describing various perturbative GR problems in the future.


\acknowledgments
We thank Stephen Angus, Seungjoon Hyun, Hyeonjoon Shin, Hyun Seok Yang and Sang-Heon Yi for discussions and valuable comments.
This work is supported by appointment to the JRG Program at the APCTP through the Science and Technology Promotion Fund and Lottery Fund of the Korean Government. KL is also supported by the National Research Foundation of Korea(NRF) grant funded by the Korean government(MSIT) No.2021R1F1A1060947 and the Korean Local Governments of Gyeongsangbuk-do Province and Pohang City. KC is supported by the NRF grant funded by the Korean government (MSIT) 2022R1F1A1068489.

\appendix
\newpage

\section{Derivation of the Results}

We derived the results in sections \ref{Sec:2} and \ref{Sec:3} with explicit expressions.

\subsection{Perturbations of pure gravity with tensor density}\label{App:A.1}
For efficient calculation, we start by introducing abbreviations replacing the square brackets in \eqref{Gmunu}, which is grouped by the structure of indices,
\begin{equation}
\begin{aligned}
  X^{\mu\nu}_{\rho\sigma} &=
  - \partial_{\rho}\partial_{\sigma}\sigma^{\mu\nu}
  - \partial_{\rho}\sigma^{\mu\kappa} \partial_{\sigma} \sigma_{\kappa\lambda} \sigma^{\lambda\nu}\,,
  \\
  Y_{\mu\nu} &= \frac{1}{4}\partial_{\mu}\sigma^{\rho\sigma}\partial_{\nu}\sigma_{\rho\sigma}
  + (D-2) \partial_{\mu}\hat{d} \partial_{\nu}\hat{d}\,,
  \\
  Z^{\mu\nu} &=
    \frac{1}{2} \partial_{\rho}\sigma^{\rho\sigma} \partial_{\sigma}\sigma^{\mu\nu}
  - \frac{1}{2} \partial_{\rho}\sigma^{\sigma\mu} \partial_{\sigma}\sigma^{\rho\nu}\,,
  \\
  W &= \partial_{\mu} \big( \sigma^{\mu\nu}\partial_{\nu}\hat{d} \big) \,.
\end{aligned}\label{EH_action_aux}
\end{equation}
Then we may denote the EoM $\mathcal{G}^{\mu\nu}$ in terms of $\sigma$ \eqref{Gmunu} in a simple form
\begin{equation}
\begin{aligned}
  \mathcal{G}^{\mu\nu}=&
  - \frac{1}{2} \sigma^{\rho\sigma} X^{\mu\nu}_{\rho\sigma}
  + \sigma^{\rho(\mu} X^{|\sigma|\nu)}_{\rho\sigma}
  + \sigma^{\mu\rho}\sigma^{\nu\sigma} Y_{\rho\sigma}
  + Z^{\mu\nu}
  + \sigma^{\mu\nu} W
  \,.
\end{aligned}\label{Gmunu_XYZW}
\end{equation}
The Lagrangian in terms of $\sigma$ \eqref{EH_action_sigma} can be written using the contraction of $\mathcal{G}^{\mu\nu}$ with $\sigma_{\mu\nu}$, similar to that of Ricci scalar
\begin{equation}
  L_{\rm EH} = \sigma_{\mu\nu} G^{\mu\nu}\,,
\label{}\end{equation}
where
\begin{equation}
\begin{aligned}
  L_{\rm EH} &=
    X^{\mu\nu}_{\nu\mu}
  - \frac{1}{2} \sigma_{\mu\nu} \sigma^{\rho\sigma} X^{\mu\nu}_{\rho\sigma}
  + \sigma^{\mu\nu} Y_{\mu\nu}
  + \sigma_{\mu\nu} Z^{\mu\nu}
  + D W
  \\
  &= \frac{1}{2} X^{\mu\nu}_{\nu\mu} + \sigma^{\mu\nu} Y_{\mu\nu} + \partial_{\mu}\Big( 2\sigma^{\mu\nu} \partial_{\nu} \hat{d} - \frac{1}{2} \partial_{\nu} \sigma^{\mu\nu}\Big)\,.
\end{aligned}\label{Lag_aux}
\end{equation}
The last term above is a total derivative, and we may ignore it. Thus the EoM $\mathcal{G}^{\mu\nu}$ and the action are represented by $X^{\mu\nu}_{\rho\sigma}$, $Y_{\mu\nu}$, $Z^{\mu\nu}$ and $W$ only.

Interestingly perturbations of the abbreviations are given by the building blocks \eqref{building_blocks0} and \eqref{building_blocks} to all orders in $h$
\begin{equation}
\begin{aligned}
  X^{\mu\nu}_{\rho\sigma} &= \partial_{\rho}\partial_{\sigma}h^{\mu\nu} +\sum_{n=0}^{\infty} \big[\partial_{\rho}h h^{n} \partial_{\sigma} h\big]^{\mu\nu}\,,
  \\
  Y_{\rho\sigma} & =
  \sum_{n=0}^{\infty} \sum_{m=0}^{n} \bigg[ \frac{1}{4(D-2)} \tr\big[h^{m} \partial_{\rho} h\big]  \tr\big[h^{n-m} \partial_{\sigma} h\big]
  - \frac{1}{4}\tr\big[h^{m}\partial_{\rho}h h^{n-m} \partial_{\sigma} h\big]\bigg]\,,
  \\
  Z^{\mu\nu} &= \frac{1}{2} \partial_{\rho} h^{\rho\sigma} \partial_{\sigma} h^{\mu\nu}
  - \frac{1}{2} \partial_{\rho} h^{\sigma\mu} \partial_{\sigma} h^{\rho\nu}\,,
  \\
  W &= \frac{1}{2(D-2)} \bigg(
  	  (\eta-h)^{\mu\nu}\sum_{n=0}^{\infty} \bigg[ \tr\big[h^{n} \partial_{\mu}\partial_{\nu} h\big]
  	+ \sum_{m=0}^{n} \tr\big[h^{m}\partial_{\mu}h h^{n-m} \partial_{\nu} h\big]\bigg]
  	\\&\qquad\qquad\qquad
  	- \partial_{\mu}h^{\mu\nu}  \sum_{n=0}^{\infty} \tr\big[ h^{n} \partial_{\nu} h\big]
  \bigg)
  \,.
\end{aligned}\label{pert_auxiliary1}
\end{equation}

We can explicitly find the general $n$-th order terms for these fields in $h$. If we denote the $n$-th order terms of \eqref{pert_auxiliary1} as $[X_{n}], [Y_{n}], [Z_{n}]$ and $[W_{n}]$, each field is expanded as
\begin{equation}
\begin{aligned}
  X^{\mu\nu}_{\rho\sigma} = \sum_{n=1}^{\infty} [X_{n}]^{\mu\nu}_{\rho\sigma} \,,
  \qquad
  Y_{\rho\sigma} = \sum_{n=2}^{\infty} [Y_{n}]_{\rho\sigma} \,,
  \qquad
  Z^{\mu\nu} = [Z_{2}]^{\mu\nu}\,,
  \qquad
  W = \sum_{n=1}^{\infty} [W_{n}]\,,
\end{aligned}\label{}
\end{equation}
where their explicit expressions are as follows. The terms for $X^{\mu\nu}_{\rho\sigma}$ are
\begin{equation}
\begin{aligned}
 & [X_{1}]^{\mu\nu}_{\rho\sigma} = \partial_{\rho}\partial_{\sigma}h^{\mu\nu}
 \\
 &[X_{n}]^{\mu\nu}_{\rho\sigma} = \big[\partial_{\rho}h h^{n-2} \partial_{\sigma} h\big]^{\mu\nu} \,, \quad n\geq 2\,.
\end{aligned}\label{}
\end{equation}
For $Y_{\rho\sigma}$, the general expression when $n\geq2$ is
\begin{equation}
\begin{aligned}
  \big[Y_{n}\big]_{\rho\sigma} = \sum_{m=0}^{n-2} \frac{1}{4}\bigg[
   \frac{1}{D-2} \tr\big[h^{m} \partial_{\rho} h\big]  \tr\big[h^{n-m-2} \partial_{\sigma} h\big]
  - \tr\big[h^{m}\partial_{\rho}h h^{n-m-2} \partial_{\sigma} h\big]\bigg]\,.
\end{aligned}\label{}
\end{equation}
Since $Z^{\mu\nu}$ is precisely the second order, we omit it. Finally, the terms for $W$ are given by
\begin{equation}
\begin{aligned}
  & [W_{1}] =  \frac{1}{2(D-2)} \Box h
  \\
  & [W_{2}] =  \frac{1}{2(D-2)} \Big(\tr\big[h\Box h\big] -h^{\mu\nu} \partial_{\mu}\partial_{\nu}h + \tr \big[\partial^{\mu}h \partial_{\mu}h \big]
  	- \partial_{\mu} h^{\mu\nu} \partial_{\nu}h
  \Big)\,,
\end{aligned}\label{}
\end{equation}
and for $n \geq3$
\begin{equation}
\begin{aligned}
  \big[W_{n}\big] =  \frac{1}{2(D-2)} \Bigg( &
  	  \tr\big[h^{n-1} \Box h\big]
  	- h^{\mu\nu} \tr\big[h^{n-2} \partial_{\mu}\partial_{\nu} h\big]
  	+ \sum_{m=0}^{n-2} \tr\big[h^{m}\partial^{\mu}h h^{n-m-2} \partial_{\mu} h\big]
  	\\&
  	- \sum_{m=0}^{n-3} h^{\mu\nu}\tr\big[h^{m}\partial_{\mu}h h^{n-m-3} \partial_{\nu} h\big]
  	- \partial_{\mu} h^{\mu\nu} \tr\big[h^{n-2}\partial_{\nu}h\big]
  \Bigg)\,.
\end{aligned}\label{}
\end{equation}

Note that we can define a tensor $A_{\mu\nu}{}^{\rho}$ and its trace $A_{\mu}$ from the structures of the perturbations above
\begin{equation}
  A_{\mu\nu}{}^{\rho} = \sum_{n=0}^{\infty} (h^{n})_{\nu\sigma}\partial_{\mu} h^{\sigma\rho}\,,
\label{Q_field}\end{equation}
and
\begin{equation}
  A_{\mu} = \sum_{n=0}^{\infty} \tr\big[h^{n}\partial_{\mu}h\big]= \sum_{n=0}^{\infty}(h^{n})_{\nu\rho}\partial_{\mu} h^{\nu\rho}\,.
\label{tr_Q_field}\end{equation}
Then the perturbations of $X^{\mu\nu}_{\rho\sigma}$, $Y_{\mu\nu}$, $Z^{\mu\nu}$ and $W$ reduce to a compact form
\begin{equation}
\begin{aligned}
  X^{\mu\nu}_{\rho\sigma} &=
    \partial_{\rho}\partial_{\sigma}h^{\mu\nu}
  + \partial_{\rho}h^{\mu\kappa} A_{\sigma\kappa}{}^{\nu}\,,
  \\
  Y_{\rho\sigma} & =
 \frac{1}{4(D-2)} A_{\rho} A_{\sigma}
  - \frac{1}{4} A_{\rho\mu}{}^{\nu} A_{\sigma\nu}{}^{\mu}\,,
  \\
  Z^{\mu\nu} &=
    \frac{1}{2} \partial_{\rho} h^{\rho\sigma} \partial_{\sigma} h^{\mu\nu}
  - \frac{1}{2} \partial_{\rho} h^{\sigma\mu} \partial_{\sigma} h^{\rho\nu}\,,
  \\
  W &= \frac{1}{2(D-2)} \partial_{\mu}\Big((\eta-h)^{\mu\nu} A_{\nu} \Big)
  \,.
\end{aligned}\label{pert_auxiliary2}
\end{equation}

We can rearrange $L_{\rm EH}$ order by order in $h$, such as $L_{\rm EH} = \sum_{n=2}^{\infty} L_{n}$, where $L_{n}$ is the $n$-th order terms\footnote{ In fact, $L_{1}$ is nonvanishing, $L_{1} = \frac{1}{2} \partial_{\mu}\partial_{\nu}h^{\mu\nu}$, which is a total derivative. }. The quadratic order action is given by
\begin{equation}
\begin{aligned}
  L_{2} &= \frac{1}{2} [X_{2}]^{\mu\nu}_{\nu\mu} + \eta^{\mu\nu} [Y_{2}]_{\mu\nu}
  \\
  &= -\frac{1}{4} \partial^{\rho} h^{\mu\nu} \partial_{\rho} h_{\mu\nu}
  + \frac{1}{2} \partial^{\rho}h^{\mu\nu} \partial_{\mu} h_{\nu\rho}
  + \frac{1}{4(D-2)} \partial^{\mu}h \partial_{\mu} h\,,
\end{aligned}\label{2nd_L}
\end{equation}
and we find the general term of the expansion of the Lagrangian for $n \geq 3$
\begin{equation}
\begin{aligned}
 L_{n} &=
       \frac{1}{2} [X_{n}]^{\mu\nu}_{\nu\mu}
     + \eta^{\mu\nu} [Y_{n}]_{\mu\nu}
     - h^{\mu\nu} [Y_{n-1}]_{\mu\nu}
     \,, \qquad \text{for}~n\geq 3\,.
\end{aligned}\label{general_L}
\end{equation}
Their explicit expressions in terms of $h$ are
\begin{equation}
\begin{aligned}
  L_{2} &= - \frac{1}{4} \partial^{\rho} h^{\mu\nu} \partial_{\rho} h_{\mu\nu}
  + \frac{1}{2} \partial^{\rho}h^{\mu\nu} \partial_{\mu} h_{\nu\rho}
  + \frac{1}{4(D-2)} \tr\big[\partial^{\mu}h\big] \tr\big[\partial_{\mu} h\big]\,,
  \\
  L_{n} &=  \sum_{m=0}^{n-2}\bigg[ - \frac{1}{4} \tr\big[ h^{m} \partial^{\mu}h h^{n-m-2} \partial_{\mu} h\big]
	+ \frac{1}{4(D-2)} \tr\big[h^{m}\partial^{\mu}h\big] \tr\big[h^{n-m}\partial_{\mu}h\big] \bigg]
  \\&\quad
	+  h^{\mu\nu} \sum_{m=0}^{n-3} \bigg[ \frac{1}{4}  \tr\big[ h^{m} \partial_{\mu}h h^{n-m-3} \partial_{\nu} h\big]
	- \frac{1}{4(D-2)} \tr\big[h^{m}\partial_{\mu}h\big] \tr\big[h^{n-m-3}\partial_{\nu}h\big] \bigg]
	\\&\quad \
	+ \frac{1}{2} \big[\partial_{\nu}h h^{n-2} \partial_{\mu}h\big]^{\mu\nu} \,, \quad \text{for}~n\geq3\,.
\end{aligned}\label{n_th_order_Lag_sigma}
\end{equation}

Similarly, we may expand the EoM $\mathcal{G}^{\mu\nu}$ order by order in $h$, $\mathcal{G}^{\mu\nu} = \sum_{n>0} \mathcal{G}^{\mu\nu}_{n}$, where $\mathcal{G}^{\mu\nu}_{n}$ is the $n$-th order terms
\begin{equation}
\begin{aligned}
  \mathcal{G}^{\mu\nu}_{1} &=
    \eta^{\rho(\mu} [X_{1}]^{|\sigma|\nu)}_{\rho\sigma}
  - \frac{1}{2} \eta^{\rho\sigma} [X_{1}]^{\mu\nu}_{\rho\sigma}
  + \eta^{\mu\nu} [W_{1}]
  \\
  \mathcal{G}^{\mu\nu}_{2} &=
    \eta^{\rho(\mu} [X_{2}]^{|\sigma|\nu)}_{\rho\sigma}
  - h^{\rho(\mu} [X_{1}]^{|\sigma|\nu)}_{\rho\sigma}
  - \frac{1}{2} \eta^{\rho\sigma} [X_{2}]^{\mu\nu}_{\rho\sigma}
  + \frac{1}{2} h^{\rho\sigma} [X_{1}]^{\mu\nu}_{\rho\sigma}
  \\&\quad
  + \eta^{\mu\rho} \eta^{\nu\sigma} [Y_{2}]_{\rho\sigma}
  + Z_{2}^{\mu\nu}
  + \eta^{\mu\nu} [W_{2}]
  - h^{\mu\nu} [W_{1}]\,,
\end{aligned}\label{pert_G_order_123}
\end{equation}
and the general term for $n\geq 3$ is
\begin{equation}
\begin{aligned}
  \mathcal{G}^{\mu\nu}_{n} &=
    \eta^{\rho(\mu} [X_{n}]^{|\sigma|\nu)}_{\rho\sigma}
  - h^{\rho(\mu} [X_{n-1}]^{|\sigma|\nu)}_{\rho\sigma}
  - \frac{1}{2} \eta^{\rho\sigma} [X_{n}]^{\mu\nu}_{\rho\sigma}
  + \frac{1}{2} h^{\rho\sigma} [X_{n-1}]^{\mu\nu}_{\rho\sigma}
  \\&\quad
  +  [Y_{n}]^{\mu\nu}
  - 2\eta^{\rho(\mu} h^{\nu)\sigma} [Y_{n-1}]_{\rho\sigma}
  + h^{\mu\rho} h^{\nu\sigma} [Y_{n-2}]_{\rho\sigma}
  + \eta^{\mu\nu} [W_{n}]
  - h^{\mu\nu} [W_{n-1}]\,.
\end{aligned}\label{pert_G_order_n}
\end{equation}
The explicit form of each term is written as
\begin{equation}
\begin{aligned}
  \mathcal{G}^{\mu \nu}_{1} &=
  	\partial^{(\mu}\partial_{\rho}h^{\nu)\rho}
  - \frac{1}{2} \Box h^{\mu\nu}
  + \frac{1}{2(D-2)} \eta^{\mu\nu} \Box h\,,
  \\
  \mathcal{G}^{\mu \nu}_{2} &= -h^{\rho(\mu}
  	\partial_{\rho}\partial_{\sigma}h^{\nu)\sigma}
	+ \big[\partial^{(\mu} h  \partial_{\rho}h\big]^{\rho|\nu)}
  + \frac{1}{2} h^{\rho\sigma} \partial_{\rho}\partial_{\sigma}h^{\mu\nu}
	- \frac{1}{2} \big[\partial^{\rho}h \partial_{\rho} h\big]^{\mu\nu}
	\\&\quad
  + \frac{1}{2} \partial_{\rho}h^{\rho\sigma} \partial_{\sigma}h^{\mu\nu}
  - \frac{1}{2} \partial_{\sigma}h^{\rho\mu} \partial_{\rho}h^{\sigma\nu}
  + \frac{1}{4(D-2)} \partial^{\mu} h \partial^{\nu} h
  - \frac{1}{4} \tr\big[\partial^{\mu}h \partial^{\nu} h\big]
  \\&\quad
  + \frac{\eta^{\mu\nu}}{2(D-2)}\Big[
  	  \tr\big[h \Box h\big]
	- h^{\rho\sigma} \partial_{\rho}\partial_{\sigma}h
	+ \tr\big[\partial^{\rho}h \partial_{\rho}h\big]
	- \partial_{\rho} h^{\rho\sigma} \partial_{\sigma}h\Big]
    - \frac{h^{\mu\nu}}{2(D-2)} \Box h\,,
\end{aligned}\label{n_th_order_Gmunu1}
\end{equation}
\begin{equation}
\begin{aligned}
  \mathcal{G}^{\mu \nu}_{3} &= \big[\partial^{(\mu} h h \partial_{\sigma}h\big]^{|\sigma|\nu)}
	- \frac{1}{2} \big[\partial^{\rho}h h \partial_{\rho} h\big]^{\mu\nu}
	+ \frac{1}{2(D-2)} \tr\big[h \partial^{(\mu} h\big] \partial^{\nu)} h
 \\&\quad
	-h^{\rho(\mu}  \big[\partial_{\rho} h  \partial_{\sigma}h\big]^{|\sigma|\nu)}
	+ \frac{1}{2} h^{\rho\sigma} \big[\partial_{\rho} h \partial_{\sigma} h\big]^{\mu\nu}
   - \frac{1}{2(D-2)} h^{\rho(\mu} \partial_{\rho} h \partial^{\nu)} h
  \\&\quad
  - \frac{1}{2} \tr\big[\partial^{(\mu}h h \partial^{\nu)} h\big]
  + \frac{1}{2} h^{\rho(\mu} \tr\big[\partial_{\rho}h \partial^{\nu)} h\big]
  \\&\quad
  + \frac{1}{2(D-2)}\eta^{\mu\nu} \bigg[\tr\big[h^{2}\Box h\big]
	-h^{\rho\sigma} \tr\big[h\partial_{\rho}\partial_{\sigma}h\big]
	\\&\qquad\qquad\qquad\qquad\quad
  	+ 2 \tr\big[\partial^{\rho}h h \partial_{\rho}h\big]
	-h^{\rho\sigma} \tr\big[\partial_{\rho}h \partial_{\sigma}h\big]
  	- \partial_{\rho} h^{\rho\sigma} \tr\big[h\partial_{\sigma}h\big]\
  \bigg]
	\\&\quad
  - \frac{1}{2(D-2)}h^{\mu\nu} \bigg[
  	  \tr\big[h\Box h\big]
	-h^{\rho\sigma} \partial_{\rho}\partial_{\sigma}h
  	  + \tr\big[\partial^{\rho}h \partial_{\rho}h\big]
  	- \partial_{\rho} h^{\rho\sigma} \partial_{\sigma}h\
  \bigg]\,,
\end{aligned}\label{n_th_order_Gmunu1}
\end{equation}
and we have the general term for $n\geq4$,
\begin{equation}
\begin{aligned}
  \mathcal{G}^{\mu \nu}_{n} &=
    \big[\partial^{(\mu} h h^{n-2} \partial_{\sigma}h\big]^{\sigma|\nu)}
	-h^{\rho(\mu} \big[\partial_{\rho} h h^{n-3} \partial_{\sigma}h\big]^{\sigma|\nu)}
  \\&\quad
  - \frac{1}{2} \big[\partial^{\rho}h h^{n-2} \partial_{\rho} h\big]^{\mu\nu}
  + \frac{1}{2} h^{\rho\sigma}\big[\partial_{\rho}h h^{n-3} \partial_{\sigma} h\big]^{\mu\nu}
  \\&\quad
  + \frac{1}{4} \sum_{m=0}^{n-2}\Bigg[ \frac{\tr\big[h^{m} \partial^{\mu} h\big] \tr\big[h^{n-m-2} \partial^{\nu} h\big]}{D-2}
  	{-} \tr\big[h^{m}\partial^{\mu}h h^{n-m-2} \partial^{\nu} h\big]
  \Bigg]
	\\&\quad
  - \frac{1}{2}h^{\rho(\mu}  \sum_{m=0}^{n-3}\Bigg[ \frac{\tr\big[h^{m} \partial_{\rho} h\big] \tr\big[h^{n-m-3} \partial^{\nu)} h\big]}{D-2}
  	{-} \tr\big[h^{m}\partial_{\rho}h h^{n-m-3} \partial^{\nu)} h\big]
  \Bigg]
	\\&\quad
  + \frac{1}{4}h^{\mu\rho} h^{\nu\sigma}
  \sum_{m=0}^{n-4}\Bigg[
  	  \frac{\tr\big[h^{m} \partial_{\rho} h\big] \tr\big[h^{n-m-4} \partial_{\sigma} h\big]}{D-2}
  	{-} \tr\big[h^{m}\partial_{\rho}h h^{n-m-4} \partial_{\sigma} h\big]
  \Bigg]
  \\&\quad
  + \frac{\eta^{\mu\nu}}{2(D-2)} \bigg[ \  \tr\big[h^{n-1}\Box h\big]
	+ \sum_{m=0}^{n-2}\tr\big[h^{m}\partial^{\rho}h h^{n-m-2} \partial_{\rho}h\big]
	- \partial_{\rho} h^{\rho\sigma} \tr\big[h^{n-2}\partial_{\sigma}h\big]
  	  \\&\qquad\qquad\qquad\qquad\quad
	-h^{\rho\sigma} \tr\big[h^{n-2}\partial_{\rho}\partial_{\sigma}h\big]
	-h^{\rho\sigma} \sum_{m=0}^{n-3}\tr\big[h^{m}\partial_{\rho}h h^{n-m-3} \partial_{\sigma}h\big] \bigg]
	\\&\quad
  - \frac{h^{\mu\nu}}{2(D-2)} \bigg[
  \tr\big[h^{n-2}\Box h\big]
	+ \sum_{m=0}^{n-3}\tr\big[h^{m}\partial^{\rho}h h^{n-m-3} \partial_{\rho}h\big]
	- \partial_{\rho} h^{\rho\sigma} \tr\big[h^{n-3}\partial_{\sigma}h\big]\
  	  \\&\qquad\qquad\qquad\qquad\qquad
	-h^{\rho\sigma} \tr\big[h^{n-3}\partial_{\rho}\partial_{\sigma}h\big]
	-h^{\rho\sigma} \sum_{m=0}^{n-4}\tr\big[h^{m}\partial_{\rho}h h^{n-m-4} \partial_{\sigma}h\big] \bigg]\,.
\end{aligned}\label{n_th_order_Gmunu2}
\end{equation}
%

\subsection{Perturbations of Riemannian geometry in metric}\label{App:A.2}
We now derive the perturbations of Riemannian geometry in the metric formulation. As in the tensor density formalism, we start by introducing a set of abbreviations by replacing the square brackets in the Ricci tensor in terms of metric $g$ appearing in \eqref{Ricci_tensor_g}
\begin{equation}
\begin{aligned}
  \check{X}^{\mu\nu}_{\rho\sigma} &=
  - \partial_{\rho}\partial_{\sigma} g^{\mu\nu}
  + 2\partial_{\rho} g^{\mu\nu} \partial_{\sigma} \hat{d}
  - \partial_{\rho}g^{\mu\kappa} \partial_{\sigma}g_{\kappa\lambda} g^{\lambda\nu}\,,
  \\
  \check{Y}_{\mu\nu} &=  \frac{1}{4} \partial_{\mu} g^{\rho\sigma} \partial_{\nu}g_{\rho\sigma}
  + 2 \partial_{\mu}\partial_{\nu} \hat{d}\,,
  \\
  \check{Z}^{\mu\nu} &=
    \frac{1}{2} \partial_{\rho}g^{\rho\sigma} \partial_{\sigma} g^{\mu\nu}
  - \frac{1}{2} \partial_{\rho} g^{\sigma\mu} \partial_{\sigma} g^{\rho\nu} \,.
\end{aligned}\label{abbreviation_g}
\end{equation}
Unlike the previous case, it is sufficient to define three fields only. 

We may recast the Ricci curvatures as follows: 
\begin{equation}
\begin{aligned}
  R^{\mu\nu} &=
  - \frac{1}{2} g^{\rho\sigma} \check{X}^{\mu\nu}_{\rho\sigma}
  + g^{\rho(\mu} \check{X}^{|\sigma|\nu)}_{\rho\sigma}
  + g^{\mu\rho} g^{\nu\sigma} \check{Y}_{\rho\sigma}
  + \check{Z}^{\mu\nu}\,,
\end{aligned}\label{Ricci_tensor_aux}
\end{equation}
and
\begin{equation}
\begin{aligned}
  R &=
  - \frac{1}{2} g_{\mu\nu} g^{\rho\sigma} \check{X}^{\mu\nu}_{\rho\sigma}
  + \check{X}^{\mu\nu}_{\nu\mu}
  + g^{\mu\nu} \check{Y}_{\mu\nu}
  + g_{\mu\nu} \check{Z}^{\mu\nu}
  \\
  &=
  \frac{1}{2} \check{X}^{\mu\nu}_{\nu\mu}
  	+ g^{\mu\nu} \check{Y}_{\mu\nu}
  	- e^{2\hat{d}} \partial_{\mu} \bigg[\partial_{\nu}\big(e^{-2\hat{d}} g^{\mu\nu}\big) - \frac{1}{2} e^{-2\hat{d}}\partial_{\nu}g^{\mu\nu} \bigg]\,.
\end{aligned}\label{Ricci_scalar_aux}
\end{equation}
Then the Einstein-Hilbert action is given, up to total derivatives, by
\begin{equation}
  S_{\rm EH} = \frac{1}{2\kappa^{2}} \int \mathrm{d}^{D} x e^{-2\hat{d}} \bigg[\frac{1}{2} \check{X}^{\mu\nu}_{\nu\mu} + g^{\mu\nu} \check{Y}_{\mu\nu}\bigg]\,.
\label{EH_action_R_aux}\end{equation}
Thus perturbations of the abbreviations produce perturbations of the Ricci curvatures.

It is straightforward to check that perturbations of $\check{X}^{\mu\nu}_{\rho\sigma}$, $\check{Y}_{\mu\nu}$, and $\check{Z}^{\mu\nu}$ are generated by the building blocks for the metric formulation in \eqref{building_blocks_g}. Substituting perturbations of the building blocks \eqref{building_blocks_g_pertub}, we have
\begin{equation}
\begin{aligned}
  \check{X}^{\mu\nu}_{\rho\sigma} &=
    \partial_{\rho}\partial_{\sigma} h^{\mu\nu}
  + \frac{1}{2} \partial_{\rho}h^{\mu\nu} \sum_{n=0}^{\infty} \tr\big[ h^{n} \partial_{\sigma} h\big]
  + \sum_{n=0}^{\infty} \big[\partial_{\rho}h h^{n} \partial_{\sigma} h\big]^{\mu\nu}\,,
  \\
  \check{Y}_{\mu\nu} &=
  - \sum_{n=0}^{\infty} \Bigg[
      \frac{1}{2} \tr\Big[ h^{n} \partial_{\mu}\partial_{\nu} h\Big]
  	+ \frac{3}{4} \sum_{m=0}^{n} \tr\Big[ h^{m} (\partial_{\mu}h) h^{n-m} (\partial_{\nu} h)\Big]
  \Bigg]\,,
  \\
  \check{Z}^{\mu\nu} &=
    \frac{1}{2} \partial_{\rho} h^{\rho\sigma} \partial_{\sigma} h^{\mu\nu}
  - \frac{1}{2} \partial_{\rho} h^{\sigma\mu} \partial_{\sigma} h^{\rho\nu} \,.
\end{aligned}\label{perturb_auxiliary_g}
\end{equation}
We may also expand them order by order in $h$
\begin{equation}
  \check{X}^{\mu\nu}_{\rho\sigma} = \sum_{n=1}^{\infty} \big[\check{X}_{n}\big]^{\mu\nu}_{\rho\sigma}\,,
  \qquad
  \check{Y}_{\rho\sigma} = \sum_{n=1}^{\infty} \big[\check{Y}_{n}\big]_{\rho\sigma}\,,
  \qquad
  \check{Z}^{\mu\nu}_{\rho\sigma} = \big[\check{Z}_{2}\big]^{\mu\nu}\,,
\label{}\end{equation}
where $\check{X}_{n}$ are
\begin{equation}
\begin{aligned}
  \big[\check{X}_{1}\big]^{\mu\nu}_{\rho\sigma} &= \partial_{\rho}\partial_{\sigma}h^{\mu\nu}\,,
  \\
  \big[\check{X}_{n}\big]^{\mu\nu}_{\rho\sigma} &=
    \frac{1}{2} \partial_{\rho}h^{\mu\nu} \tr\big[ h^{n-2} \partial_{\mu} h\big]
  + \big[\partial_{\rho}h h^{n-2} \partial_{\sigma} h\big]^{\mu\nu}\,, \quad \text{for} \geq2 \,,
\end{aligned}\label{}
\end{equation}
and $\check{Y}_{n}$ are
\begin{equation}
\begin{aligned}
  \big[\check{Y}_{1}\big]_{\mu\nu} &= -\frac{1}{2} \partial_{\mu} \partial_{\nu} h\,,
  \\
  \big[\check{Y}_{n}\big]_{\mu\nu} &= -
      \frac{1}{2} \tr\big[ h^{n-1} \partial_{\mu}\partial_{\nu} h\Big]
  	- \frac{3}{4} \sum_{m=0}^{n-2} \tr\Big[ h^{m} (\partial_{\mu}h) h^{n-m-2} (\partial_{\nu} h)\Big]\,,
  \quad \text{for} \geq2 \,.
\end{aligned}\label{}
\end{equation}
Since the order of $\check{Z}$ is already fixed as two, we do not need to find its general form.

We may also rearrange the perturbations of $R^{\mu\nu}$ order-by-order in $h$, $R^{\mu\nu} = \sum_{n=1}^{\infty} R^{\mu\nu}_{n}$, where the terms at each order are
\begin{equation}
\begin{aligned}
  R_{1}^{\mu\nu} =&
    - \frac{1}{2} \eta^{\rho\sigma} [\check{X}_{1}]^{\mu\nu}_{\rho\sigma}
  + \eta^{\rho(\mu} [\check{X}_{1}]^{|\sigma|\nu)}_{\rho\sigma}
  + \eta^{\mu\rho} \eta^{\nu\sigma} [\check{Y}_{1}]_{\rho\sigma}
  \\
  R_{2}^{\mu\nu} =&
  - \frac{1}{2} \Big(
  	  \eta^{\rho\sigma} [\check{X}_{2}]^{\mu\nu}_{\rho\sigma}
  	- h^{\rho\sigma} [\check{X}_{1}]^{\mu\nu}_{\rho\sigma}
  \Big)
  + \eta^{\rho(\mu} [\check{X}_{2}]^{|\sigma|\nu)}_{\rho\sigma}
  - h^{\rho(\mu} [\check{X}_{1}]^{|\sigma|\nu)}_{\rho\sigma}
  \\&
  + \eta^{\mu\rho} \eta^{\nu\sigma} [\check{Y}_{2}]_{\rho\sigma}
  - 2\eta^{\rho(\mu} h^{\nu)\sigma} [\check{Y}_{1}]_{\rho\sigma}
  + [\check{Z}_{2}]^{\mu\nu}\,,
\end{aligned}\label{}
\end{equation}
and we have the general term for $n\geq3$
\begin{equation}
\begin{aligned}
  R_{n}^{\mu\nu} =&
  - \frac{1}{2} \Big(
  	  \eta^{\rho\sigma} [\check{X}_{n}]^{\mu\nu}_{\rho\sigma}
  	- h^{\rho\sigma} [\check{X}_{n-1}]^{\mu\nu}_{\rho\sigma}
  \Big)
  + \eta^{\rho(\mu} [\check{X}_{n}]^{|\sigma|\nu)}_{\rho\sigma}
  - h^{\rho(\mu} [\check{X}_{n-1}]^{|\sigma|\nu)}_{\rho\sigma}
  \\&
  + \eta^{\mu\rho} \eta^{\nu\sigma} [\check{Y}_{n}]_{\rho\sigma}
  - 2\eta^{\rho(\mu} h^{\nu)\sigma} [\check{Y}_{n-1}]_{\rho\sigma}
  + h^{\mu\rho} h^{\nu\sigma} [\check{Y}_{n-2}]_{\rho\sigma}\,.
\end{aligned}\label{}
\end{equation}

Substituting the perturbations of $\check{X}^{\mu\nu}_{\rho\sigma}$, $\check{Y}_{\mu\nu}$, and $\check{Z}^{\mu\nu}$ in the above into \eqref{Ricci_scalar_aux}, we can derive the order-by-order expansion of the Ricci scalar $R = \sum_{n=1}^{\infty} R_{n}$, where
\begin{equation}
\begin{aligned}
  R_{1} =&\
  \partial_{\mu} \partial_{\nu} h^{\mu \nu} - \square h \,,
\end{aligned}\label{}
\end{equation}
and for $n\geq 2$
\begin{equation}
\begin{aligned}
  R_{n} =&\
    \frac{1}{2} \big[\partial_{\nu} h h^{n-2} \partial_{\mu}h\big]^{\mu\nu}
  + \partial_\mu h^{\mu\nu} \operatorname{tr}[h^{n-2} \partial_\nu h]
  + \tr\big[h^{n-1} \Box h \big]
  + h^{\mu\nu} \tr\big[h^{n-2} \partial_\mu \partial_\nu h \big]
  \\&
  - \sum_{p=0}^{n-2} \bigg(
  	  \frac{5}{4} \tr\big[h^p \partial^{\mu} h h^{n-p-2} \partial_{\mu} h \big]
  	+ \frac{1}{4} \tr[h^{p} \partial^\mu h] \tr[h^{n-p-2} \partial_\mu h]
  \bigg)
  \\&
  + \sum_{p=0}^{n-3} h^{\mu \nu} \bigg(
    \frac{5}{4}\tr \big[h^p \partial_{\mu} h h^{n-p-3} \partial_\nu h \big]
  + \frac{1}{4} \tr[h^{p} \partial_\mu h] \tr[h^{n-p-3} \partial_{\nu}h]
  \bigg) \,.
\end{aligned}\label{}
\end{equation}

Similarly, perturbations of the Ricci tensor $R^{\mu\nu}$ are
\begin{equation}
\begin{aligned}
  R_{1}^{\mu\nu} &=- \frac{1}{2} \Box h^{\mu\nu}
  + \partial^{(\mu}\partial_{\rho}h^{\nu)\rho}
  -  \frac{1}{2} \tr\big[\partial^{\mu}\partial^{\nu}h\big]\,,
  \\
  R_{2}^{\mu\nu} &= \frac{1}{2} h^{\rho\sigma} \partial_{\rho}\partial_{\sigma} h^{\mu\nu}
  - \frac{1}{2} \big[\partial^{\rho}h  \partial_{\rho}h\big]^{\mu\nu}
  - \frac{1}{4}  \partial^{\rho}h^{\mu\nu} \partial_{\rho}h
  - h^{\rho(\mu} \partial_{\rho}\partial_{\sigma}h^{\nu)\sigma}
  + \frac{1}{2} \partial_{\rho}h^{\rho(\mu} \partial^{\nu)}h
  \\&
  - \frac{1}{2} \tr\big[h \partial^{\mu}\partial^{\nu}h\big]
  + h^{\rho(\mu} \partial^{\nu)}\partial_{\rho}h
  - \frac{3}{4} \tr\big[\partial^{\mu} h \partial^{\nu}h\big]
  + \frac{1}{2} \partial_{\rho} h^{\rho\sigma} \partial_{\sigma}h^{\mu\nu}
  - \frac{1}{2} \partial_{\rho} h^{\sigma\mu} \partial_{\sigma}h^{\rho\nu}\,,
  \\
  R_{3}^{\mu\nu} &=
  - \frac{1}{2} \big[\partial^{\rho}h h \partial_{\rho}h\big]^{\mu\nu}
  - \frac{1}{4} h^{\rho\sigma} \partial_{\rho}h^{\mu\nu} \partial_{\sigma}h
  + \frac{1}{2} h^{\rho\sigma} \big[\partial_{\rho}h \partial_{\sigma}h\big]^{\mu\nu}
  + \frac{1}{4} h^{\rho\sigma} \partial_{\sigma}h^{\mu\nu}\partial_{\rho}h
  \\&
  + \big[\partial^{(\mu}h h \partial_{\rho}h\big]^{|\rho|\nu)}
  	 + \frac{1}{2}  \partial_{\rho}h^{\rho(\mu} \tr\big[h\partial^{\nu)}h\big]
  - \frac{1}{2} \tr\big[h^{2}\partial^{\mu}\partial_{\nu}h\big] - \frac{3}{2} \tr\big[h\partial^{(\mu} h \partial^{\nu)}h\big]
  \\&
  - h^{\rho(\mu} \big[\partial_{\rho}h \partial_{\sigma}h\big]^{|\sigma|\nu)}
  - \frac{1}{2} h^{\rho(\mu}  \partial_{\sigma}h^{\nu)\sigma} \partial_{\rho}h
  +  h^{\rho (\mu} \tr\big[h \partial_{\rho}\partial^{\nu)}h\big]
  \\&
  + \frac{3}{2} h^{\rho (\mu} \tr\big[\partial_{\rho} h \partial^{\nu)}h\big]
  - \frac{1}{2} h^{\mu\rho}h^{\nu\sigma} \partial_{\rho}\partial_{\sigma}h\,,
\end{aligned}\label{general_n_Rmunu1}
\end{equation}
and we have the general term for $n\geq4$
\begin{equation}
\begin{aligned}
  R_{n}^{\mu\nu} &=- \frac{1}{2} \big[\partial^{\rho}h h^{n-2} \partial_{\rho}h\big]^{\mu\nu}
  	 - \frac{1}{4} \partial^{\rho}h^{\mu\nu} \tr\big[h^{n-2}\partial_{\rho}h\big]
  + \frac{1}{2} h^{\rho\sigma} \big[\partial_{\rho}h h^{n-3} \partial_{\sigma}h\big]^{\mu\nu}
  \\&\quad
  + \frac{1}{4} h^{\rho\sigma} \partial_{\sigma}h^{\mu\nu} \tr\big[h^{n-3}\partial_{\rho}h\big]
  + \big[\partial^{(\mu}h h^{n-2} \partial_{\rho}h\big]^{|\rho|\nu)}
  + \frac{1}{2} \partial_{\rho}h^{\rho(\mu} \tr\big[h^{n-2}\partial^{\nu)}h\big]
  \\&\quad
  - h^{\rho(\mu} \big[\partial_{\rho}h h^{n-3} \partial_{\sigma}h\big]^{|\sigma|\nu)}
  	- \frac{1}{2} h^{\rho(\mu} \partial_{\sigma}h^{\nu)\sigma} \tr\big[h^{n-3}\partial_{\rho}h\big]
  \\&\quad
   - \frac{1}{2} \tr\big[h^{n-1}\partial^{\mu}\partial^{\nu}h\big]
  + h^{\rho(\mu} \tr\big[h^{n-2}\partial_{\rho}\partial^{\nu)}h\big]
  - \frac{1}{2} h^{\mu\rho}h^{\nu\sigma} \tr\big[h^{n-3}\partial_{\rho}\partial_{\sigma}h\big]
  \\&\quad
  - \frac{3}{4} \sum_{m=0}^{n-2} \tr\big[h^{m}\partial^{\mu} h h^{n-m-2} \partial^{\nu}h\big]
  + \frac{3}{2} h^{\rho(\mu} \sum_{m=0}^{n-3} \tr\big[h^{m}\partial_{\rho} h h^{n-m-3} \partial^{\nu)}h\big]
  \\&\quad
  - \frac{3}{4} h^{\mu\rho}h^{\nu\sigma}\sum_{m=0}^{n-4} \tr\big[h^{m}\partial_{\rho} h h^{n-m-4} \partial_{\sigma}h\big]\,.
\end{aligned}\label{general_n_Rmunu2}
\end{equation}

\subsubsection{In the other perturbation conventions}
We now present perturbations in the other conventions. We will discuss two examples: the conventional perturbations, $g_{\mu\nu} = \eta_{\mu\nu} + h'_{\mu\nu}$ and the exponential perturbations $g^{\mu\nu} = \big[e^{-h'}\big]^{\mu\nu}$. We shall denote the explicit form of $\check{X}^{\mu \nu}_{\rho \sigma}$ and $\check{Y}_{\mu\nu}$. First, the usual metric perturbations \eqref{usual_metric_pert} are
\begin{equation}
\begin{aligned}
  \check{X}^{\mu \nu}_{\rho \sigma} =&\
    \partial_{\rho} \partial_{\sigma} h^{\mu \nu}
  + \frac{1}{2} \partial_{\rho} h^{\mu \nu} \partial_{\sigma} h
  - 2 [ \partial_{(\rho} h \partial_{\sigma)} h]^{\mu \nu}
  - [ \partial_{\rho} \partial_{\sigma} h h]^{\mu \nu}
    - [h \partial_{\rho} \partial_{\sigma} h ]^{\mu \nu}
  \\&
  + [ \partial_{\rho} \partial_{\sigma} h h^2]^{\mu \nu}
  + [h^2 \partial_{\rho} \partial_{\sigma} h]^{\mu \nu}
  + 2 [ h \partial_{(\rho} h \partial_{\sigma)} h]^{\mu \nu}
  + [h \partial_{\rho} \partial_{\sigma} h h]^{\mu \nu}
  \\&
  - [h \partial_{(\rho} h]^{\mu \nu} \partial_{\sigma)} h
  - \frac{1}{2} [\partial_{\rho} h h]^{\mu \nu} \partial_{\sigma} h
  + 2 [\partial_{(\rho} h \partial_{\sigma)} h h]^{\mu \nu}
  + 2 [\partial_{(\rho} h h \partial_{\sigma)} h]^{\mu \nu}
  \\&
  - [\partial_{\rho} \partial_{\sigma} h h^3]^{\mu \nu}
  - [h^3 \partial_{\rho} \partial_{\sigma} h]^{\mu \nu}
  - [h \partial_{\rho} \partial_{\sigma} h h^2]^{\mu \nu}
  - [h^2 \partial_{\rho} \partial_{\sigma} h h]^{\mu \nu}
  \\&
  - 2 [h^2 \partial_{(\rho} h \partial_{\sigma)} h]^{\mu \nu}
  - 2 [ \partial_{(\rho} h \partial_{\sigma)} h h^2]^{\mu \nu}
  - 2 [\partial_{(\rho} h h^2 \partial_{\sigma)} h]^{\mu \nu}
  + \frac{1}{2} \partial_{\sigma} h [h^2 \partial_{\rho} h]^{\mu \nu}
  \\&
  + \frac{1}{2} \partial_{\sigma} h [\partial_{\rho} h h^2]^{\mu \nu}
  + \frac{1}{2} \partial_{\rho} h^{\mu \nu} \tr[h^2 \partial_{\sigma} h]
  - 2 [h \partial_{(\rho} h h \partial_{\sigma)} h]^{\mu \nu}
  - 2 [h \partial_{(\rho} h \partial_{\sigma)}h h]^{\mu \nu}
  \\&
  - 2 [ \partial_{(\rho} h h \partial_{\sigma)}h h]^{\mu \nu}
  + \frac{1}{2} [ h \partial_{\rho} h]^{\mu \nu} \tr[h \partial_{\sigma} h]
  + \partial_{(\rho} h [h \partial_{\sigma)}h h]^{\mu \nu} + \mathcal{O}(h^5)\,,
\end{aligned}\label{}
\end{equation}
and
\begin{equation}
\begin{aligned}
  \check{Y}_{\mu \nu} =&
  - \frac{1}{2} \tr[ \partial_{\mu} \partial_{\nu} h]
  + \frac{1}{4} \tr[ \partial_{\mu} h \partial_{\nu} h]
  + \frac{1}{2} \tr[ h \partial_{\mu} \partial_{\nu} h]
  - \frac{1}{2} \tr[h^2 \partial_{\mu} \partial_{\nu} h]
  \\&
  - \frac{1}{2} \tr[ h \partial_{\mu} h \partial_{\nu} h]
  + \frac{1}{2} \tr[ h^3 \partial_{\mu} \partial_{\nu} h]
  + \frac{1}{2} \tr[ \partial_{\mu} h \partial_{\nu} h h^2]
  + \frac{1}{4} \tr [ \partial_{\mu} h h \partial_{\nu} h h]  + \mathcal{O}(h^5)\,.
\end{aligned}\label{}
\end{equation}

Next, we consider the exponential metric perturbations \eqref{exp_metric_pert}
\begin{equation}
\begin{aligned}
  \check{X}^{\mu \nu}_{\rho \sigma} =& \
     \partial_{\rho} \partial_{\sigma} h^{\mu \nu}
   - [\partial_{(\rho} h \partial_{\sigma)} h]^{\mu \nu}
   - \frac{1}{2} [\partial_{\rho} \partial_{\sigma} h h]^{\mu \nu}
   - \frac{1}{2} [h \partial_{\rho} \partial_{\sigma} h]^{\mu \nu}
   + \frac{1}{6} [\partial_{\rho} \partial_{\sigma} h h^2]^{\mu \nu}
   \\&
   + \frac{1}{6} [h^2 \partial_{\rho} \partial_{\sigma} h]^{\mu \nu}
   + \frac{1}{6} [h \partial_{\rho} \partial_{\sigma} hh]^{\mu \nu}
   - \frac{1}{4} \tr[\partial_{\sigma} h] [h\partial_{\rho} h]^{\mu \nu}
   - \frac{1}{4} \tr[\partial_{\sigma} h] [\partial_{\rho} h h]^{\mu \nu}
   \\&
   + \frac{1}{3} [h \partial_{(\rho} h \partial_{\sigma)} h]^{\mu \nu}
   + \frac{1}{3} [\partial_{(\rho} h \partial_{\sigma)} h h]^{\mu \nu}
   + \frac{1}{3} [ \partial_{(\rho} h  h \partial_{\sigma)} h]^{\mu \nu}
   \\&
   - \frac{1}{24} [\partial_{\rho} \partial_{\sigma} h h^3]^{\mu \nu}
   - \frac{1}{24} [h^3 \partial_{\rho} \partial_{\sigma} h]^{\mu \nu}
   - \frac{1}{24} [h  \partial_{\rho} \partial_{\sigma} h h^2]^{\mu \nu}
   - \frac{1}{24} [h^2 \partial_{\rho} \partial_{\sigma} h h]^{\mu \nu}
   \\&
   + \frac{1}{12} \tr[\partial_{\sigma} h] [h^2 \partial_{\rho} h]^{\mu \nu}
   + \frac{1}{12} \tr[\partial_{\sigma} h] [\partial_{\rho} h h^2]^{\mu \nu}
   + \frac{1}{12} \tr[\partial_{\sigma} h] [ h \partial_{\rho} h h]^{\mu \nu}
   \\&
   - \frac{1}{12} [ h^2 \partial_{(\rho} h \partial_{\sigma)}h h]^{\mu \nu}
   - \frac{1}{12} [ \partial_{(\rho} h \partial_{\sigma)} h h^2]^{\mu \nu}
   - \frac{1}{12} [\partial_{(\rho} h h^2 \partial_{\sigma)} h]^{\mu \nu}
   \\&
   - \frac{1}{12} [ h \partial_{(\rho} h h \partial_{\sigma)}h ]^{\mu \nu}
   - \frac{1}{12} [ h \partial_{(\rho} h \partial_{\sigma)} h h]^{\mu \nu}
   - \frac{1}{12} [\partial_{(\rho} h h \partial_{\sigma)}h h]^{\mu \nu}   + \mathcal{O}(h^5)\,,
\end{aligned}\label{}
\end{equation}
and
\begin{equation}
\begin{aligned}
 \check{Y}_{\mu \nu} =&  -\frac{1}{2} \tr[ \partial_{\mu} \partial_{\nu} h]
- \frac{1}{4} \tr[ \partial_{\mu} h \partial_{\nu} h]
- \frac{1}{4} \tr[ h \partial_{\mu} \partial_{\nu} h]
+ \frac{1}{4} \tr[h \partial_{\mu} \partial_{\nu} h]   \\
& - \frac{1}{24} \tr[ \partial_{\mu} h \partial_{\nu} h h^2]
+ \frac{1}{24} \tr[ \partial_{\mu} h h \partial_{\nu} h h] + \mathcal{O}(h^5)\,.
\end{aligned}\label{}
\end{equation}
%

\subsection{Einstein-Dilaton Theory}

We now introduce a set of abbreviations $X^{\mu\nu}_{\rho\sigma}$, $\tilde{Y}_{\mu\nu}$, $Z^{\mu\nu}$ and $\tilde{W}$ for abbreviating the square brackets in $\tilde{G}^{\mu\nu}$ \eqref{tilde_Gmunu}. As the notation indicates, these are similar to the previous fields in \eqref{EH_action_aux}, and $\tilde{Y}_{\mu\nu}$ and $\tilde{W}$ are on-shell equivalent to $Y_{\mu\nu}$ and $W$, respectively,
\begin{equation}
\begin{aligned}
  Y_{\mu\nu} &\to \tilde{Y}_{\mu\nu} = \frac{1}{4}\partial_{\mu}\sigma^{\rho\sigma}\partial_{\nu}\sigma_{\rho\sigma}
  + \partial_{\mu} d \big(\sigma^{\rho\sigma} \partial_{\nu}\sigma_{\rho\sigma}\big)
  - (D-2) \partial_{\mu}d \partial_{\nu}d\,,
  \\
  W &\to \tilde{W} = \partial_{\mu} \big( \sigma^{\mu\nu}\partial_{\nu}d\big) \,,
\end{aligned}\label{aux_tilde}
\end{equation}
while $X^{\mu\nu}_{\rho\sigma}$ and $Z^{\mu\nu}$ are the same as before because they do not contain $\hat{d}$. Using them, we may rewrite the action \eqref{pert_action_1} as
\begin{equation}
  \tilde{L}_{\rm ED} = \frac{1}{2} X^{\mu\nu}_{\nu\mu} + \sigma^{\mu\nu} \tilde{Y}_{\mu\nu} \,,
\label{ED_action_XY}\end{equation}
and also its EoM
\begin{equation}
\begin{aligned}
  \tilde{G}^{\mu\nu}=&\
    \sigma^{\rho(\mu} X^{|\sigma|\nu)}_{\rho\sigma}
  - \frac{1}{2} \sigma^{\rho\sigma} X^{(\mu\nu)}_{\rho\sigma}
  + \sigma^{\mu\rho}\sigma^{\nu\sigma} \tilde{Y}_{(\rho\sigma)}
  + Z^{\mu\nu}
  + \sigma^{\mu\nu} \tilde{W}
  \,.
\end{aligned}\label{tildeGmunu}
\end{equation}

Next, let us consider perturbations of $X^{\mu\nu}_{\rho\sigma}$, $\tilde{Y}_{\mu\nu}$, $Z^{\mu\nu}$ and $\tilde{W}$ arising from our second formulation of GR. Since the perturbations of $X^{\mu\nu}_{\rho\sigma}$ and $Z^{\mu\nu}$ are the same as in the previous case, we will focus on $\tilde{Y}_{\mu\nu}$ and $\tilde{W}$ only. We also denote perturbations of $d$ as
\begin{equation}
  d= d_{0} + f\,,
\label{}\end{equation}
where $d_{0}$ is a trivial background value, $d_{0}= 0$, and $f$ is linear fluctuation. Then $\tilde{Y}_{\mu\nu}$ and $\tilde{W}$ are expanded as
\begin{equation}
\begin{aligned}
  \tilde{Y}_{\mu\nu} & =
  \sum_{n=0}^{\infty} \bigg[ \tr\big[h^{n} \partial_{(\mu} h\big]  \partial_{\nu)}f
  - \frac{1}{4}\sum_{m=0}^{n} \tr\big[h^{m}\partial_{\mu}h h^{n-m} \partial_{\nu} h\big]\bigg]
  - (D-2)\partial_{\mu}f \partial_{\nu} f
  \\
  \tilde{W} &= \partial_{\rho} \Big( (\eta-h)^{\rho\sigma} \partial_{\sigma} f\Big)= (\eta-h)^{\rho\sigma} \partial_{\rho} \partial_{\sigma} f- \partial_{\rho} h^{\rho\sigma} \partial_{\sigma} f \,.
\end{aligned}\label{pert_auxiliary_2}
\end{equation}
The perturbative expansion of $\tilde{W}$ terminates at the quadratic order, and $\tilde{Y}_{\rho\sigma}$ is expanded by
\begin{equation}
\begin{aligned}
   & [\tilde{Y}_{2}]_{\rho\sigma} = \partial_{(\rho} h \partial_{\sigma)} f
   - \frac{1}{4} \tr\big[\partial_{\rho}h \partial_{\sigma} h\big]
   - (D-2) \partial_{\rho}f \partial_{\sigma} f
   \\
   & [\tilde{Y}_{n}]_{\rho\sigma} = \tr\big[h^{n-2} \partial_{(\rho} h\big]  \partial_{\sigma)}f
  - \frac{1}{4}\sum_{m=0}^{n-2} \tr\big[h^{m}\partial_{\rho}h h^{n-m-2} \partial_{\sigma} h\big]\,,~~ \text{for~} n\geq 3\,.
\end{aligned}\label{}
\end{equation}

Their general $n$-th order terms can be obtained similarly by treating $h_{\mu\nu}$ and $f$ as linear-order fluctuations. This is simply achieved by replacing $[Y_{n}]_{\mu\nu} \to [\tilde{Y}_{n}]_{\mu\nu}$ and $[W_{n}] \to [\tilde{W}]_{n}$ in the expansion of $G^{\mu\nu}$ \eqref{pert_G_order_123} and \eqref{pert_G_order_n}.

Finally, we consider perturbations of the action \eqref{ED_action_sigma_d}. Substituting the perturbations of the abbreviations above into the action \eqref{ED_action_sigma_d}, we have perturbations of the $\tilde{L}_{\rm ED}$
\begin{equation}
\begin{aligned}
  \tilde{L}_{\rm ED} =&
  \sum_{n=0}^{\infty} \bigg[
    - \frac{1}{4} (\eta-h)^{\mu\nu} \sum_{m=0}^{n} \tr\big[ h^{m} \partial_{\mu}h h^{n-m} \partial_{\nu} h\big]
  	+ \frac{1}{2} \big[\partial_{\nu}h h^{n} \partial_{\mu}h\big]^{\mu\nu}
  \\&\qquad\
  	+ (\eta-h)^{\mu\nu} \partial_{\mu} f\ \tr\big[h^{n}\partial_{\nu}h\big]
  \bigg]
  - (D-2) (\eta-h)^{\mu\nu} \partial_{\mu}f \partial_{\nu}f\,.
\end{aligned}\label{pert_action_1}
\end{equation}
We can rearrange $\tilde{L}_{\rm EH}$ order by order in $h_{\mu\nu}$ and $f$,  $\tilde{L}_{\rm EH} = \sum_{n=2}^{\infty} \tilde{L}_{n}$, where
\begin{equation}
\begin{aligned}
  \tilde{L}_{2} &= \frac{1}{2} [X_{2}]^{\mu\nu}_{\nu\mu} + \eta^{\mu\nu} [\tilde{Y}_{2}]_{\mu\nu}\,,
  \\
  \tilde{L}_{3} &= \frac{1}{2} [X_{3}]^{\mu\nu}_{\nu\mu}
  + \eta^{\mu\nu} [\tilde{Y}_{3}]_{\mu\nu}
  - h^{\mu\nu} [\tilde{Y}_{2}]_{\mu\nu}\,,
\end{aligned}\label{tilde_L_order_23}
\end{equation}
and for $n\geq 4$ we have the general term of the perturbative expansion
\begin{equation}
\begin{aligned}
  \tilde{L}_{n} =&\
   \frac{1}{2} [X_{n}]^{\mu\nu}_{\nu\mu} + \eta^{\mu\nu} [\tilde{Y}_{n}]_{\mu\nu} - h^{\mu\nu} [\tilde{Y}_{n-1}]_{\mu\nu}  \,.
\end{aligned}\label{tilde_L_order_n}
\end{equation}
At this level, the structure of the above perturbations are the same as the pure gravity case. As discussed in section \ref{Sec:2.1}, their difference arises in solving the EoM of $d$\footnote{Using the perturbations of EoM of $d$ \eqref{pert_EoM_d}, $\partial_{\mu} f$ has the on-shell value as
\begin{equation}
  \partial_{\mu}f = \partial_{\mu}\psi + \frac{1}{2(D-2)}(\eta-h)^{\mu\nu} \sum_{n=0}^{\infty} \tr \big[h^{n} \partial_{\nu}h\big]\,,
\label{}\end{equation}
where $\psi$ is fluctuations of the dilaton $\phi$. If we ignore $\psi$, it gives the perturbations of pure GR.
}.
 The explicit forms of \eqref{tilde_L_order_23} and \eqref{tilde_L_order_n} are given by
\begin{equation}
\begin{aligned}
  \tilde{L}_{2} &= - \frac{1}{4} \tr\big[ \partial^{\mu}h \partial_{\mu} h\big]
  	+ \frac{1}{2} \big[\partial_{\nu}h \partial_{\mu}h\big]^{\mu\nu}
  	+ \partial^{\mu} f\ \tr\big[\partial_{\mu}h\big]
  - (D-2) \partial^{\mu}f \partial_{\mu}f\,,
	\\
  \tilde{L}_{3} &= - \frac{1}{4} \tr\big[ \partial^{\mu}h h \partial_{\mu} h\big]
	- \frac{1}{4} \tr\big[ h \partial^{\mu}h \partial_{\mu} h\big]
	+ \frac{1}{4} h^{\mu\nu} \tr\big[ \partial_{\mu}h \partial_{\nu} h\big]
  	+ \frac{1}{2} \big[\partial_{\nu}h h \partial_{\mu}h\big]^{\mu\nu}
  \\&\qquad\
  	+ \partial^{\mu} f\ \tr\big[h\partial_{\mu}h\big]
	-h^{\mu\nu} \partial_{\mu} f\ \tr\big[\partial_{\nu}h\big]
  + (D-2) h^{\mu\nu} \partial_{\mu}f \partial_{\nu}f \,,
\end{aligned}\label{}
\end{equation}
and for $n\geq4$,
\begin{equation}
\begin{aligned}\label{}
  \tilde{L}_{n} &=  - \frac{1}{4} \sum_{m=0}^{n-2} \tr\big[ h^{m} \partial^{\mu}h h^{n-m-2} \partial_{\mu} h\big]
	+ \frac{1}{4} h^{\mu\nu} \sum_{m=0}^{n-3} \tr\big[ h^{m} \partial_{\mu}h h^{n-m-3} \partial_{\nu} h\big]
	\\&\qquad\
  	+ \frac{1}{2} \big[\partial_{\nu}h h^{n-2} \partial_{\mu}h\big]^{\mu\nu}
  	+ \partial^{\mu} f\ \tr\big[h^{n-2}\partial_{\mu}h\big]
	-h^{\mu\nu} \partial_{\mu} f\ \tr\big[h^{n-3}\partial_{\nu}h\big] \,.
\end{aligned}\label{}
\end{equation}

We now consider perturbations of $\tilde{G}^{\mu\nu}$ and $F$ defined in \eqref{tildeGmunu} and \eqref{EoM_F}, respectively. Substituting the perturbations of the supplementary fields \eqref{pert_auxiliary_2}, we obtain the expansion order by order in $h$ and $f$,
\begin{equation}
  \tilde{G}^{\mu\nu} = \sum_{n>0} \tilde{G}^{\mu\nu}_{n}\,,
  \qquad
  F = \sum_{n>0} F_{n}\,,
\label{}\end{equation}
where the linear order terms are
\begin{equation}
\begin{aligned}
  \tilde{G}^{\mu \nu}_{1} &= \partial^{(\mu}\partial_{\rho}h^{\nu)\rho}
  - \frac{1}{2} \Box h^{\mu\nu}
  + \eta^{\mu\nu} \Box f\,,
  \\
  F_{1} &= 2(D-2) \Box f
  - \partial^{\mu} \big( \tr \big[\partial_{\mu}h\big] \big) \,.
\end{aligned}
\end{equation}
The second- and third-order terms are
\begin{equation}
\begin{aligned}
  \tilde{G}^{\mu \nu}_{2} &=  \big[\partial^{(\mu} h \partial_{\rho}h\big]^{\rho|\nu)}
- h^{\rho(\mu}
  	\partial_{\rho}\partial_{\sigma}h^{\nu)\sigma}
   - \frac{1}{2} \big[\partial^{\rho}h \partial_{\rho} h\big]^{\mu\nu}
  + \frac{1}{2} h^{\rho\sigma} \partial_{\rho}\partial_{\sigma}h^{\mu\nu}
  \\&\quad
  - \frac{1}{4} \tr\big[\partial^{\mu}h \partial^{\nu} h\big]
  + \partial^{(\mu} h \partial^{\nu)}f
  - (D-2) \partial^{\mu}f \partial^{\nu} f
  - h^{\mu\nu} \Box f
  - \frac{1}{2} \partial_{\sigma}h^{\rho\mu} \partial_{\rho}h^{\sigma\nu}
  \\&\quad
  - \eta^{\mu\nu} \Big[ h^{\rho\sigma} \partial_{\rho} \partial_{\sigma} f + \partial_{\rho} h^{\rho\sigma} \partial_{\sigma} f  \Big]\,,
  \\
  F_{2} &= - 2(D-2) \partial_{\mu}\big( h^{\mu\nu} \partial_{\nu} f \big)
  - \partial^{\mu} \big( \tr \big[h \partial_{\mu}h\big] \big)
  + \partial_{\mu} \big( h^{\mu\nu} \tr \big[\partial_{\nu}h\big] \big)
  + \frac{1}{2} \partial_{\rho}h^{\rho\sigma} \partial_{\sigma}h^{\mu\nu}\,.
\end{aligned}
\end{equation}
and
\begin{equation}
\begin{aligned}
  \tilde{G}^{\mu \nu}_{3} &= \big[\partial^{(\mu} h h \partial_{\rho}h\big]^{\rho|\nu)}
  - h^{\rho(\mu} \big[\partial_{\rho} h \partial_{\sigma}h\big]^{\sigma|\nu)}
  - \frac{1}{2} \big[\partial^{\rho}h h \partial_{\rho} h\big]^{\mu\nu}
  + \frac{1}{2} h^{\rho\sigma} \big[\partial_{\rho}h \partial_{\sigma} h\big]^{\mu\nu}
  \\&\quad
  - \frac{1}{2} \tr\big[h \partial^{(\mu}h \partial^{\nu)} h\big]
  + \frac{1}{2}  h^{\rho(\mu} \tr\big[\partial_{\rho}h \partial^{\nu)} h\big]
  + \tr\big[h \partial^{(\mu} h\big]  \partial^{\nu)}f
  - h^{\rho(\mu} \partial_{\rho} h \partial^{\nu)}f
  \\&\quad
  - h^{\rho(\mu} \partial^{\nu)} h \partial_{\rho}f
  + 2 (D-2) h^{\rho(\mu} \partial_{\rho}f \partial^{\nu)} f
  + h^{\mu\nu} \Big[ h^{\rho\sigma} \partial_{\rho} \partial_{\sigma} f + \partial_{\rho} h^{\rho\sigma} \partial_{\sigma} f  \Big]\,,
  \\
  F_{3} &= - \partial^{\mu} \big( \tr \big[h^{2} \partial_{\mu}h\big] \big)
	+ \partial_{\mu} \big( h^{\mu\nu} \tr \big[h \partial_{\nu}h\big] \big) \,.
\end{aligned}
\end{equation}
For $n\geq4$, we have the general terms $\tilde{G}^{\mu \nu}_{n}$ and $F_{n}$
\begin{equation}
\begin{aligned}
  \tilde{G}^{\mu \nu}_{n} &=
    \big[\partial^{(\mu} h h^{n-2} \partial_{\rho}h\big]^{\rho|\nu)}
  - h^{\rho(\mu} \big[\partial_{\rho} h h^{n-3} \partial_{\sigma}h\big]^{\sigma|\nu)}
  - \frac{1}{2} \big[\partial^{\rho}h h^{n-2} \partial_{\rho} h\big]^{\mu\nu}
  \\&\quad
  + \frac{1}{2} h^{\rho\sigma} \big[\partial_{\rho}h h^{n-3} \partial_{\sigma} h\big]^{\mu\nu}
  + \tr\big[h^{n-2} \partial^{(\mu} h\big]  \partial^{\nu)}f
  + h^{\mu\rho} h^{\nu\sigma} \tr\big[h^{n-4} \partial_{(\rho} h\big]  \partial_{\sigma)}f
  \\&\quad
  - \frac{1}{4} \sum_{m=0}^{n-2} \tr\big[h^{m}\partial^{\mu}h h^{n-m-2} \partial^{\nu} h\big]
  + \frac{1}{2}  h^{\rho(\mu} \sum_{m=0}^{n-3} \tr\big[h^{m}\partial_{\rho}h h^{n-m-3} \partial^{\nu)} h\big]
 \\&\quad
  - \frac{1}{4}  h^{\mu\rho} h^{\nu\sigma} \sum_{m=0}^{n-4} \tr\big[h^{m}\partial_{\rho}h h^{n-m-4} \partial_{\sigma} h\big]
  \\&\quad
  - h^{\rho(\mu} \Big(
  	  \tr\big[h^{n-3} \partial_{\rho} h\big]  \partial^{\nu)}f
  	+ \tr\big[h^{n-3} \partial^{\nu)} h\big]  \partial_{\rho}f
  \Big)\,,
  \\
  F_{n} &= - \partial^{\mu} \big( \tr \big[h^{n-1} \partial_{\mu}h\big] \big)
	+ \partial_{\mu} \big( h^{\mu\nu} \tr \big[h^{n-2} \partial_{\nu}h\big] \big) \,.
\end{aligned}\label{}
\end{equation}
%

\section{Expansion in Anonther Convention}\label{Sec:B}

Let us consider the perturbations in different conventions. Substituting the mapping \eqref{mapping_usual} into $X^{\mu\nu}_{\rho\sigma}$, $Y_{\mu\nu}$, $Z^{\mu\nu}$ and $W$, we calculate the perturbations up to fourth order (here we drop the primes in $h'$ for simplicity)
\begin{equation}
\begin{aligned}
  X^{\mu \nu}_{\rho \sigma} =&
   \ \partial_{\rho} \partial_{\sigma} h^{\mu \nu}
  - \big[\partial_{\sigma} h \partial_{\rho} h\big]^{\mu \nu}
  - \big[\partial_{\rho} \partial_{\sigma} h h \big]^{\mu \nu}
  - \big[ h \partial_{\rho} \partial_{\sigma} h \big]^{\mu \nu}
  + \big[\partial_{\rho} \partial_{\sigma} h h^{2} \big]^{\mu \nu}
  \\&
  + \big[h^2 \partial_{\rho} \partial_{\sigma} h \big]^{\mu \nu}
  + \big[h \partial_{\sigma} h \partial_{\rho} h \big]^{\mu \nu}
  + \big[h \partial_{\rho} \partial_{\sigma} h h \big]^{\mu \nu}
  + \big[\partial_{\sigma} h \partial_{\rho} h h \big]^{\mu \nu}
  \\&
  + \big[\partial_{\sigma} h h \partial_{\rho} h \big]^{\mu \nu}
  - \big[h \partial_{\sigma} h \partial_{\rho} h \big]^{\mu \nu}
  - \big[\partial_{\sigma} h h \partial_{\rho} h \big]^{\mu \nu}
  - \big[\partial_{\rho} \partial_{\sigma} h h^3 \big]^{\mu \nu}
  \\&
  - \big[h^3 \partial_{\rho} \partial_{\sigma} h \big]^{\mu \nu}
  - \big[ h \partial_{\rho} \partial_{\sigma} h h^2 \big]^{\mu \nu}
  - \big[h^2 \partial_{\rho} \partial_{\sigma} h h\big]^{\mu \nu}
  - \big[h^2 \partial_{\sigma} h  \partial_{\rho} h\big]^{\mu \nu}
  \\&
  - \big[\partial_{\sigma} h \partial_{\rho} h h^2\big]^{\mu \nu}
  - \big[\partial_{\sigma} h h^2 \partial_{\rho} h\big]^{\mu \nu}
  - \big[h \partial_{\sigma} h h \partial_{\rho} h\big]^{\mu \nu}
  + \mathcal{O}(h^5)\,,
\end{aligned}\label{}
\end{equation}
\begin{equation}
\begin{aligned}
  Y_{\mu \nu} =&
  -\frac{1}{4} \tr\big[\partial_{\mu} h \partial_{\nu} h\big]
  + \frac{1}{4(D-2)} \partial_{\mu}h \partial_{\nu}h
  + \frac{1}{2} \tr\big[ h \partial_{\mu} h \partial_{\nu} h\big]
   \\&
  - \frac{1}{2(D-2)} \tr\big[h\partial_{(\mu} h\big] \partial_{\nu)}h
  - \frac{1}{2} \tr\big[ h^2 \partial_{\mu} h \partial_{\nu} h\big]
  - \frac{1}{4} \tr\big[h \partial_{\mu} h  h \partial_{\nu} h\big]
  \\&
  + \frac{1}{2(D-2)} \tr\big[h^{2}\partial_{(\mu} h\big] \partial_{\nu)}h
  + \frac{1}{4(D-2)} \tr[h \partial_{\mu} h ] \tr[h \partial_{\nu} h]  + \mathcal{O}(h^5)\,,
\end{aligned}\label{}
\end{equation}
\begin{equation}
\begin{aligned}
 Z_{\mu \nu} =&\
    \frac{1}{2} \partial_{\rho} h_{\mu \nu}  \partial_{\sigma} h^{\rho \sigma}
  - \frac{1}{2} \partial_{\rho} h_{\mu \sigma} \partial_{\sigma} h_{\nu \rho}
  - \frac{1}{2} \partial_{\rho}h_{\mu \nu} [ h \partial_{\sigma} h]^{(\rho \sigma)}
  + \partial^{\rho} h_{\sigma(\mu} [h \partial_{\sigma} h]_{\nu)\rho}
  \\&
  - [h \partial_{\sigma} h]_{(\mu \nu)} \partial_{\rho} h^{\rho \sigma}
  + \partial^{\rho} h_{\sigma(\mu} [ \partial^{\sigma} h h]_{\nu) \rho}
  - [ h \partial^{\rho} h]_{\sigma (\mu} [h \partial^{\sigma} h]_{\nu)\rho}
   \\&
  - [h \partial^{\sigma} h h]_{\rho(\mu} \partial^{\rho}h_{\nu)\sigma}
  + [ h \partial^{\rho} h]_{(\mu \nu)} [h \partial^{\sigma} h]_{\rho \sigma}
  + [ h \partial^{\rho} h]_{(\mu \nu)} [h \partial^{\sigma} h]_{\sigma \rho} \\
   &
    + \frac{1}{2} \partial_{\rho} h^{\rho \sigma} [ h \partial_{\sigma} h h ]_{\mu \nu}
    - \frac{1}{2} [ h \partial^{\rho} h]_{\mu \sigma} [h \partial^{\sigma} h ]_{\nu \rho} - \partial^{\rho} h_{\sigma (\mu} [\partial^{\sigma} h h^2]_{\nu) \rho} \\
    & - \frac{1}{2} [ h \partial^{\rho} h]_{\sigma \mu} [ h \partial^{\sigma} h]_{\rho \nu}
    + \frac{1}{2} \partial_{\rho} h_{\mu \nu} [ h^2 \partial_{\sigma} h]^{(\rho \sigma)}
    + \frac{1}{4} \partial_{\rho} h_{\mu \nu} [ h \partial_{\sigma} h]^{\rho \sigma} \\
   & - \partial^{\rho} h_{\sigma(\mu} [h^2 \partial^{\sigma} h]_{\nu)\rho}
   + [h^2 \partial^{\sigma} h]_{(\mu \nu)} \partial^{\rho} h_{\rho \sigma}
  + \mathcal{O}(h^5)\,,
\end{aligned}\label{}
\end{equation}
\begin{equation}
\begin{aligned}
  W =\
  \frac{1}{2(D-2)} &\bigg[ \square h
  - \tr[ \partial_{\mu} h \partial^{\mu} h]
  - \tr[ h \square h]
  - \tr[\partial_{\mu} h] \partial_{\nu} h^{\mu \nu}
  - h^{\mu \nu} \tr[\partial_{\mu} \partial_{\nu} h]
  \\&
  + \tr[h^2 \square h]
  + \big[h^{2}\big]^{\mu \nu} \tr[ \partial_{\mu} \partial_{\nu} h]
  + h^{\mu \nu} \tr[\partial_{\mu} h \partial_{\nu} h]
  + 2 \tr[h \partial^{\mu} h \partial_{\mu} h]
  \\&
  + 2 \partial_{\mu} h [h \partial_{\nu} h]^{(\mu \nu)}
  + 2\tr[h \partial_{\mu} h]  \partial_{\nu} h^{\mu \nu}
  + h^{\mu \nu} \tr[ h \partial_{\mu} \partial_{\nu} h]
  \\&
  - \tr [h^3 \square h]
  - \big[h^3\big]^{\mu \nu} \tr[ \partial_{\mu} \partial_{\nu} h]
  - \big[h^{2}\big]^{\mu \nu} \Big(\tr[ \partial_{\mu} h \partial_{\nu} h] + \tr[ h \partial_{\mu} \partial_{\nu} h] \Big)
  \\&
  - h^{\mu \nu} \tr[ h^2 \partial_{\mu} \partial_{\nu} h]
  - \partial_{\mu} h \Big(2\big[ h^2 \partial_{\nu} h \big]^{(\mu \nu)} + [ h \partial_{\nu} h h]^{\mu \nu} \Big)
  - 2 \tr[h^2 \partial^{\mu} h \partial_{\mu} h]
  \\&
  -2 \tr[ h \partial_{\mu} h] [h\partial_{\nu} h]^{(\mu \nu)}
  - \tr[ h \partial^{\mu} h  h \partial_{\mu} h]
  - \tr[ h^2 \partial_{\nu} h] \partial_{\mu} h^{\mu \nu}
  \\&
   - 2 h^{\mu \nu} \tr[ h \partial_{\mu} \partial_{\nu} h] + \mathcal{O}(h^5)
   \bigg] \,.
\end{aligned}\label{}
\end{equation}

Let us consider metric perturbations in the exponential form \eqref{exponential_perturbations}. We also present the expansion of $X^{\mu\nu}_{\rho\sigma}$, $Y_{\mu\nu}$, $Z^{\mu\nu}$ and $W$ up to fourth order in fluctuations (here we drop the primes in $h'$ for simplicity)
\begin{equation}
\begin{aligned}
 X^{\mu \nu}_{\rho \sigma} =&\
    \partial_{\rho} \partial_{\sigma} h^{\mu \nu}
  + [\partial_{[\rho} h \partial_{\sigma]} h]^{\mu \nu}
  - \frac{1}{2} [ \partial_{\rho} \partial_{\sigma} h h ]^{\mu \nu}
  - \frac{1}{2} [h \partial_{\rho} \partial_{\sigma} h]^{\mu \nu}
  + \frac{1}{6} [ \partial_{\sigma} h \partial_{\rho} h]^{\mu \nu}
  \\&
  + \frac{1}{6} [ \partial_{\rho} \partial_{\sigma} h h^2]^{\mu \nu}
  + \frac{1}{6} [ h^2 \partial_{\rho} \partial_{\sigma} h ]^{\mu \nu}   + \frac{1}{6} [ h \partial_{\sigma} h \partial_{\rho} h]^{\mu \nu}
  - \frac{1}{3} [ h \partial_{\rho} h \partial_{\sigma} h]^{\mu \nu}
  \\&
  + \frac{1}{6} [h \partial_{\rho} \partial_{\sigma} h h]^{\mu \nu}
  - \frac{1}{3} [ \partial_{\rho} h \partial_{\sigma} h h]^{\mu \nu}
  + \frac{1}{3} [\partial_{(\rho} h h \partial_{\sigma)} h]^{\mu \nu}
  + \frac{1}{8} [ h^2 \partial_{\rho} h \partial_{\sigma} h]^{\mu \nu}
  \\&
  + \frac{1}{8} [ \partial_{\rho} h \partial_{\sigma} h h^2]^{\mu \nu}
  - \frac{1}{24} [ \partial_{\sigma} h \partial_{\rho} h h^2]^{\mu \nu}
  + \frac{1}{12} [ \partial_{[\rho} h h^2 \partial_{\sigma]} h]^{\mu \nu}
  \\&
  - \frac{1}{24} [ h \partial_{\sigma} h h \partial_{\rho} h]^{\mu \nu}
  - \frac{1}{8} [h \partial_{\rho} h h \partial_{\sigma} h]^{\mu \nu}
  - \frac{1}{24} [h \partial_{\sigma} h \partial_{\rho} h h]^{\mu \nu}
  + \frac{5}{24} [h \partial_{\rho} h \partial_{\sigma} h h]^{\mu \nu}
  \\&
  - \frac{1}{8} [\partial_{\rho} h h \partial_{\sigma} h h]^{\mu \nu}
  - \frac{1}{24} [\partial_{\sigma} h h \partial_{\rho} h h]^{\mu \nu}
  - \frac{1}{24} [ \partial_{\rho} \partial_{\sigma} h h^3]^{\mu \nu}
  - \frac{1}{24} [ h^3 \partial_{\rho} \partial_{\sigma} h]^{\mu \nu}
  \\&
  - \frac{1}{24} [h \partial_{\rho} \partial_{\sigma} h h^2]^{\mu \nu}  - \frac{1}{24} [h^2 \partial_{\rho} \partial_{\sigma} h h]^{\mu \nu}- \frac{1}{24} [h^2 \partial_{\sigma} h \partial_{\rho} h]^{\mu \nu}  + \mathcal{O}(h^5)\,,
\end{aligned}\label{}
\end{equation}
\begin{equation}
\begin{aligned}
 Y_{\mu \nu} & =
  - \frac{1}{4} \tr[\partial_{\mu} h \partial_{\nu} h]
  + \frac{1}{4(D-2)} \tr[\partial_{\mu} h] \tr[\partial_{\nu}h]
  - \frac{1}{24} \tr[ h^2 \partial_{\mu} h \partial_{\nu} h]
  \\&\quad
  + \frac{1}{24} \tr[ \partial_{\mu} h h \partial_{\nu} h h]
  + \mathcal{O}(h^5)\,,
\end{aligned}\label{}
\end{equation}
\begin{equation}
\begin{aligned}
  Z_{\mu \nu} =&\
    \frac{1}{2} \partial_{\rho} h_{\mu \nu}  \partial_{\sigma} h^{\rho \sigma}
  - \frac{1}{2} \partial^{\rho} h_{\mu \sigma} \partial^{\sigma} h_{\nu \rho}
  - \frac{1}{2} \partial^{\rho} h_{\mu \nu} [ h \partial^{\sigma} h]_{(\rho \sigma)}
  + \frac{1}{2} \partial^{\rho} h_{\sigma(\mu} [h \partial^{\sigma} h]_{\nu)\rho}
  \\&
  - \frac{1}{2} [h\partial^{\sigma} h]_{(\mu \nu)} \partial^{\rho} h_{\rho \sigma}
  + \frac{1}{2} [h \partial^{\sigma} h]_{\rho(\mu} \partial^{\rho} h_{\nu)\sigma}
  - \frac{1}{4} [h \partial^{\rho} h]_{\sigma(\mu} [ h \partial^{\sigma} h]_{\nu)\rho}
  \\&
  - \frac{1}{6} \partial^{\rho} h_{\sigma(\mu} [h \partial^{\sigma} h h]_{\nu)\rho}
  + \frac{1}{4} [h \partial^{\rho} h]_{(\mu \nu)} [ h \partial^{\sigma} h]_{\rho \sigma}
  + \frac{1}{4} [h \partial^{\rho} h]_{(\mu \nu)} [h \partial^{\sigma} h]_{\sigma \rho}
  \\&
  + \frac{1}{12} \partial_{\rho} h^{\rho \sigma} [ h \partial_{\sigma} h h]_{\mu \nu}
  - \frac{1}{8} [h \partial^{\rho} h]_{\mu \sigma} [h \partial^{\sigma} h]_{\nu \rho}
  - \frac{1}{6} [h^2 \partial^{\sigma} h]_{\rho (\mu} \partial^{\rho} h_{\nu)\sigma}
  \\&
  - \frac{1}{8} [h\partial^{\rho} h]_{\sigma \mu} [h \partial_{\sigma} h]_{\rho \nu}
  + \frac{1}{6} [h^2 \partial^{\sigma} h]_{(\rho \sigma)} \partial^{\rho} h_{\mu \nu}
  + \frac{1}{12} \partial^{\rho} h_{\mu \nu} [h \partial^{\sigma} h h]_{\rho \sigma}
  \\&
  - \frac{1}{6} \partial^{\rho} h_{\sigma(\mu} [h^2 \partial^{\sigma} h]_{\nu) \rho}
  + \frac{1}{6} \partial^{\rho}h_{\rho \sigma} [h^2 \partial^{\sigma} h]_{(\mu \nu)}
  + \mathcal{O}(h^5)\,,
\end{aligned}\label{}
\end{equation}
\begin{equation}
\begin{aligned}
2(D-2) W & = \square h
- h^{\mu \nu} \tr[\partial_{\mu \nu} h]
- \tr[\partial_{\mu} h] \partial_{\nu} h^{\mu \nu} \\
& + \frac{1}{2} (h^2)^{\mu \nu} \tr[\partial_{\mu} \partial_{\nu} h]
+ \frac{1}{2} \tr[\partial_{\mu} h] [h \partial_{\nu} h]^{\mu \nu}
+ \frac{1}{2} \tr[\partial_{\mu} h] [h \partial_{\nu} h]^{\nu \mu}  \\
& - \frac{1}{6} (h^3)^{\mu \nu} \tr[\partial_{\mu \nu} h]
- \frac{1}{6} \tr[\partial_{\mu} h] [h^2 \partial_{\nu} h]^{\mu \nu}
- \frac{1}{6} \tr[\partial_{\mu} h] [h^2 \partial_{\nu} h]^{ \nu\mu }  \\
& - \frac{1}{6} \tr[\partial_{\mu} h] [h \partial_{\nu} h]^{\mu \nu}   + \mathcal{O}(h^5) \,.
\end{aligned}\label{}
\end{equation}
%

\section{Explicit Form of the Graviton Off-Shell Currents}\label{App:C}
In this appendix, we present the graviton currents $J^{\mu \nu}_{\mathcal{P}}$ explicitly up to rank-4 by solving the off-shell recursion using the perturbiner method. We derive the rank-2 in a general helicity, and we employ the spinor-helicity basis for the rank-3 and rank-4 currents.

Here is our convention of the choice of reference momenta: for the one opposite helicity case is
\begin{equation}
\begin{aligned}
  (1-,2+,3+,4+,\cdots) &: \quad \begin{cases} \text{for}~ \epsilon^{1-}_{\mu}\,,~ q_{\mu} = k^{2}_{\mu}
  \\
  \text{for}~ \epsilon^{i+} \,,~ q_{\mu} = k^{1}_{\mu}\,,\quad i\geq 2
   \end{cases}
\end{aligned}\label{one_opposite}
\end{equation}
and for the two opposite helicity case 
\begin{equation}
\begin{aligned}
     (1-,2-,3+,4+,\cdots) & :  \quad \begin{cases} \text{for}~ \epsilon^{1-}_{\mu}\,,~ q^{1}_{\mu} = k^{3}_{\mu}
  \\
  \text{for}~ \epsilon^{2-}_{\mu}\,,~ q^{2}_{\mu} = k^{3}_{\mu} \\
  \text{for}~ \epsilon^{i+} \,,~ q_{\mu} = k^{1}_{\mu}\,,\quad i\geq 3
   \end{cases}
\end{aligned}\label{two_opposite}
\end{equation}

We further define a set of abbreviations for a compact notation as follows:
\begin{equation}
\begin{aligned}
    &P_{ijk} =  k_i \cdot \epsilon_j (k_i \cdot \epsilon_k + k_j \cdot \epsilon_k) \,,
    \qquad
  P^{A}_{ijk} = (k_i \cdot \epsilon_j) (k_{j} \cdot \epsilon_k) - (k_{i} \cdot \epsilon_{k})(k_{k} \cdot \epsilon_j) \,,
  \\& 
  Q_{ijk} = (k_i \cdot \epsilon_j)(\epsilon_i \cdot \epsilon_k)
  \qquad
  S_{ijk} = (k_{i} \cdot \epsilon_{k})+ (k_{j} \cdot \epsilon_k) 
  \\&
  T_{ijk} = \frac{(k_i \cdot \epsilon_k)^2}{s_{ik}} + \frac{(k_j \cdot \epsilon_k)^2}{s_{jk}}\,,
  \qquad
  T_{ijk,l} = \frac{\left(k_i\cdot \epsilon _k\right)^2}{s_{ik}}+\frac{\left(s_{jk}+s_{jl}\right) \left(k_j\cdot \epsilon _k\right)^2}{s_{jk} s_{jkl} }\,,
\end{aligned}\label{}
\end{equation}
%

\subsection{Solutions of the recursion}

We now present the explicit form of the currents up to rank 4, which corresponds to the 5-point graviton scattering amplitudes.

\paragraph{Rank-2}~
\\
The rank $2$ current $J^{\mu\nu}_{ij}$ is easily determined by solving the recursion relation \eqref{recursions} with the initial condition \eqref{initial_condition}
 \begin{equation}
\begin{aligned}
 J^{\mu \nu}_{ij}
  &= \frac{1}{4 s_{ij}}
  \Bigg[ 
  (\epsilon_{i} {\cdot}  \epsilon_{j})^2  \Big(5 s_{ij}\eta^{\mu\nu} +k_{i}^{(\mu} k_{j}^{\nu)}\Big)
  -4 \Big((k_{i} {\cdot} \epsilon_{j}) \epsilon_{i}^{\mu} - (k_{j} {\cdot} \epsilon_{i}) \epsilon_{j}^{\mu}\Big)\Big((k_{i} {\cdot} \epsilon_{j}) \epsilon_{i}^{\nu} - (k_{j} {\cdot} \epsilon_{i}) \epsilon_{j}^{\nu}\Big)
  \\&\qquad\quad 
  - (\epsilon_{i} {\cdot} \epsilon_{j}) \left( 2 (k_{i} {\cdot} \epsilon_{j}) k_{j}^{(\mu} \epsilon_{i}^{\nu)}
  + 2 (k_{j} {\cdot} \epsilon_{i}) k_{i}^{(\mu} \epsilon_{j}^{\nu)}
  + s_{ij} \epsilon_{i}^{(\mu} \epsilon_{j}^{\nu)} -4\eta^{\mu \nu} (k_{i} {\cdot} \epsilon_{j}) (k_{j} {\cdot} \epsilon_{i})\right) \Bigg]
\end{aligned}\label{}
\end{equation}
From the scattering amplitude point of view, for non-vanishing scattering amplitudes for higher rank, at least two negative helicity condition is required. Hence for rank $3$ and rank $4$ graviton currents, we will focus on one minus and two minus helicity conditions.

\paragraph{Rank-3}~
\\
Without loss of generality, we present two currents which are relevant for the four-point scattering amplitude, $J^{\mu\nu}_{1^{-}2^{+}3^{+}}$ and $J_{1^{-}2^{-}3^{+}}$ associated with the conventions in \eqref{one_opposite} and \eqref{two_opposite} respectively. Note that $J^{\mu\nu}_{1^{+}2^{+}3^{+}}$ and $J^{\mu\nu}_{1^{-}2^{-}3^{-}}$ do not contribute to the amplitude.
\begin{equation}
\begin{aligned}
  & s_{123} J_{\mu \nu}^{1^{-}2^{+}3^{+}}
  \\&\qquad
  = \frac{s_{13}+s_{23}}{s_{13}s_{23}} \left(k_{3} \cdot \epsilon_{1}\right)^{2} \Big(\left(k_{2} \cdot \epsilon_{3}\right) \epsilon_{2}^{\mu}-\left(k_{3} \cdot \epsilon_{2}\right) \epsilon_{3}^{\mu}\Big)\Big(\left(k_{2} \cdot \epsilon_{3}\right) \epsilon_{2}^{\nu}-\left(k_{3} \cdot \epsilon_{2}\right) \epsilon_{3}^{\nu}\Big)\,.
\end{aligned}\label{}
\end{equation}
and
\begin{equation}
\begin{aligned}
    & s_{123} J_{\mu \nu}^{1^{-}2^{-}3^{+}}
  \\&\qquad
  =\frac{\left(s_{12}+s_{23}\right)}{s_{12} s_{23}} \left(k_{2} \cdot \epsilon_{3}\right)^{2} \Big(\left(k_{1} \cdot \epsilon_{2}\right) \epsilon_{1}^{\mu}-\left(k_{2} \cdot \epsilon_{1}\right) \epsilon_{2}^{\mu}\Big)\Big(\left(k_{1} \cdot \epsilon_{2}\right) \epsilon_{1}^{\nu}-\left(k_{2} \cdot \epsilon_{1}\right) \epsilon_{2}^{\nu}\Big)\,,
\end{aligned}\label{}
\end{equation}

\paragraph{Rank-4}~
\\
Again we consider two helicity choices: one-minus and two-minus helicities. The rest of them are trivial. Here we present the contributions to the five-point amplitudes only.

First, the one-minus case, $J^{\mu\nu}_{1^{-} 2^{+} 3^{+} 4^{+}}$ consists with

\begin{equation}
\begin{aligned}
  s_{1234} J_{\mu\nu}^{1234}|_{\epsilon_{2}^{\mu} \epsilon_{2}^{\nu}}
  & = - \frac{\big(P^{A}_{234}\big)^2 }{s_{134}} \bigg[T_{341} + \frac{S_{341}^2(s_{134} + s_{234})}{s_{34} s_{234}}\bigg]  
  \\&\quad
  - \frac{P_{243}^2}{s_{24}}\bigg[ T_{341,2}  + \frac{ S_{341}^2  }{s_{234}} \bigg] 
  - \frac{P_{234}^2}{s_{23}} \bigg[T_{431,2} +\frac{S_{431}^2}{s_{234}}\bigg]\,,
\end{aligned}\label{}
\end{equation}
\begin{equation}
\begin{aligned}
  s_{1234} J_{\mu\nu}^{1234}|_{\epsilon_{2}^{\mu} \epsilon_{3}^{\nu}}&=
  \frac{P_{342} P_{234}^{A}}{s_{134}}   \bigg[  T_{431} + \frac{S_{341}^2(s_{134}+s_{234})}{s_{34}s_{234}} \bigg]
   + \frac{P_{243} P_{324}^{A}}{s_{24}} \bigg[T_{341,2} + \frac{S_{341}^2 }{  s_{234}} \bigg]
  \\&\quad
   + \frac{ P_{324} P_{234} }{ s_{23}} \bigg[T_{431,2} + \frac{S_{341}^2  }{  s_{234}} \bigg]   \,,
\end{aligned}\label{}
\end{equation}
\begin{equation}
\begin{aligned}
  s_{1234} J_{\mu\nu}^{1234}|_{\epsilon_{2}^{\mu} \epsilon_{4}^{\nu}}
  &= \frac{P_{432}P_{243}^A }{ s_{134}} \Bigg[T_{341} + \frac{S_{341}^2 (s_{134} + s_{234})}{s_{34} s_{234}}   \Bigg]
  + \frac{P_{234}P^{A}_{423}}{s_{23}} \bigg[T_{431,2} + \frac{S_{341}^2}{  s_{234}} \bigg] 
  \\&\quad
  + \frac{P_{423}P_{243}}{ s_{24}} \bigg[ T_{341,2} + \frac{ S_{341}^2}{s_{234}} \bigg]\,,
\end{aligned}\label{}
\end{equation}
\begin{equation}
\begin{aligned}
  s_{1234} J_{\mu\nu}^{1234}|_{\epsilon_{3}^{\mu} \epsilon_{3}^{\nu}}
  & =
   -\frac{P_{324}^2 }{ s_{23}} \bigg[ T_{431,2} + \frac{S_{341}^2 }{s_{234}} \bigg] 
  - \frac{P_{324}^{A}{}^2}{s_{24} } \bigg[ T_{341,2} + \frac{S_{431}^2}{s_{234}} \bigg]
 \\&\quad 
 - \frac{P_{342}^2}{s_{134}} \bigg[T_{341}+ \frac{S_{341}^2 (s_{134}+s_{234})}{ s_{34}s_{234}}\bigg]  \,,
\end{aligned}\label{}
\end{equation}
\begin{equation}
\begin{aligned}
  s_{1234} J_{\mu\nu}^{1234}|_{\epsilon_{3}^{\mu} \epsilon_{4}^{\nu}}
  & = \frac{P_{423} P_{342}^A}{s_{24}}\Bigg[ T_{341,2}+\frac{S_{341}^2}{s_{234}} \Bigg] 
  + \frac{ P_{324}P_{432}^A }{ s_{23} } \Bigg[ T_{431,2} + \frac{S_{341}^2}{s_{234}} \Bigg]
  \\&\quad
  +  \frac{P_{432} P_{342}}{ s_{134}} \bigg[ T_{341} + \frac{S_{341}^2(s_{134}+s_{234})}{s_{34}s_{234}} \bigg] \,,
\end{aligned}\label{}
\end{equation}
\begin{equation}
\begin{aligned}
  s_{1234} J_{\mu\nu}^{1234}|_{\epsilon_{4}^{\mu} \epsilon_{4}^{\nu}}
  & =
  -\frac{P_{423}^2 }{ s_{24}}
  \bigg[T_{341,2}+ \frac{S_{341}^2 }{s_{234}} \bigg] 
  - \frac{P_{423}^{A}{}^2}{s_{23}} \bigg[ T_{431,2} + \frac{S_{341}^2}{s_{234}} \bigg]
  \\&\quad 
  - \frac{P_{432}^2 }{ s_{134}}
  \bigg[ T_{341} + \frac{S_{341}^2(s_{134}+s_{234})}{s_{34}s_{234}} \bigg] \,.
\end{aligned}\label{}
\end{equation}

Next, we consider two minus helicity case, $J^{\mu\nu}_{1^{-} 2^{-} 3^{+} 4^{+}}$:
\begin{equation}
\begin{aligned}
 s_{1234} J^{\mu \nu}_{1^{-}2^{-} 3^+4^+}\big|_{\epsilon_{2}^{\mu} \epsilon_{2}^{\nu}}&=
  - \frac{P_{243}^2 }{s_{124}} \bigg[T_{241} + \frac{S_{241}^2(s_{124} + s_{234})}{s_{24} s_{234}}\bigg]  
  \\&\quad
  - \frac{\big(P_{234}^{A}\big)^2}{s_{34}}\bigg[ T_{241,3}  + \frac{S_{241}^2}{s_{234}} \bigg] 
  - \frac{P_{234}^2}{s_{23}}\bigg[ T_{421,3} +\frac{S_{241}^2}{s_{234}}\bigg]\,,
\end{aligned}\label{}
\end{equation}
 \begin{equation}
\begin{aligned}
 s_{1234} J^{\mu \nu}_{1^{-}2^{-} 3^+4^+}{}|_{\epsilon_{2}^{\mu} k_{3}^{\nu}}& = 
 Q_{432} \Bigg[ \frac{P^{A}_{234}}{s_{34}} \bigg(\frac{S_{241}^{2}}{s_{234}}+ T_{241,3}\bigg)
 - \frac{P_{243} }{s_{124}} \bigg( T_{241} + \frac{S_{241}^2(s_{124}+s_{234})}{s_{24}s_{234}} \bigg)
  \Bigg]\,,
\end{aligned}\label{}
\end{equation}
 \begin{equation}
\begin{aligned}
 s_{1234} J^{\mu \nu}_{1^{-}2^{-} 3^+4^+}\big|_{\epsilon_{2}^{\mu} k_{4}^{\nu}} 
 &=
  - \frac{P_{234} Q_{234}}{s_{23}} \Bigg[ T_{421,3} +\frac{S_{241}^2}{s_{234}} \Bigg]
  - \frac{P^{A}_{243} Q_{432}}{s_{34}} \Bigg[ T_{241,3} + \frac{S_{241}^2}{s_{234}} \Bigg] 
  \\&\quad
  - \frac{S_{243}^2 Q_{244}}{s_{124}} \Bigg[ \frac{S_{241}^2(s_{124}+s_{234})}{s_{24}s_{234}} +T_{241} \Bigg] \,,
\end{aligned}\label{}
\end{equation}
 \begin{equation}
\begin{aligned}
 s_{1234} J^{\mu \nu}_{1^{-}2^{-} 3^+4^+}{}|_{k_{2}^{\mu} k_{3}^{\nu}} &= \frac{Q_{432}^2}{ 2}
 \Bigg[ \frac{S_{241}^2 }{s_{24}} \left( \frac{s_{24} + s_{34}}{ s_{34} s_{234}} + \frac{1}{s_{124}}  \right)
  + \frac{(k_2 \cdot \epsilon_1)^2 (s_{34} + s_{124})}{s_{12}s_{34} s_{124}}
  \\& \qquad\qquad
 +\left(\frac{1}{s_{14}s_{134}}+\frac{1}{s_{14}s_{124}}+\frac{1}{s_{34} s_{134}}\right) (k_4 \cdot \epsilon_1)^2 
 \Bigg]\,,
\end{aligned}\label{}
\end{equation}
 \begin{equation}
\begin{aligned}
 s_{1234} J^{\mu \nu}_{1^{-}2^{-} 3^+4^+}{}|_{k_{2}^{\mu} k_{4}^{\nu}}  &= \frac{\left(Q_{243}+Q_{432}\right)^2}{2 s_{124}} \left(\frac{S_{241}^2 \left(s_{124}+s_{234}\right)}{s_{24} s_{234}}+T_{241}\right)
 \\&\quad
 +\frac{Q_{412} T_{243,1}}{2 s_{14}}+\frac{Q_{214} T_{423,1}}{2 s_{12}}+\frac{\left(Q_{214}+Q_{412}\right){}^2 T_{243}}{2 s_{234}}\,,
\end{aligned}\label{}
\end{equation}
\begin{equation}
\begin{aligned}
 s_{1234} J^{\mu \nu}_{1^{-}2^{-} 3^+4^+}{}\big
 |_{k_{3}^{\mu} k_{4}^{\nu}}  
  &= \frac{Q_{234}^{2}}{2}\Bigg[\frac{S_{241}^2}{s_{24}} \left(\frac{s_{23}+s_{24}}{s_{23} s_{234}}+\frac{1}{s_{124}}\right) 
  +  \frac{\left(k_4\cdot \epsilon _1\right)^2}{s_{14}}\left(\frac{1}{s_{124}}+\frac{1}{s_{23}}\right) 
  \\&\qquad\qquad
  +\left( \frac{1}{s_{12}s_{124}}+\frac{1}{s_{12}s_{123}}+\frac{1}{s_{23} s_{123}}\right) \left(k_2\cdot \epsilon _1\right)^2 
  \Bigg]\,.
\end{aligned}\label{}
\end{equation}
%

\newpage
\bibliography{references}
\bibliographystyle{JHEP}

\end{document}